\documentclass{aa}
\usepackage{graphicx}
\usepackage{txfonts}
\usepackage{natbib}
\bibpunct{(}{)}{;}{a}{}{,}

\newcommand{\micron}{$\mu$m~}
\newcommand{\lbol}{$L_{\rm bol}$}

\def\mathstacksym#1#2#3#4#5{\def#1{\mathrel{\hbox to 0pt{\lower#5\hbox{#3}\hss} \raise #4\hbox{#2}}}}
\mathstacksym\gta{$>$}{$\sim$}{1.5pt}{3.5pt} 
\mathstacksym\lta{$<$}{$\sim$}{1.5pt}{3.5pt} 
\begin{document}
\title{Resolved 24.5 micron emission from massive young stellar objects\thanks{Based on data collected at the Subaru telescope, which is operated by the National Astronomical Observatory of Japan.}}
\authorrunning{W.J. de Wit et al.}
\titlerunning{Subaru imaging of MYSOs}
\author{W.J. de Wit\inst{1}, M.G. Hoare\inst{1}, T. Fujiyoshi\inst{2}, R.D. Oudmaijer\inst{1}, M. Honda\inst{3}, H. Kataza\inst{4}, T. Miyata\inst{5}, Y. K. Okamoto\inst{6}, T. Onaka\inst{7}, S. Sako\inst{5}, T. Yamashita\inst{2}}
                                %
%
%
%
%
                                %
\offprints{W.J. de Wit, \email{w.j.m.dewit@leeds.ac.uk}}
\institute{School of Physics \& Astronomy, University of Leeds, Woodhouse Lane, Leeds LS2 9JT, UK\and
  Subaru Telescope, National Astronomical Observatory of Japan, National Institutes of Natural Sciences, 650 North A'ohoku Place, Hilo, HI 96720, USA\and
  Department of Information Science, Kanagawa University, 2946 Tsuchiya, Hiratsuka, Kanagawa, 259-1293, Japan\and
  Department of Infrared Astrophysics, Institute of Space and Astronautical Science, Japan Aerospace Exploration Agency, Sagamihara, Kanagawa, 229-8510, Japan\and
  Institute of Astronomy, University of Tokyo, Osawa 2-21-1, Mitaka, Tokyo 181-0015, Japan\and
  Faculty of Science, Ibaraki University, 2-1-1 Bunkyo, Mito, Ibaraki, 310-8512, Japan\and
  Department of Astronomy, Graduate School of Science, University of Tokyo, Bunkyo-ku, Tokyo 113-0022, Japan
 }          
        
\date{Received date; accepted date}
\abstract
{Massive young stellar objects (MYSO) are surrounded by massive dusty envelopes, whose physical structure and geometry
are determined by  the star formation process.}
{Our principal aim is to establish the density structure of MYSO envelopes on scales of $\sim$1000\,AU. This constitutes an increase
of a factor $\sim$10 in angular resolution compared to similar studies performed in the (sub)mm.}
{We have obtained diffraction-limited (0.6\arcsec) 24.5\,\micron images (field of view of $40\arcsec\times30\arcsec$) of 14 well-known massive star formation 
regions with the COMICS instrument 
mounted on the 8.2 meter Subaru telescope. We construct azimuthally averaged intensity profiles 
of the resolved MYSO envelopes and build spectral energy distributions (SEDs) from archival data and the COMICS 24.5\,\micron flux 
density. The SEDs range from near-infrared to millimeter wavelengths.
Self-consistent 1-D radiative transfer models 
described by a density dependence of the form $n(r) \propto r^{-p}$ are used to 
simultaneously compare the intensity profiles and SEDs to model predictions.}
{The images reveal the presence of 
discrete MYSO sources 
which are resolved on arcsecond scales, and, to
first-order, the observed emission is circular on the sky. 
For many sources, the spherical models are capable of satisfactorily
reproducing the 24.5\,\micron intensity profile, the 24.5\,\micron
flux density, the 9.7\,\micron silicate absorption feature, and the
submm emission. They are described by density distributions with
$p =1.0\pm0.25$. Such distributions are shallower than those found on
larger scales probed with single-dish (sub)mm studies. Other sources
have density laws that are shallower/steeper than $p=1.0$ and there is
evidence that these are viewed near edge-on or near face-on 
respectively. In these cases spherical models fail to provide good fits to the data.
The images  also reveal a diffuse component tracing somewhat larger 
scale structures, particularly visible in the 
regions S\,140, AFGL\,2136, IRAS\,20126+4104, Mon\,R2, and Cep\,A.
}
{{We find a flattening of the MYSO density law going from scales probed with single-dish submm observations down to 
scales of $\sim$1000\,AU probed with the observations presented here. We propose that this may be evidence of 
rotational support of the envelope. This finding will be explored further in a future paper using 2-D axisymmetric radiative transfer models.}}
\maketitle
\section{Introduction}
\label{intro}
The early stages in the lives of massive stars are identified with
highly luminous yet very cool objects deeply embedded in molecular
clouds. Their emission is characterised by a steeply rising spectral
energy distribution (SED) peaking around 100\,$\mu$m, which features 
strong silicate absorption. These properties are indicative of radiation
reprocessed by a dusty circumstellar envelope. Measured bolometric
luminosities are such that if a single main sequence star resides at the heart of
the dusty envelope, it should have the ability to ionise its surroundings,
yet only little (if any) recombination radiation is observed. This
radio-quiet appearance of massive young stellar objects (MYSO) is unlike that of
ultra-compact (UC) \ion{H}{ii} regions (e.g. Hoare et
al. 2007\nocite{2007prpl.conf..181H}), and the latter can therefore be considered as a
successor phase. The reason may be that the MYSO is actively
accreting material from its surrounding environment quenching the development 
of an \ion{H}{ii} region. The ongoing accretion would also give rise to
bipolar outflow activity ubiquitously observed in massive star forming
regions. It is clear that MYSOs are prime observational targets for addressing 
outstanding issues in our current understanding of massive star formation.

In this paper we address the radial structure of the MYSO dust envelope.
It is determined by the forces
that operate during the onset and the subsequent evolution of the
initial molecular core. The radial density profile is
predicted to be a powerlaw with a value for the power index that
depends on the dominant physics. The exact power index can be extracted from the 
observables by the use of radiative transfer models.  Spherical envelope models may be
assumed for the dust that dominates the emission at wavelengths larger than
$30\,\mu$m. At these wavelengths the SED of MYSOs (but also UC\ion{H}{ii}s) are observed
to be remarkably similar, arguing for little deviation from spherical 
symmetry (Chini et al 1986; Churchwell, Wolfire \& Wood
1990\nocite{1990ApJ...354..247C}; Henning et al 1990\nocite{1990A&A...227..542H}; Hoare et al. 1991\nocite{1991MNRAS.251..584H}; G\"{u}rtler et
al. 1991\nocite{1991A&A...252..801G}; Wolfire \& Churchwell 1994\nocite{1994ApJ...427..889W}; Faison et
al. 1998\nocite{1998ApJ...500..280F}; Hatchell et al. 2000\nocite{2000A&A...357..637H};
Van der Tak et al. 2000\nocite{2000ApJ...537..283V}; Mueller et al. 2002\nocite{2002ApJS..143..469M};
Beuther et al. 2002\nocite{2002ApJ...566..945B}).
At shorter wavelengths ($\lambda <30\,\mu$m) the geometry of the envelope
becomes important (Yorke \& Sonnhalter 2002\nocite{2002ApJ...569..846Y}; Whitney et al. 2003).
For example, under
favourable inclinations, mid-IR radiation can be 
observed to originate directly from the surface of cavity walls that are
sculpted by the polar outflows. In this case, the
mid-IR photons are emitted by warm dust particles that have a clear
line-of-sight to the star (e.g. De Buizer
2007\nocite{2007ApJ...654L.147D}). At even shorter wavelengths, near-IR photons from the (generally monopolar)
reflection nebulae (e.g. Alvarez et al. 2004, 2005\nocite{2005A&A...440..569A}) may originate
either from the stellar surface, an inner dust truncation structure or
from an accretion disk. They can scatter and escape through existing
inhomogeneities in the spherical envelope (e.g. G\"{u}rtler et
al. 1991; Henning et al. 2000\nocite{2000A&A...353..211H}) and still
suffer extinction from any foreground molecular cloud material
(e.g. De Buizer, Osorio \& Calvet 2005\nocite{2005ApJ...635..452D}).

Here, we aim to constrain the radial density distribution 
on scales of 1000\,AU using resolved 24.5\,\micron emission. This
constitutes an increase of a factor 10 in angular resolution compared to similar
studies performed in the (sub)mm. We present diffraction-limited images at 24.5\,$\mu$m, the longest mid-IR
wavelength amenable to high resolution imaging from the ground with
large telescopes. This long mid-IR wavelength maximises the possibility of
resolving the envelope emission because, due to the nature of the
temperature gradients in the optically thick emitting region, the size
of the emission region gets larger with increasing wavelength to the
power of about 1.5, whilst the diffraction limit of the telescope only
increases linearly with wavelength. We have selected a set of 14
well-known MYSO and imaged these at 24.5\,$\mu$m with the 8m Subaru
telescope.  The images have an angular resolution set by the
telescope's diffraction limit of 0.6\arcsec, corresponding to 
linear scales of $\sim$1000\,AU for the average distance of 1.5\,kpc to 
our target MYSOs. Most previous 10\,\micron and 20\,\micron imaging of MYSOs have
been carried out on 4 m class telescopes where the (radio-quiet)
objects are invariably unresolved (e.g. Mottram et al. 2007\nocite{2007A&A...476.1019M}; De
Buizer et al. 2000\nocite{2000ApJS..130..437D}, 2005\nocite{2005ApJS..156..179D}). Some work on 8-10\,m class
telescopes at 8-22\,\micron has begun to resolve a few sources such as
BN and source n in Orion (Shuping et al. 2004\nocite{2004AJ....128..363S}; Greenhill et
al. 2004\nocite{2004ApJ...605L..57G}), but no detailed modelling of these data has been
carried out.

We analyse the resolved emission in
conjunction with the SED in terms of spherical dust radiative transfer
models as calculated by DUSTY, and using background literature information for each
individual case. We present simultaneous model fits to the 24.5\,\micron
intensity profile and the SED, that stretches from the near-IR to the
mm wavelengths. It includes the silicate 9.7\,\micron absorption profile
thanks to ISO-SWS spectra for nearly the whole sample. The approach of simultaneously analysing the
intensity profiles and SEDs follows van der Tak et al. (2000),
Hatchell et al. (2000), Beuther et al. (2002), Mueller et al. (2002),
Hatchell \& van der Tak (2003\nocite{2003A&A...409..589H}), Williams
et al. (2005\nocite{2005A&A...434..257W}).
Most of the quoted work is exclusively aimed at (sub)mm wavelengths, and
thus probing linear scales about a factor 10 larger than in the
mid-IR. Constraints on the density distributions from (sub)mm and
mid-IR are therefore highly complementary as they give insight into
the evolution of the density distribution as function of radius.

This paper is organized as follows. 
Our observational data were taken with the Japanese 8.2 meter Subaru
telescope on Hawaii in conjunction with the COMICS instrument. We give
an overview of the instrument and detail the observations in
Sect.\,\ref{sec-obs}.  We present and discuss the morphology
seen on the COMICS images in the subsections to Sect.\,\ref{mormod} and Sect.\,\ref{compl}, alongside the
simultaneous modelling of the intensity profile and SED.  We summarise the modelling part highlighting
certain trends in Sect.\,\ref{summ}. We
discuss our findings and their consequences in Sect.\,\ref{discus} and conclude in
Sect.\,\ref{concl}.

\begin{table}
   {
     \begin{center}
       \caption[]{COMICS 24.5\,$\mu$m observing log of standard stars (P = PSF standard; S = standard). 
Standard stars are further divided into Cohen (C) and Sloan (L) standards. 
Upper part of the table are observations performed with the Q24.5-OLD filter, the lower part denote the observations 
performed with the Q24.5-NEW filter.  Filter
characteristics are given in Table~\ref{2filters}.}
       \begin{tabular}{ccrcr}
         \hline
         \hline
	 Date       & Object             & Integration & ${\overline {\rm AM}}$ & S/N  \\
	 (UT)       &                    & (sec)       &                        &      \\
	 \hline
	 2002/12/15 & Asteroid \#511 (P) &  502        & 1.8                    &  900 \\
	 2002/12/15 & Asteroid \#51 (P)  &  402        & 2.0                    &  160 \\
	 2003/06/17 & $\mu$ Cep (SL)     &  402        & 1.3                    & 1800 \\
	 2003/06/20 & $\mu$ Cep (SL)     &  402        & 1.3                    & 1000 \\
	 2003/07/14 & $\mu$ Cep (SL)     &  101        & 1.5                    &  180 \\
	 2003/11/12 & $\alpha$ Tau (SL)  & 1204        & 1.2                    &  800 \\
	 2004/01/08 & $\alpha$ Tau (SL)  &  402        & 1.0                    &  830 \\
	 2004/05/05 & $\mu$ Cep (SL)     &  201        & 1.4                    &  790 \\
	 \hline
	 2004/06/08 & $\alpha$ Her (SL)  &  100        & 1.2                    &  630 \\
	 2004/06/08 & $\mu$ Cep (SL)     &  602        & 1.3                    & 3000 \\
	 2004/06/08 & $\alpha$ Sco (SL)  &  201        & 1.9                    & 1200 \\
	 2004/07/14 & $\mu$ Cep  (SL)    &  401        & 1.3                    & 1500 \\
	 2005/07/27 & $\alpha$ Her (SL)  &  201        & 2.6                    &  160 \\
	 2005/07/27 & $\gamma$ Aql (SC)  &  201        & 1.3                    &   30 \\
	 2005/12/13 & $\alpha$ Tau (SL)  &  401        & 1.3                    &  430 \\
	 2005/12/21 & $\alpha$ Tau (SL)  & 1203        & 1.4                    &  620   \\
	 \hline								       
       \end{tabular}
     \end{center}
   }
   \label{obspsf}
\end{table}

\begin{table*}
  {
    \begin{center}
      \caption[]{COMICS 24.5\,$\mu$m observing details. Integration is
the duration of on-source integration in seconds, and ${\overline {\rm
AM}}$ is the average airmass. The signal to noise in Col.\,5
is calculated as peak flux divided by the background standard
deviation. Observations after 2004/06/08 are performed with Q24.5-NEW filter. The J2000 coordinates in Col.\,6 and 7 correspond to the target MYSO in each
image, with its reference given in Col.\,8. The offsets in Col.\,9 correspond to the identification of known sources (Col.\,10) 
in the images. The one uncertain identification is marked with a colon. 
The uncertainty of the flux densities given in Col.\,11 is on the order of 10\%. The Mon R2 and W3 regions are relatively large and have been mosaiced using five, respectively four images (two long and two short exposures).}
      \begin{tabular}{llllllllllr}
        \hline
        \hline
	Region      & Date       & Integration & ${\overline {\rm AM}}$ & S/N  & R.A.        & Dec.         & Ref. & offset   & Source & Flux density \\
	            & (UT)       &   (s)       &                        &      & (h,m,s)     & ($\degr, \arcmin, \arcsec$) & & (\arcsec)&  ID    & (Jy)          \\
	(1)         & (2)        & (3)         & (4)                    & (5)  & (6)         & (7)                         &(8)& (9)    & (10)   & (11)\\
	\hline                   
	\object{S140}        & 2004/06/08 & 602         & 1.4                    & 2000 & 22:19:18.3& +63:18:49.3  & 1 & (0,0) &  IRS1  & 1170     \\  
	            &&&&&                                                                  &              &   & (2,14)&  IRS2S  &  5      \\  
	            &&&&&                                                                  &              &   & (0,18)&  IRS2N: &  170      \\  	
	            &&&&&                                                                  &              &   & (10,3)&  IRS3  &  180      \\  
	\object{M8E}         & 2004/06/08 &  802        & 1.4                    & 1100 & 18:04:53.3& $-$24:26:42.3& 2 & (0,0) &  IR    &  210      \\  
             	    &&&&&                                                                  &              &   & (-6,5)&  HII   &   30      \\  
	            &&&&&                                                                  &              &   &(-6.5,0.5) &  MIRS1 &    5       \\  
	\object{AFGL\,2136}  & 2005/07/27 & 201         & 2.5                    & 170  & 18:22:26.5& $-$13:30:12.0& 3 & (0,0) &  IR    &  140      \\  
	\object{AFGL\,2591}  & 2004/05/05 & 603         & 1.1                    & 2000 & 20:29:24.9& +40:11:20.3  & 4  & (0,0) &  IR    &  870      \\  
		    &&&&&                                                                  &              &   & (-5,-3) &  HII   &  170      \\
	            &&&&&                                                                  &              &   & (-7,9) &  MIRS1 &   20      \\
	\object{NGC\,2264}   & 2002/12/15 & 1306        & 1.9                    & 1100 & 06:41:10.1& +09:29:34.0  & 5 & (0,0) &  IRS1  &  330      \\  
	\object{S255}        & 2003/11/12 & 1404        & 1.0                    & 1000 & 06:12:54.1& +17:59:25.1  & 6 & (0,0) &  IRS3  &  170      \\  
	            &&&&&                                                                  &              &   &(-2.5,0.5) &  IRS1  &   20      \\  
	\object{AFGL\,5180}  & 2005/12/20 & 1003        & 1.2                    & 100  & 06:08:53.3& +21:38:30.5  & 4 & (0,0) &  IRS1  &  490      \\  
		    &&&&&                                                                  &              &   &(12,-4)&  HII   &  210      \\
	            &&&&&                                                                  &              &   & (2,2)     &  MIRS3 &   35       \\  
	\object{IRAS\,20126} & 2005/07/27 &  401        & 1.3                    & 60   & 20:14:26.1& +41:13:32.5  & 7 & (0,0) &  IR    &   60	   \\
	\object{Mon R2}      & 2005/12/13 & 100         & 1.2        	        &      & 06:07:47.8& $-$06:22:54.7& 1 & (0,0) &  IRS3  & 1150	   \\
           	    &&&&&                                                                  &              &   & (-31,3)      &  IRS2  &   40	   \\
                    &&&&&                                                                  &              &   & (-33,17)      &  IRS5  &   70	   \\
	\object{AFGL\,437}   & 2005/12/21 & 401         & 1.3                    & 80   & 03:07:24.6& +58:30:44.4  & 8 & (0,0) &  S     &   30	   \\
                    &&&&&                                                                  &              &   & (0,10)&  N     &   30	   \\ 
                    &&&&&                                                                  &              &   & (-6,6)&  W     &  200	   \\
	\object{AFGL\,4029}  & 2005/12/21 & 1203        & 1.3                    & 80   & 03:01:31.3& +60:29:12.9  & 9 & (0,0) &  IRS1  &   70	   \\
	            &&&&&                                                                  &              &    & (24,1)&  IRS2  &   70       \\
	\object{AFGL\,961}   & 2003/11/12 & 1204        & 1.0                    & 620  & 06:34:37.7& +04:12:44.4  & 10& (0,0) &  E     &  250	   \\
	            &&&&&                                                                  &              &   & (-5,-2)&  W     &   60	   \\
        \object{W3}          & 2005/12/21 &  501, 100       & 1.4          &      & 02:25:41.4& +62:06:21.8  & 11& (0,0) &  IRS5  & 1300	   \\
	            &&&&&                                                                  &              &   & (3,-7)&  IRS6  &   70      \\
	            &&&&&                                                                  &              &   & (-1,-12)&  IRS7  &  160	   \\
	            &&&&&                                                                  &              &   &(-2,11)  &  MIRS1 &   15	   \\
	\object{Cep A}       & 2004/07/13 &  602        & 1.4                    & 70   &  22:56:18.0&  +62:01:49.5& 12& (0,0) & 	     &  440	   \\
	\hline				   
      \end{tabular}			   
      \label{tabflux}			   
    \end{center}			   
}					   
{\small (1) Hackwell et al. (1982)\nocite{1982ApJ...252..250H}; (2) Simon et al. (1984); (3) Kastner et al. (1992); (4) Tamura et al. (1991); (5) Thompson et al. (1998); 
(6) Longmore et al. (2006); (7) Sridharan et al. (2005); (8) Wynn-Williams et al. (1981); (9) Zapata et al. (2001); (10) Castelaz et al. (1985)\nocite{1985AJ.....90.1113C}; 
(11) van der Tak (2005); (12) comparison with {\it Spitzer} images, see Fig.\,\ref{cepa}\newline
}
\end{table*}

\section{Observations and data reduction}
\label{sec-obs}
\subsection{Observations with the COMICS instrument}
\label{sec-comobs}
Tables\,\ref{obspsf} and \ref{tabflux} summarise the 24.5\,$\mu$m
observations of the 14 target MYSOs and the standard stars. All measurements
were made using the mid-infrared imaging spectrometer
COMICS (Kataza et al.\ 2000\nocite{2000SPIE.4008.1144K}) at the
Cassegrain focus of the 8.2 meter Subaru Telescope on
Mauna Kea, Hawaii. Imaging mode of COMICS
utilises a Raytheon 320$\times$240 Si:As IBC array,
which is cooled by a Sumitomo 4-K Gifford-McMahon type
cryo-cooler but usually operates at around 7$\sim$8~K
because of the self-heating. The camera provides
over-sampled diffraction-limited images at 24.5~$\mu$m
with a pixel size of 0.13$\times$0.13~arcsec$^{2}$ and a
field of view of approximately 40$\times$30~arcsec$^{2}$.

\begin{figure}[t]
\includegraphics[width=84mm]{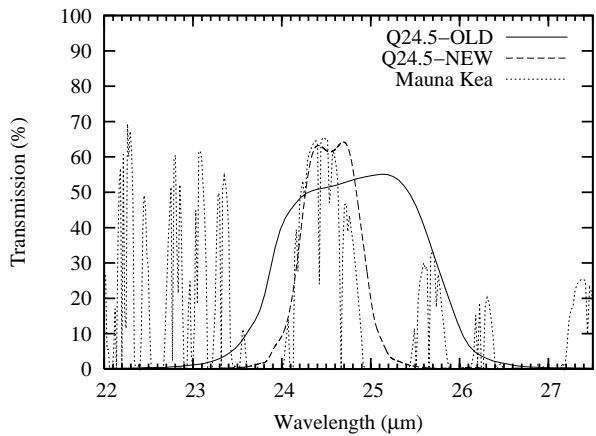}
\caption{Transmission curves for the Q24.5-OLD filter (solid line),
Q24.5-NEW (dashed line), and the atmosphere above Mauna Kea
(dotted line). The filter transmission spectra have been
made available by the manufacturer. The atmospheric transmission
curve has been calculated using USF HITRAN-PC for a standard `US
tropical model' with an H$_{2}$O partial pressure of 1.35~mmHg at
4200~m looking through the atmosphere at a zenith angle of
30$^{\circ}$ (Tomono 2000).}
\label{trans}
\end{figure}

We used two different 24.5\,$\mu$m filters which are both manufactured
by Infrared Multilayer Laboratory, University of Reading. The
characteristics of these filters are summarised in
Table~\ref{2filters}, and the transmission curves, along
with an atmospheric transmission model above Mauna Kea,
are shown in Figure~\ref{trans}.
As can be clearly seen
in the plot, the new filter (Q24.5-NEW) is a much better
fit to the small atmospheric window at 24.5~$\mu$m.
As a result, the whole array can be read out with the
new filter, whilst only a part of the array is read out
with the old one since radiation from the sky quickly
saturates the well. 

Chop-subtracted frames were stacked using the {\sc imalign} task (with
cubic-spline interpolation) in the {\sc iraf}\footnote{{\sc iraf} is
distributed by the National Optical Astronomy Observatories, which are
operated by the Association of Universities for Research in Astronomy,
Inc., under cooperative agreement with the National Science
Foundation.} data reduction package. Flux calibration was achieved
against either Cohen (Cohen et al. 1995\nocite{1995AJ....110..275C};
1999\nocite{1999AJ....117.1864C}) or Sloan (Sloan et
al. 2003\nocite{2003ApJS..147..379S}) standards. We selected stars at
similar airmasses to the target MYSOs whenever possible; however, in
some cases when this was not feasible, we scaled the standard flux to
the airmass of the relevant object by the atmospheric extinction
relationships measured on 2002/12/15 UT (0.57 mag per airmass) and on
2003/11/12 UT (0.56) for the Q24.5-OLD filter. Note that these
extinction values would probably only apply to this specific filter,
along particular lines of sight, and at the times of observations.  In
some other instances (e.g.\ 2005/12/20 UT), even this airmass
correction was not possible due to lack of appropriate data, and we
reluctantly accepted the airmass mismatch as additional
uncertainty. We estimate the overall uncertainty in flux calibration
to be of the order of 10\%. The largest contribution usually comes
from the absolute calibration uncertainty in the standard flux
templates. Details of the target observations and extracted fluxes are given
in Table\,\ref{tabflux}.  Images are not astrometrically
calibrated. Positional offsets of the various sources in our images are with respect to the
brightest 24.5\,\micron source, generally identifiable with the brightest
MYSO in each region presented in this paper.

\begin{figure}[t]
  \centering
    \includegraphics[height=8.5cm,width=8.5cm]{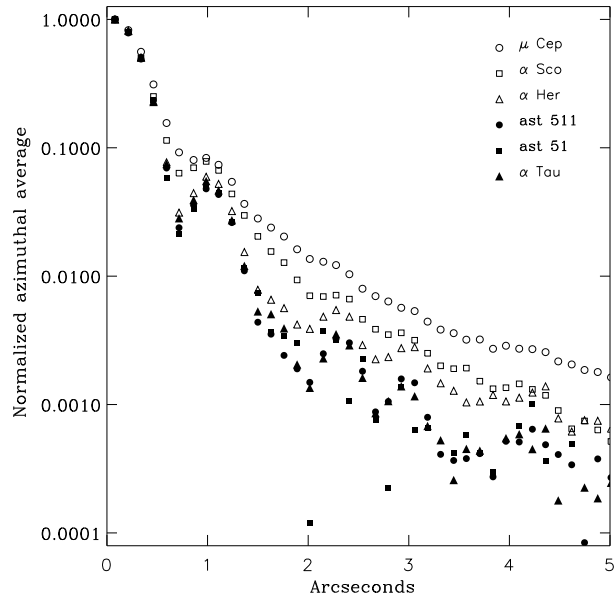}
  \caption[]{Comparison of azimuthally averaged intensity profiles of the
observed standard stars. Standard stars represented by open symbols are apparently surrounded by extended envelopes (see text).}
  \label{fig-PSF}
\end{figure}

\begin{table}
\centering
\caption{Characteristics of two 24.5\,$\mu$m filters.
$\Delta\lambda$ is measured from the 50\%
transmission cut-on to cut-off wavelengths and $\lambda_{0}$
is the half-way point between the two. Measurements made
available by the manufacturer.}\label{2filters}
\begin{tabular}{@{}lccc@{}}
\hline
\hline
ID        & $\lambda_{0}$ & $\Delta\lambda$ & Peak transmission \\
          & ($\mu$m)     & ($\mu$m)       & (\%) \\
\hline
Q24.5-OLD & 24.47                   & 1.91                      & 64 \\
Q24.5-NEW & 24.56                   & 0.75                      & 64 \\
\hline
\end{tabular}
\end{table}

\subsection{Point spread function reference stars}
\label{sec-22}
We spend a few words on the point spread function (PSF) reference objects, as they are crucial in 
our analysis of comparing model images to the resolved MYSO emission (see Sect.\,\ref{mormod}). All standard 
star observations are given in Table\,\ref{obspsf}.  To test the validity of both the PSF standards and the flux standards
as a model for the instrumental PSF, we investigate their azimuthally averaged intensity 
profiles in Fig.\,\ref{fig-PSF}. 
The figure makes clear that three of the flux standard stars
are actually extended, and therefore do not qualify as
a PSF standard. They are the following objects:

\noindent {\bf $\mu$ Cep} is one of the largest stars in our Galactic
neighbourhood. The star was observed a total of six times with COMICS. 
Detailed analysis shows this star to be extended in
such degree and complexity, that we report on this star in a separate
publication (de Wit et al. 2008\nocite{2008ApJ...685L..75D}).\newline
{\bf $\alpha$ Sco} is a supergiant star undergoing mass-loss. Mid-IR
images at 12.5\,\micron and at 20.8\,\micron clearly show an extended distribution of 
circumstellar dust (Marsh et al. 2001\nocite{2001ApJ...548..861M}).\newline
{\bf $\alpha$ Her} is an irregular variable M star. Although the star mimics a PSF
in the inner part of the profile (Fig.\,\ref{fig-PSF}), its extent becomes apparent at radii $>2\arcsec$. Interestingly, interferometric
observations of the star have been interpreted as showing a mass loss event
in the period 1989-1992. The material has an expansion speed of approximately
170 mas $\rm yr^{-1}$ (Tatebe et al. 2007\nocite{2007ApJ...658L.103T}), and it 
should have reached a distance of 2.5\arcsec~from the star at the time of the observations. This distance is comparable
to the distance where the excess emission becomes apparent.\newline

As a consequence of the above discoveries, most observations do not
have a concomitant PSF reference observation. In addition, the instrumental PSF 
shows a non-symmetric pattern due to the mirror support structure that rotates on the sky, 
making the interpretation of non-symmetric structure in the partially resolved sources
taken at different times difficult. A comparison of
the azimuthally averaged intensity profile of the six $\mu$ Cep observations taken over a
period of more than 1 year shows however very few differences (see de
Wit et al. 2008). This provides confidence in the stability of the
(azimuthally averaged) COMICS PSF, and we therefore construct a single reference PSF from the
remaining three genuine point sources ($\alpha$ Tau, and asteroids 51
and 511), which we use throughout the remainder of this paper.

\begin{table*}
  {
    \begin{center}
      \caption[]{Overview of the modelling results. Graphical
 representation of the models are given in the figures to each
 subsection.  Models \#1 are the best-fitting models. Models \#2 and
 \#3 are the best fitting models for radial density powerlaws that
 bracket the best-fitting density powerlaw of model \#1 (except fot AFGL\,2591). Note that for
 steeper density profiles to fit, larger dust optical depths are
 required. Distances are taken from the listed references (Col.\,2),
 and observed luminosities are given in Col.\,3. Their respective
 references are given in Col.\,4. The columns from Col.\,6 and onwards
 are model parameters. Col.\,7 is the power index of the radial
 density powerlaw of the model calculation, i.e. $n=n_{0}\,(r/r_{\rm
 subl})^{-p}$. Col.\,8 is the dust type (see
 Sect.\,\ref{anmet}). Col.\,11 is the ratio of the outer radius to the 
dust sublimation  radius ($r_{\rm subl}$). Col.\,15 is the angular
size on the sky of the inner dust free region.}
      \begin{tabular}{lccrrcrlcrrrrrr}
        \hline
        \hline
	MYSO          &  Distance   & $L$              & ref  &mod  &$L$                & $p$  & dust & $T_{\rm subl}$&$A_{V}$& $R_{\rm out}/r_{\rm subl}$ & $r_{\rm subl}$   & $n_{0}$ & $M_{\rm tot}$     & $\Theta$ \\
	              &  (kpc)      & ($\rm L_{\odot}$)&  &\#   &($\rm L_{\odot}$)  &    &      &      (K)       &           &                          &  (AU)      & ($\rm cm^{-3}$) &($\rm M_{\odot}$) & (mas) \\
	   (1)        &  (2)        & (3)              &  (4)&(5)  & (6)               & (7)& (8)  &      (9)       & (10)       & (11)                      & (12)           &  (13)           & (14) & (15)\\
	
	\hline   
	\object{S140\,IRS1}    &   0.91      & $2.0\,10^{4}$& 1, 15 &1  &$1.3\,10^{4}$ & 1.0  & ISM    &1500  &      62 &       1250 &       32 &  $2.8\,10^{7}$ &      $  32$ &        69\\
	              &             &              &      &2  &$7.6\,10^{3}$ & 0.5  & ISM2   &1000  &      18 &       750  &       63 &  $5.5\,10^{5}$ &      $  39$ &       138 \\
	              &             &              &      &3  &$1.2\,10^{4}$ & 1.5  & ISM2   &1000  &     104 &       1000 &       90 &  $6.0\,10^{7}$ &      $  43$ &       199 \\
	\object{M8E-IR}        &   1.8       & $3.5\,10^{3}$& 2    &1  &$8.0\,10^{3}$ & 1.25 & ISM    &1000  &      64 &       3000 &       70 &  $2.7\,10^{7}$ &      $ 300$ &       78 \\
	              &             &              &      &2  &$1.1\,10^{4}$ & 1.0  & ISM    &1000  &      64 &       3000 &       81 &  $1.0\,10^{7}$ &      $1100$ &       90 \\
	              &             &              &      &3  &$6.0\,10^{3}$ & 1.5  & ISM    &1000  &      64 &       5000 &       61 &  $5.5\,10^{7}$ &      $ 130$ &       68 \\
	\object{AFGL\,2136}    &   2.0       & $7.0\,10^{4}$& 3, 16 &1  &$2.5\,10^{4}$ & 1.0  & ISM    &1000  &      70 &       1000 &      124 &  $8.3\,10^{6}$ &      $ 370$ &      123 \\
                      &             &              &      &2  &$2.7\,10^{4}$ & 0.5  & ISM    &1000  &      38 &       750  &      120 &  $6.0\,10^{5}$ &      $ 300$ &      119 \\
                      &             &              &      &3  &$3.0\,10^{4}$ & 1.5  & ISM    &1000  &     120 &       2000 &      139 &  $4.5\,10^{7}$ &      $ 330$ &      139 \\
	\object{AFGL\,2591}    &   1.0       & $2.0\,10^{4}$& 4    &1  &$7.1\,10^{3}$ & 1.5  & ISM    &1000  &     170 &       5000 &       67 &  $1.3\,10^{8}$ &      $ 430$ &      134\\
		      &   	    &              &      &2  &$4.0\,10^{3}$ & 1.0  & ISM    &1000  &      90 &       2000 &       49 &  $2.4\,10^{7}$ &      $ 270$ &       98 \\
	              &             &              &      &3  &$1.4\,10^{4}$ & 1.0  & ISM    &1500  &      82 &       3000 &       33 &  $3.1\,10^{7}$ &      $ 240$ &       66 \\
	\object{NGC\,2264\,IRS1} &   0.8       & $3.5\,10^{3}$& 5, 17 &1  &$1.5\,10^{3}$ & 1.5  & ISM    &1000  &      90 &       5000 &       31 &  $1.5\,10^{8}$ &      $ 480$ &       77 \\
		      &   	    &              &      &2  &$1.1\,10^{3}$ & 1.25 & ISM    &1000  &     100 &       2000 &       27 &  $1.2\,10^{8}$ &      $ 330$ &       66 \\
	              &             &              &      &3  &$1.1\,10^{3}$ & 1.0  & ISM    &1000  &      40 &       2000 &       26 &  $2.1\,10^{7}$ &      $ 320$ &       64 \\
	\object{S255\,IRS3}   &   2.5       & $2.6\,10^{4}$& 6, 18 &1  &$2.4\,10^{4}$ & 1.25 & ISM    &1000  &      70 &       5000 &      120 &  $1.7\,10^{7}$ &      $2300$ &        96 \\
	              &             &              &      &2  &$2.7\,10^{4}$ & 1.0  & ISM    &1000  &      58 &       2000 &      127 &  $6.0\,10^{6}$ &      $1200$ &       102\\
	              &             &              &      &3  &$3.5\,10^{4}$ & 1.5  & ISM    &1000  &     176 &       5000 &      149 &  $6.0\,10^{7}$ &      $2200$ &       119\\
	\object{AFGL\,5180}    &   1.5       & $1.1\,10^{4}$& 7    &1  &$7.0\,10^{3}$ & 1.0  & ISM    &1000  &     200 &       1250 &       66 &  $4.3\,10^{7}$ &      $ 450$ &       88 \\
		      &   	    &              &      &2  &$6.3\,10^{3}$ & 0.5  & ISM    &1000  &     100 &       1250 &       59 &  $2.5\,10^{6}$ &      $ 530$ &       79 \\
		      &   	    &              &      &3  &$4.2\,10^{3}$ & 0.0  & ISM    &1000  &     100 &       1250 &       43 &  $1.9\,10^{5}$ &      $ 460$ &       57 \\
	\object{IRAS\,20126}   &   1.7       & $1.3\,10^{4}$& 8, 19 &1  &$9.2\,10^{3}$ & 0.0  & ISMz2  &1000  &     116 &       1000 &       51 &  $2.3\,10^{5}$ &      $ 470$ &       60 \\
	              &             &              &      &2  &$7.9\,10^{3}$ & 0.5  & ISMz2  &1000  &     130 &       1250 &       53 &  $3.7\,10^{6}$ &      $ 560$ &       61 \\
	              &             &              &      &3  &$7.9\,10^{3}$ & 1.0  & ISM2   &1000  &     184 &       1250 &       73 &  $3.6\,10^{7}$ &      $  50$ &        85 \\
	\object{Mon\,R2\,IRS3} &   0.83      & $6.5\,10^{3}$& 9, 20 &1  &$1.3\,10^{4}$ & 1.0  & ISM    &1500  &      80 &       1000 &       31 &  $3.8\,10^{7}$ &      $  26$ &        74 \\
	              &             &              &      &2  &$8.0\,10^{3}$ & 0.5  & ISM2   &1000  &      16 &        750 &       64 &  $4.6\,10^{5}$ &      $  36$ &       154 \\
                      &             &              &      &3  &$1.3\,10^{4}$ & 1.5  & ISM2   &1000  &     128 &        750 &       93 &  $6.9\,10^{7}$ &      $  37$ &       223 \\
	\object{AFGL\,437S}    &   2.7       & $2.4\,10^{4}$& 10, 21&1  &$4.2\,10^{3}$ & 0.0  & ISM    &1000  &      62 &       1000 &       42 &  $1.5\,10^{5}$ &      $ 180$ &       31 \\
                      &             &              &      &2  &$6.5\,10^{3}$ & 0.5  & ISM    &1000  &      82 &        750 &       60 &  $2.6\,10^{6}$ &      $ 160$ &       44 \\
                      &             &              &      &3  &$6.9\,10^{3}$ & 1.0  & ISM    &1000  &     198 &        750 &       65 &  $4.6\,10^{7}$ &      $ 170$ &       48 \\
	\object{AFGL\,4029}    &   2.2       & $2.0\,10^{4}$& 11, 22&1 & $1.1\,10^{4}$ & 0.0 & ISMz  &1500 &       82 &       1250 &       17 &   $4.0\,10^{5}$  &        57 &       15 \\
                      &             &              &      &2   & $1.1\,10^{4}$ & 0.5 & ISMz  &1000 &       86 &       1250 &       54 &  $2.3\,10^{6}$   &       385 &       49 \\
                      &             &              &      &3   & $1.2\,10^{4}$ & 1.0 & ISMz   &1000 &     130 &       3000 &       63 &  $2.6\,10^{7}$  &      1355 &       57 \\
	\object{AFGL\,961E}    &   1.4       & $6.0\,10^{3}$& 12, 23& 1 &     $5.5\,10^{3}$ & 0.5 & ISMz & 1000 &       32 &       750 &       38 &   $1.6\,10^{6}$ &         25 &       54 \\
		      &   	    &              &      &2  & $5.6\,10^{3}$ & 0.0 & ISM2 & 1500 &       28 &       750 &       18 &   $2.1\,10^{5}$ &          8 &       26 \\
		      &   	    &              &      &3  &  $6.1\,10^{3}$ & 1.0 & ISMz & 1000 &       48 &      3000 &       43 &   $1.4\,10^{7}$ &        239 &       62\\
	\object{W3\, IRS5}     &   1.8       & $1.9\,10^{5}$& 13, 24&1  &     $7.0\,10^{4}$ & 1.0 & ISM  & 1500 &       144 &      750 &       73 &   $3.0\,10^{7}$ &        156 &       81    \\
		      &   	    &              &      &2  &  $8.8\,10^{4}$ & 0.5 & ISM  & 1500 &       106 &      750 &       78 &   $2.6\,10^{6}$ &        359 &       87\\
		      &   	    &              &      &3  & $6.3\,10^{4}$ & 1.5 & ISM2 & 1000 &       196 &      750 &      208 &   $5.0\,10^{7}$ &        284 &      231\\
	\object{Cep\,A}        &   0.7       & $2.5\,10^{4}$& 14, 25&   & &            &    &             &         &       &                   &    &   &              \\      
	\hline
      \end{tabular}
      \label{tabtar}

    \end{center}
}
{\small
(1) Crampton \& Fisher (1974);
(2) Simon et al. (1985); \nocite{1985ApJ...298..328S}
(3) Kastner et al. (1992); \nocite{1992ApJ...389..357K}
(4) van der Tak et al. (1999);
(5) Walker (1956)\nocite{1956ApJS....2..365W};
(6) Moffat et al (1979)\nocite{1979A&AS...38..197M};
(7) Snell et al. 1988\nocite{1988ApJ...325..853S};
(8) Cesaroni et al. (2005);
(9) Herbst \& Racine (1976); \nocite{1976AJ.....81..840H}
(10) Alvarez et al. (2004);
(11) Becker \& Fenkart (1971);\nocite{1971A&AS....4..241B}
(12) Cohen et al. (1973);\nocite{1973ApJ...185L..75C}
(13) Imai et al. (2000);\nocite{2000ApJ...538..751I}
(14) Evans et al. (1981);\nocite{1981ApJ...244..115E}
(15) Lester et al. (1986);\nocite{1986ApJ...309...80L}
(16) Kastner et al. (1994);
(17) Harvey et al. {1977);\nocite{1977ApJ...215..151H}
(18) Jaffe et al. (1984);\nocite{1984ApJ...284..637J}
(19) Cesaroni et al. (1997);\nocite{1997A&A...325..725C}
(20) Thronson et al. (1980);\nocite{1980ApJ...237...66T}
(21) Wynn-Williams (1982);\nocite{1982ARA&A..20..587W}
(22) Beichman et al. (1979);
(23) Harvey et al. (1977);
(24) Ladd et al. (1993);\nocite{1993ApJ...419..186L}
(25) Evans et al. (1981)\nocite{1981ApJ...244..115E}}
}

\end{table*}


\section{Results}
\subsection{Brief description of the images}
\label{results}
The 24.5\,$\mu$m images reveal that all principal MYSO sources in the fourteen targeted massive
star forming regions are resolved. The sources are generally discrete and have fairly simple, circular morphologies on
the sky suggesting that the emission is dominated by the circumstellar envelope. 
We compare the observed azimuthally averaged intensity profiles of the MYSO envelopes
to 1-D dust radiative transfer calculations. For each target, we attempt to reach
consistency between the spatially resolved 24.5\,\micron emission and the
emission seen at other wavelengths as represented by the SED. In
addition, we describe  the 24.5\,\micron emission morphology 
in relation to known star formation activity in the following subsection. 

\subsection{Method of analysis}
\label{anmet}
Model SEDs and images are calculated using DUSTY, a code that solves
the scaled 1-D dust radiative transfer problem (see Ivezic \& Elitzur
1997\nocite{1997MNRAS.287..799I}). We use a spherically symmetric dust
distribution illuminated by a central, unresolved star. Numerical
solutions are independent of the star's bolometric luminosity, and the
SED and images need to be scaled before making a comparison to the
observations. The bolometric luminosity is the prime stellar parameter
that sets the inner dust sublimation radius and thus the size scale of
the envelope; an increase causes the size of the emitting region to be
larger (as $r_{\rm subl} \propto \sqrt{L_{\rm bol}}$).  As a result the
intensity profile strongly depends on the $L_{\rm bol}$ assumed for the model. In practice,
we determine the model \lbol~ by minimising the difference between the model
and observed SED. The model \lbol~can differ significantly from the observed \lbol, but for
each model fit our adopted procedure finds the closest match between the two. In some case
this match is poor and results in a large difference in observed and model \lbol.
Observed \lbol~are listed for each target
in Col.\,3 of Table\,\ref{tabtar}.

Other important model parameters are the outer radius of the envelope
($R_{\rm out}$) and the total amount of dust. The latter is
parametrised in DUSTY by the total optical depth $A_{V}$. Higher values for
these two parameters tend to increase the long wavelength flux
levels. Of course, neither the total amount of dust nor the envelope's outer
radius can have any arbitrary value.  We constrain the extinction
by the 9.7\,\micron silicate absorption feature that is generally
found to be strongly in absorption among MYSOs. The required data were
in most cases provided by mid-IR spectra taken with the short
wavelength spectrometer (SWS, de Graauw et
al. 1996\nocite{1996A&A...315L..49D}) on board the ISO satellite
(Kessler et al. 1996\nocite{1996A&A...315L..27K}). The ISO data have been 
obtained from ESA's ISO data archive.\footnote{Accessible at this URL \tt{http://iso.esac.esa.int/ida/}}

The MYSO SEDs are constructed form the measured COMICS fluxes (Table\,\ref{tabflux}) and from literature data. For most targets
continuum measurements in the IR and (sub)mm are taken from the
catalogue compiled by Gezari et al. (1999\nocite{1999yCat.2225....0G}, available through
Vizier), and includes IRAS and MSX observations.  These data were
supplemented, where possible, with more recent observations, especially
the compilation by Mueller et al. (2002).  The
continuum slope of the ISO-SWS spectrum longwards of the silicate
absorption is used to complete the SED at mid-IR wavelengths. In three cases {\it Spitzer} MIPS
data are available. We extracted the photometry applying the non-linearity correction recipe
by Dale et al. (2007). Data
taken with large beams ($\gta 15\arcsec$) were generally avoided in the model fitting
procedure.  The following subsections, that are devoted to each
object, highlight the used and discarded data in the fitting
process. In the accompanying figures for each source open symbols are
used to indicate large beam data not taken into account in the
model fitting procedure, whereas filled symbols indicate the data that were
actually used in the fitting procedure. Fluxes at wavelengths larger than
1.3\,mm are excluded as they might be contaminated by free-free emission
from ionized winds (e.g. Gibb \& Hoare 2007\nocite{2007MNRAS.380..246G}).  Overall, the
constructed SEDs cover a wavelength range from about 1\,\micron to 1\,mm.

The model and observed images are compared in the fitting procedure by
means of the azimuthally averaged intensity profile, 
normalized to its peak intensity. Model images are first scaled to model
\lbol~and convolved with the instrumental PSF. The intensity profiles
are determined by binning and averaging the pixel counts in radial distance bins of size 
0.13\arcsec.  The 
centre of the profile is found by minimising the pixel scatter as function of radial
distance. We indicate the range in pixel counts found at each radius
bin by an errorbar in the presented intensity profiles.

In order to perform a systematic comparison between models and observations, 
we have generated a standard grid that consists of 120\,000 DUSTY models.
The grid probes the envelope parameter space and ranges in the following way:

\begin{itemize}
\item Five radial density profiles. The radial density profile of the
dust is described by a powerlaw of the form
$n(r) \propto n_{0}\,r^{-p}$. The five density profiles cover power
indices from $p=0.0$ to $p=2.0$ in steps of 0.5.
\item Four types of dust mixtures. The dust size
distribution is MRN (Mathis, Rumple, \& Nordsieck 1977\nocite{1977ApJ...217..425M}), and
remains untouched.  We use DL (Draine \& Lee 1984\nocite{1984ApJ...285...89D})
opacities for silicates and combine this with either DL opacities for
graphites or amorphous carbon (Zubko et al. 1996\nocite{1996MNRAS.282.1321Z}). 
Opacities are combined in a standard ISM mixture of 53\% silicate and 47\% graphite
or a ``twice ISM'' mixture, i.e. 67\% silicate and 33\% graphite. Different mixtures
may accommodate the silicate absorption profile for given total dust mass. The 
dust mixtures used have an opacity at 1.3\,mm of $\kappa$ between 0.3 -- 0.4\,g$^{-1}\,$cm$^{2}$, similar
to the values of other types of dust in the literature, e.g. Ossenkopf \& Henning (1994)\nocite{1994A&A...291..943O}.
\item Two dust sublimation temperatures ($T_{\rm subl}$) : 1000\,K and 1500\,K. The value
for this parameter is the only unscaled parameter in DUSTY. The value of 
$T_{\rm subl}$ has a large effect on the location of the dust sublimation radius 
but only a small effect on the normalized intensity profile. This is mainly due to 
a similar temperature dependence through the cloud with a very steep initial drop followed by a gradual
decline. The similar temperature stratification leads to a similar
normalized intensity profile. The different dust sublimation
temperatures produce a different balance between short and long
wavelength flux in the SED. 
\item Three scaled sizes for the envelope. The outer radius is scaled to the inner
sublimation radius. The standard model grid contains models with sizes 750, 1000 and 1250 times
the inner dust sublimation radius, which correspond to typical envelope sizes 
of about 0.1\,pc. 
\end{itemize}

For each combination of these four envelope structure
parameters, DUSTY calculates the resulting SEDs and images for an increasing total
dust mass contained in the envelope. The $A_{V}$
grid goes from 2 to 200 in steps of 2, resulting in a grand total of 120\,000 models.
The envelope parameters adopted in this grid lead to typical
temperatures found at the outer radius of 20\,K for $T_{\rm
subl}=1000\,$K for and 30\,K for $T_{\rm subl}=1500\,$K with typical densities 
at $R_{\rm out}$ of $\rm 10^{3-4}\,cm^{-3}$

The fitting procedure tries to find a model that fits the intensity
profile and SED simultaneously. The SED fit includes the silicate
absorption feature, slope of the ISO-SWS spectra, the far-IR SED peak
and the (sub)mm continuum data. We aim to match the flux levels of the integrated 
(sub)mm continuum data providing additional constraints on the model parameters.
The ISO spectra were rebinned
logarithmically in 35 wavelength bins between 2.5 and 43\,\micron
in order to reduce its weight with respect to the continuum data. The binned spectrum probes the silicate
feature at 9.7\,\micron and the change in continuum slope longwards of
it.  In short, the fit procedure initially estimates the model $L_{\rm bol}$
by matching the overall shape of the scaled model SED to the observed
one. DUSTY's output images are then accordingly scaled and convolved
with the instrumental PSF. A comparison of the model SED and 
intensity profile to the observed ones is made for all generated
models. A simple tally is performed based on a goodness-of-fit
criterion.  We apply a  least-squares criterion for the
SED fit, and a weighted least-squares criterion for the
intensity profile fit. The latter uses weights that are inversely
proportional to the square of the range in normalized intensity
at each radial distance bin.  This range is represented by
errorbars in the intensity profile plots for each MYSO target.  The
model that fits both sets of data best is the one with the highest
average ranking in the two tallies.  If no simultaneous satisfactory
fits were obtained, we changed the emphasis of the procedure in order
to get fits to intensity profile {\it and} the long wavelength range
of the SED only, i.e. dropping the requirement of fitting the near-IR and mid-IR part of the SED. In some cases no satisfactory model was found within
our standard grid, and we chose to explore the effects of changes in
envelope size. This is discussed in more detail in the subsections 
dedicated to each source.

Finally, the parameters for the underlying star are the same for each
model and are the following. $L_{\rm bol}$ of the MYSOs are indicative
of early B-type stars, thus we adopt a $T_{\rm eff}$ of
25\,000\,K. There is only a negligible effect on the model dust emission
for stars with slightly different effective temperatures.  Distances
for our sample stars were taken from the literature (see Table\,\ref{tabtar}).

All final model parameters are listed in the modelling overview Table\,\ref{tabtar}. 
For each MYSO we give the parameters for three models. Models \#1
correspond to the overall best-fitting model. Models \#2 and \#3 constitute
the best-fit for models restricted to density powerlaws with indices
that bracket the powerlaw index of the best-fitting model \#1, and are
{\it not} the second and third best-fitting model. This is
to illustrate how well the powerlaw index is constrained by the
observables as this is our main priority; in some cases, models \#2 and \#3 do not provide
acceptable fits. Therefore, the model parameters of
Table\,\ref{tabtar} cannot be considered to represent the
uncertainties on the model parameters. More detailed explanation 
can be found in the subsections dedicated to each MYSO.

\subsection{24.5\,\micron morphology and modelling}
\label{mormod}

\begin{figure}[t]
  \center{\includegraphics[height=8.5cm,width=8.5cm]{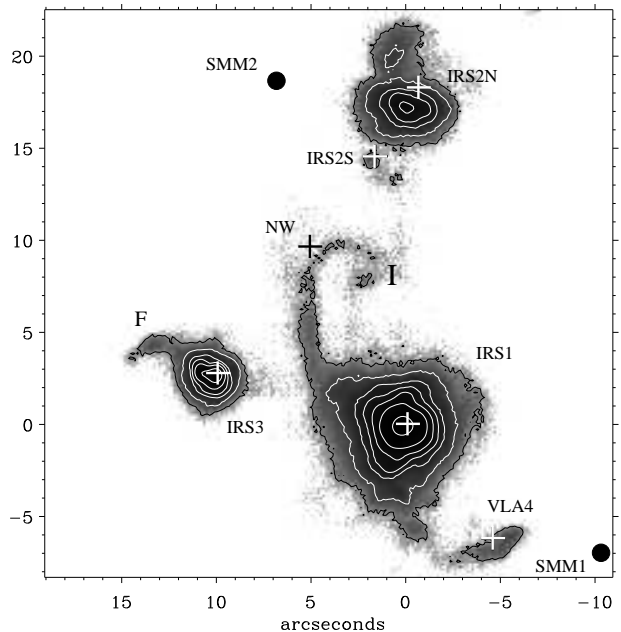}}
  \caption[]{COMICS 24.5\,\micron image of the S140 region. The image is linearly scaled. Annotated objects are discussed in the text. Crosses correspond
to radio sources (Evans et al. 1989\nocite{1989ApJ...346..212E}; Tofani et al. 1995\nocite{1995A&AS..112..299T}). Contour levels are at 
0.5\%, 1\%, 2\%, 3\%, 5\%, 10\%, and 40\% of peak flux density ($\rm 5.9\,10^{2}\,Jy\,arcsec^{-2}$). North is up, East is to the left.}
  \label{s140a}
\end{figure}

\begin{figure}[t]
  \center{\includegraphics[height=12cm,width=8.5cm]{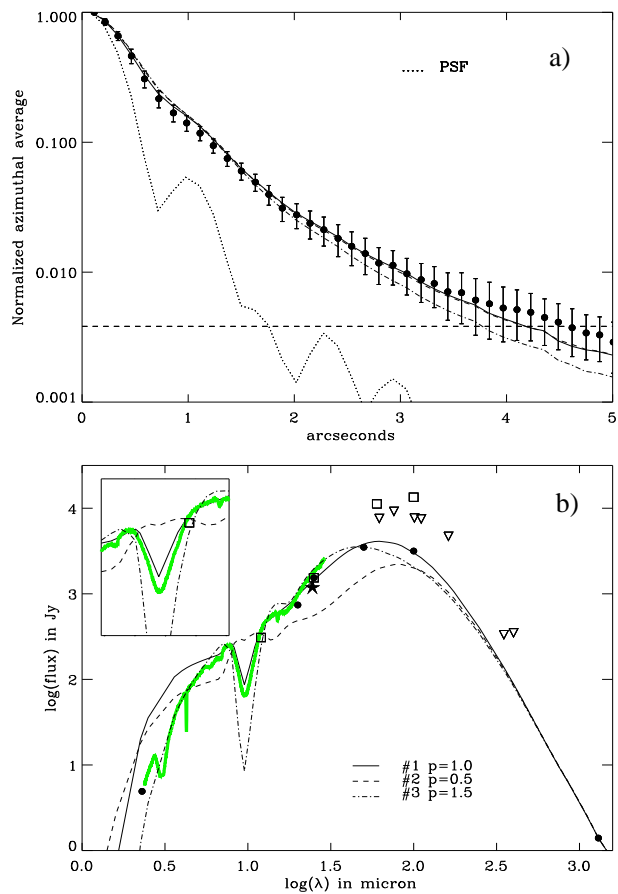}}
  \caption[]{Simultaneous model fits to the intensity profile and SED of
  S140\,IRS1. {\it Panel a)} Observed (with errorbars) and three model
  normalized azimuthal intensity profiles at
  24.5\,$\mu$m.  The code for the 
  model line styles is the same as in panel b. The errorbars indicate 
  the range in observed intensity values found at each radial distance bin.
  The horizontal dashed line
  equals three times the rms level of the normalized background
  level. {\it Panel b)} Observed (symbols) and model SEDs (various
  line styles). The COMICS flux measurement is represented by an
  asterisk. Open squares are IRAS measurements. Open triangles are
  large beam measurements discussed  in the {\it model results}
  paragraph for each MYSO target. The thick line denotes the ISO-SWS
  spectrum. Models are fitted to the filled symbols and the
  logarithmically binned ISO-SWS spectrum only. The best fit model
  (model \#1 in Table\,\ref{tabtar}) is indicated by the full line,
  models \#2, and \#3 by a dashed line, dot-dashed linestyle
  respectively. The best fitting model (\#1) has a $p=1.0$ radial
  density profile.}
  \label{s140b}
\end{figure}

\subsubsection{S140 (Figs.\,\ref{s140a} and \ref{s140b})} 

{\it Description:} At a distance of 910\,pc, S140 is a bright-rimmed cloud that forms the
interface between an \ion{H}{ii} region and the molecular cloud
L1204 (Crampton \& Fisher 1974\nocite{1974QB4.V5v14n12...}). S140 contains an IR cluster of at least
three sources (Beichmann et al. 1979\nocite{1979ApJ...232L..47B}). 
The main CO outflow in the region goes SE-NW and a monopolar reflection nebula emanates from 
IRS1 which is associated with the blue-shifted SE lobe (Hayashi et al. 1987\nocite{1987ApJ...312..327H}; see also
Hoare \& Franco 2007\nocite{2007dmsf.book...61H}). The region displays spectacular
arc-like features that are seen in high angular resolution $K$-band images
(Forrest \& Shure 1986\nocite{1986ApJ...311L..81F}). They probably constitute
material swept-up by outflow activity in the region
(Weigelt et al. 2002\nocite{2002A&A...381..905W}). Recent cm continuum
and $K$-band polarimetric imaging corroborate to suggest the presence
of a disk-like structure for IRS1 oriented in a NE-SW direction, perpendicular to the main outflow
(Hoare 2006\nocite{2006ApJ...649..856H}; Jiang et
al. 2008\nocite{2008ApJ...673L.175J}). \newline
{\it Mid-IR morphology:}
The COMICS image of the S140 region is presented in
Fig.\,\ref{s140a}. It reveals a wealth of features associated with
previously known objects, the positions of which are labelled in the
figure. Discrete peaks in mid-IR emission are found at the positions
of IRS1, IRS2, and IRS3.  The bright emission within 3--4\arcsec~of IRS1 has a fairly
symmetric profile on the sky.  The bright emission
region 17\arcsec~to the North of IRS1 coincides closely, but not exactly with 
the position of IRS2N according to the positions given in Tofani et
al. (1995\nocite{1995A&AS..112..299T}). IRS2S is identified with a point-like mid-IR
source just South of that.  IRS3 is found to be a triple system (Preibisch et
al. 2003\nocite{2001A&A...378..539P}) and our image partly resolves
the secondary object (IRS3b) at a distance of 0.75\arcsec~East of the
primary source (IRS3a).

The COMICS image also shows faint structures of diffuse emission.  An arc of
mid-IR emission concurs with the $K$-band arc labelled ``I'' by Weigelt
et al. (2002\nocite{2002A&A...381..905W}). The curved wisp at
$\sim3\arcsec$ from IRS3 corresponds to feature ``F'' of Preibisch et
al. This feature is also known to have strongly polarised
emission and strong $\rm H_{2}$ line emission, suggesting scattering 
and shocked material. 
Patches of diffuse emission are also found coincident with the radio
sources VLA4 and NW (see Evans et
al. 1989\nocite{1989ApJ...346..212E}). This is the first time that the
two radio sources are seen in the mid-IR. Previously (Evans et
al. 1989), they were found coincident with the brightest parts of very
extended near-IR nebulosity. 
Finally, we note that
the two submm emission peaks (Minchin et
al. 1995\nocite{1995A&A...298..894M}; Thompson et
al. 2006\nocite{2006A&A...453.1003T}) do not have 24.5\,\micron emission
counterparts.
Clearly the image reveals that
24.5\,\micron emission is not restricted to compact MYSO envelope
emission. The emission has a faint diffuse character when it is associated with shocked dense material.\newline
{\it Model results:} 
We focus the modelling on the main component in the region: bright MYSO IRS1.  Its azimuthally averaged
24.5\,\micron intensity profile is presented in panel a of Fig.\,\ref{s140b}. The  errorbars cover the peak to peak range in 
pixel counts measured at each radial
distance bin. Their small values indicate that IRS1 can be considered to be
symmetric to first order. We
build the IRS1 SED (Fig.\,\ref{s140b}) from literature data. Photometry obtained using image
restoration techniques applied to KAO continuum observations at 50 and
100\,\micron by Lester et al. (1986\nocite{1986ApJ...309...80L}) are preferred
over large beam data (50\arcsec) presented by Schwartz et al. (1983)\nocite{1983ApJ...271..625T} and IRAS.  
We adopt for IRS1 the IRAM measurement at 1.3\,mm (11\arcsec~HPBW) by G\"{u}rtler et
al. (1991)\nocite{1991A&A...252..801G}. Large beam far-IR and submm data 
by Schwartz et al. (1983) and G\"{u}rtler et al. (1991) are
indicated by triangles in Fig.\,\ref{s140b}.

The intensity profile and the SED can be reproduced simultaneously by
models with a $p=1.0$ radial density distribution. It is clear from Fig.\,\ref{s140b} that
the intensity profile can also be fit by model \#2, and to a somewhat lesser extent, model \#3. However, the simultaneous fit to the SED excludes 
shallow $p=0.5$ (model \#2) and steep $p=1.5$ (model \#3) models as viable alternatives. This is
especially evident from their poor reproduction of the
observed silicate absorption profile and the SED peak. On larger scales, modelling of the SED
longwards of 100\,\micron by Van der Tak et al. (2000) resulted in a $p=1.5$ density distribution. Mueller et al. (2002) find an
overall best-fit to the SED and 350\,\micron intensity profile for
spherical models with $p=1.25$. 

\begin{figure}[t]
  \center{\includegraphics[height=8.5cm,width=8.5cm]{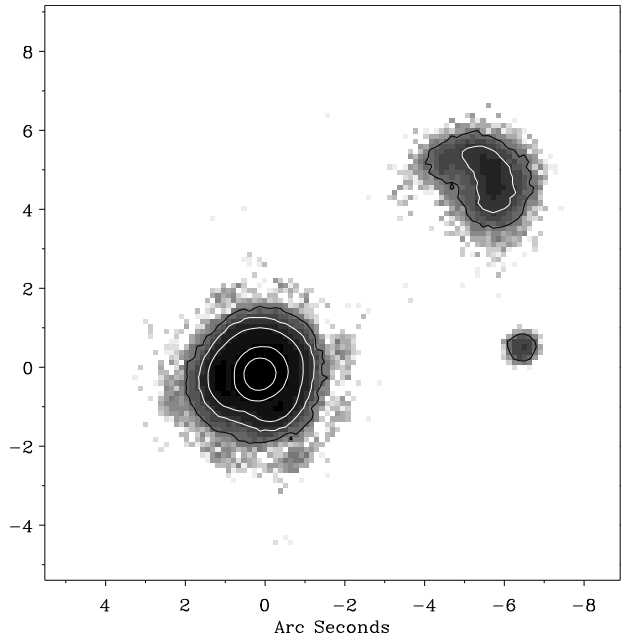}}
  \caption[]{COMICS 24.5\,\micron image of the M8E region. Relative
  positions of the MYSO and the cometary shaped \ion{H}{ii} region are
  consistent with near-IR (2MASS) and radio images (Simon et
  al. 1984). Contour levels are at 1\%, 2\%, 5\%, 10\%, and 40\% of peak flux density ($\rm 2.4\,10^{2}\,Jy\,arcsec^{-2}$). North is up, East is to the left.}
  \label{m8ea}
\end{figure}

\begin{figure}[t]
  \center{\includegraphics[height=12cm,width=8.5cm]{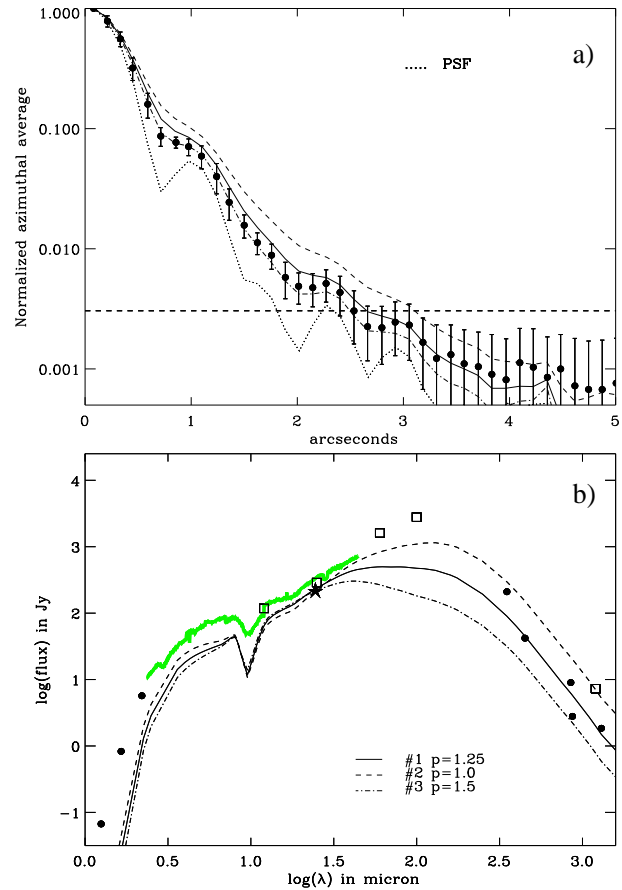}}
  \caption[]{As Fig.\,\ref{s140b}, but for M8E-IR. The SED data is described in Sect.\,\ref{m8e}. The best fitting model has $p=1.25$ radial density profile.}
  \label{m8eb}
\end{figure}

\subsubsection{M8E (Figs.\,\ref{m8ea} and \ref{m8eb})}
\label{m8e}

{\it Description:} M8E consists of a bright IR source and 
an optically thin radio source 8\arcsec~to its North-West (Simon et
al. 1984\nocite{1984ApJ...278..170S}). The region shows evidence of a bipolar
CO outflow, but it is not clear which of the two sources powers it (Simon et al. 1984). 
The IR object is resolved on scales of milli arcseconds by lunar occultation
observations at 3.8\,\micron and 10\,$\mu$m. It can be interpreted as
consisting of two physically different components (Simon et al. 1985\nocite{1985ApJ...298..328S}): a compact hot component and a broad cooler component.  
The broad component, that has an angular size of 0.1\arcsec,  does not correspond to the 30\,K molecular cloud material
causing the SED to peak at 100\,$\mu$m, but is suggested to be a transition region
between relatively hot disk material and cold cloud material. 
\newline
{\it Mid-IR morphology:} Fig.\,\ref{m8ea} shows the detection 
of M8E-IR as a compact discrete source which is symmetric
to first order.  The radio source emission is diffuse and has a cometary morphology.  Its peak emission is found at a distance of
7.7\arcsec~from M8E-IR. A third source is detected 6\arcsec~West of M8E-IR. 
\newline
{\it Model results:} The primary source of M8E-IR continuum data is
from Mueller et al. (2002) and G\"{u}rtler et al. (1991). The six
(sub)mm continuum measurements have been done with various beamsizes between 
19\arcsec~and 30\arcsec, except the 870\,\micron one that was performed with a 9\arcsec~half-power beamwidth (G\"{u}rtler et al. 1991).
For reference, we also plot the 1.2\,mm datapoint obtained by Beltr\'{a}n et
al. (2006\nocite{2006A&A...447..221B}) with a half-power beamwidth of 26\arcsec~(open
square). It is clear that the intrinsic (sub)mm flux levels are uncertain, given 
the spread in flux levels in this wavelength range. At the short wavelength region we use the measurements by Simon et al. (1985).

M8E-IR is compact and only marginally resolved. In fact, the intensity
profile is too steep for the bolometric luminosity derived
from the observed SED. Steeper radial density powerlaws (creating a more
compact cloud structure) lead to steeper intensity profiles, yet this
advantage is offset by a decrease in flux at the long wavelength
end. A commensurate increase in bolometric luminosity is required to
fit the (sub)mm, but this in turn makes the intensity profile
too shallow. Choosing a smaller outer radius is marred with the same
problem, as is increasing the dust sublimation temperature.

The model that fits the intensity profile best has a $p=1.5$
powerlaw. Density profiles as shallow as $p=1.0$ are incompatible with
the intensity profile.  In order to increase the (sub)mm flux levels
(without increasing the bolometric luminosity) we make the cloud
larger than the sizes assumed in our model grid. Depending on the
intrinsic (sub)mm flux levels, $p=1.25$ to $p=1.5$ are preferred with
an outer radius of 3000 times the dust sublimation radius, or $\sim1$\,pc. The flux levels
of the ISO-SWS spectrum cannot be reproduced by any model SED.  On scales
of 1000--10\,000\,AUs, the 350\,\micron intensity profile is best
represented by models with $p=1.75$ radial density
distributions (Mueller et al. 2002)\nocite{2002ApJS..143..469M}, i.e. 
too steep to reproduce the 24.5\,\micron intensity profile.

\begin{figure}[t]
  \center{
    \includegraphics[height=8.5cm,width=8.5cm]{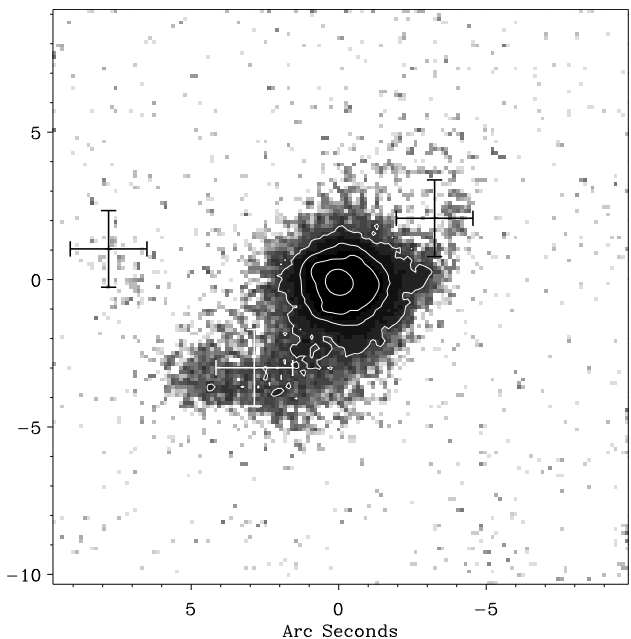}  }
  \caption[]{COMICS 24.5\,\micron image of the AFGL\,2136 region. The
    large crosses correspond to the peak emission of the three near-IR
    scattering nebulae (Kastner et al. 1994). Contour levels are at
    2\%, 5\%, 10\%, and 40\% of peak flux density ($\rm
    1.1\,10^{2}\,Jy\,arcsec^{-2}$). North is up, East is to the left.
  }
  \label{gl2136a}
\end{figure}

\begin{figure}[t]
  \center{
    \includegraphics[height=12cm,width=8.5cm]{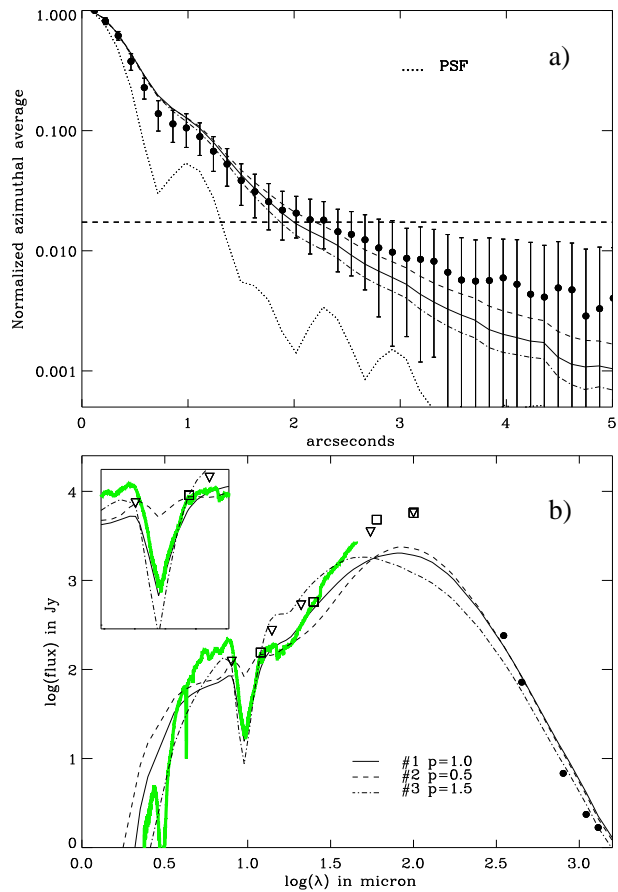}  }
  \caption[]{As Fig.\,\ref{s140b}, but for AFGL\,2136. The SED data is described in Sect.\,\ref{gl2136}. The best fitting model has $p=1.0$ radial density profile.}
  \label{gl2136b}
\end{figure}

\subsubsection{AFGL 2136 (Figs.\,\ref{gl2136a} and \ref{gl2136b})}
\label{gl2136}

{\it Description:} A three-lobed near-IR reflection nebula littered
with some 30 faint point sources is illuminated by the dominant source IRS1. 
Near-IR polarisation indicates that IRS1 is illuminating conical cavities
within a dusty envelope (Kastner, Weintraub, \& Aspin
1992\nocite{1992ApJ...389..357K}). IRS1 is the driving source of an arcminute-scale
bipolar CO outflow with a P.A. of 135\degr~(Kastner et al. 1994\nocite{1994ApJ...425..695K}).
Weak, optically thick radio emission originating from IRS1 is detected by Menten \& van
der Tak (2004\nocite{2004A&A...414..289M}).
The radio source is somewhat elongated in the South-Easterly direction.
\newline
{\it Mid-IR morphology:} The COMICS image in Fig.\,\ref{gl2136a} reveals one
dominant source, IRS1, that is resolved and symmetric. The image is not very deep, and the envelope can be
traced out to $\sim$3\arcsec, where noise starts to dominate. IRS1
displays an extended wing to the South-East coincident with the near-IR ``South''
lobe of the nebula. A faint counter wing to the North-West would correspond to
the ``West'' lobe.  There are also traces of a 24.5\,\micron counterpart to the faint
``East'' lobe (terminology from Kastner et al. 1992).  The bright South lobe extends at roughly 
the same P.A. as the outflow activity.
\newline
{\it Model results:} The SED is built using the set of (sub)mm data from
Kastner et al. (1994) obtained with the JCMT (16.8\arcsec--18.5\arcsec~HPBW), except 
the datapoint at 350\,\micron which is from Mueller et al. (2002). IRAS measurements agree closely with the Harvey et
al. (2000) data at 50 and 100\,\micron. 

The intensity profile and SED of AFGL\,2136 are best fit by a $p=1.0$
model.  None of the models that can reproduce the intensity profile do
a particular good job in reproducing the SED's short wavelength range,
i.e. $<10\,\mu$m.  At larger spatial scales, Mueller et al. (2002)
and van der Tak et al. (2000) prefer steeper models with $p=1.75$ and
$p=1.25$, respectively. Harvey et
al. (2000\nocite{2000ApJ...534..846H}) find reasonable fits to the 50
and 100\,\micron intensity profiles and the SED, by dividing a $p=1.5$
model envelope in a high and low optical depth ``hemisphere''. We also
find that certain $p=1.5$ models are capable of nicely fitting the
full SED longwards of 5\,\micron (not shown in Fig.\,\ref{gl2136b}), but
they do not reproduce the 24.5\,\micron intensity profile. The $p=1.5$
model shown in Fig.\,\ref{gl2136b} does reproduce the intensity
profile, but fits the SED worse than our preferred $p=1.0$ model.

\begin{figure}
  \center{\includegraphics[height=13.5cm,width=8.5cm]{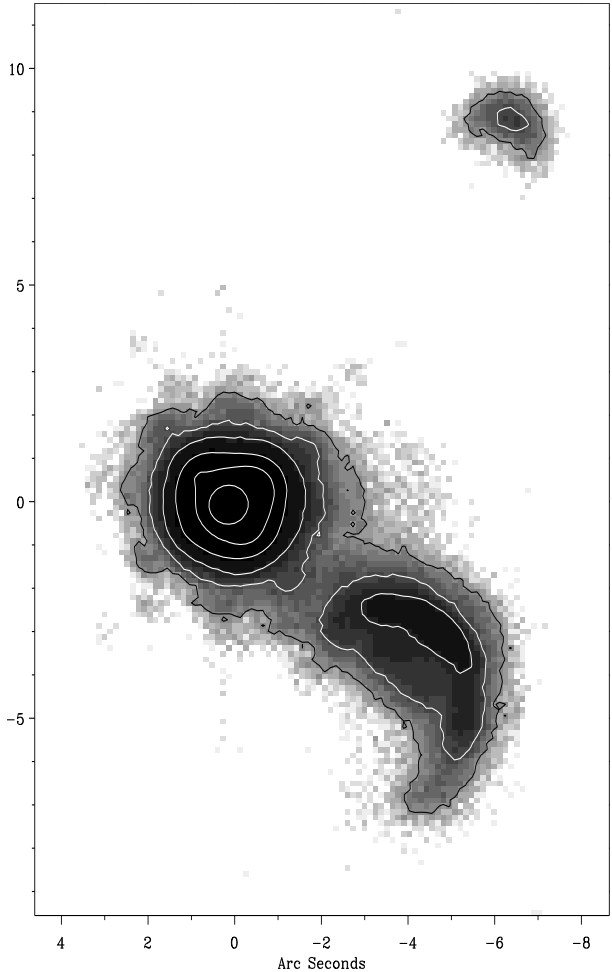}}
  \caption[]{COMICS 24.5\,\micron image of the AFGL\,2591 region. The relative positions of the MYSO and the two
cometary shaped \ion{H}{ii} regions are consistent with the 8\,GHz sources in Tofani et al. (1995\nocite{1995A&AS..112..299T}).
Contour levels are at 0.5\%, 1\%, 2\%, 5\%, 10\%, and 40\% of peak flux density ($\rm 7.7\,10^{2}\,Jy\,arcsec^{-2}$). North is up, East is to the left.
} 
  \label{gl2591a}
\end{figure}

\begin{figure}
  \center{\includegraphics[height=12cm,width=8.5cm]{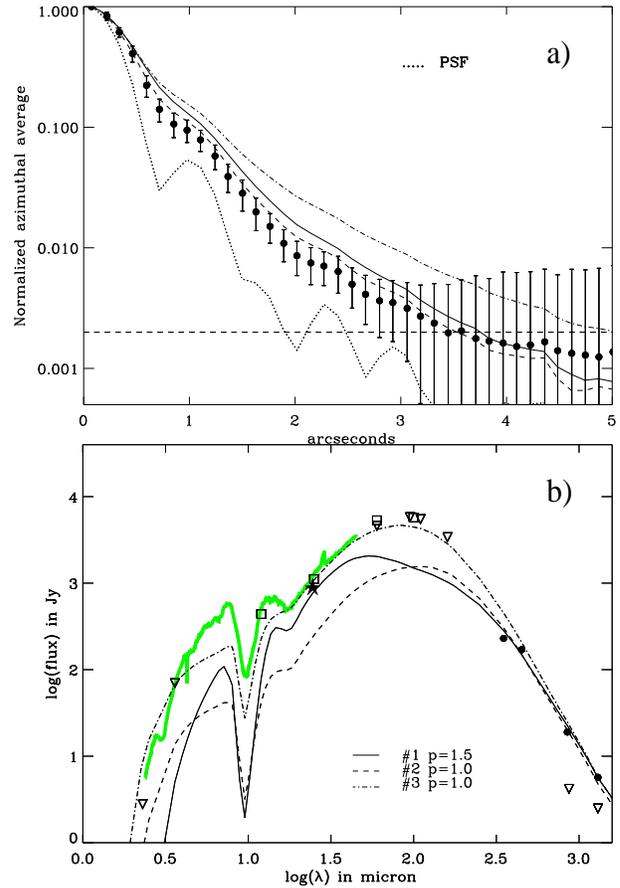}}
  \caption[]{As Fig.\,\ref{s140b}, but for AFGL\,2591\,IRS1. The SED data is described in Sect.\,\ref{gl2591}. No simultaneous model fit
    to SED and intensity profile is found.} 
  \label{gl2591b}
\end{figure}

\subsubsection{AFGL\,2591 (Figs.\,\ref{gl2591a} and \ref{gl2591b})}
\label{gl2591}

{\it Description:} Various aspects of the AFGL\,2591 envelope are
discussed in van der Tak et al. (1999\nocite{1999ApJ...522..991V})
using an extensive set of line and continuum observations. The authors
conclude that the mid-IR and far-IR emission originates from the envelope,
and that the line-of-sight to the source nearly coincides with the
opening cone carved by an East-West oriented bipolar CO outflow.  There
is ample evidence that the circumstellar material is not distributed in a spherically 
symmetric fashion and that a low-opacity pathway close to the
line of sight has important effects on the source's appearance (Preibisch et al. 2003\nocite{2003A&A...412..735P}).
\newline
{\it Mid-IR morphology:} The COMICS image presents three 
sources: bright IRS1, a cometary shaped emission feature to the
South-West, and a small emission region to the North-West. IRS1 is so bright that the
spider diffraction pattern and the detector cross-talk (the vertical
dark lane) are visible in Fig.\,\ref{gl2591a}. The
cometary shaped feature to its South-West closely follows the optically
thin \ion{H}{ii} region VLA1 (Wynn-Williams et
al. 1977\nocite{1977ApJ...211L..89W}; Trinidad et al. 2003\nocite{2003ApJ...589..386T}). 
The relative position of the source at the Northern rim with respect to the other two 
is consistent with 8\,GHz source ``n4'' from Tofani et al. (1995\nocite{1995A&AS..112..299T}).\newline 
{\it Model results:} 
In Fig.\,\ref{gl2591b}, we use JCMT submm data from Jenness et
al. (1995\nocite{1995MNRAS.276.1024J}), and CSO data from Mueller et al. (2002),
that have comparable beamsizes (18\arcsec~and 14\arcsec, respectively).  At mm wavelengths we show data from
G\"{u}rtler et al. (1991) and one datapoint at 1.3\,mm from Walker, Adams
\& Lada (1990\nocite{1990ApJ...349..515W}). The G\"{u}rtler et al. data
were taken with a beam of $\sim$10\arcsec~whereas the Walker et al. data have a 30\arcsec~beam, possibly 
explaining the flux level difference between these two sets. The far-IR KAO data of
Lada et al. (1984\nocite{1984ApJ...286..302L}) closely corresponds
to the IRAS 60\,\micron and 100\,\micron photometry.

AFGL\,2591 IRS1 presents an interesting case, as no simultaneous SED
and intensity profile fit can be obtained. The bolometric luminosity
implied by the observed SED impedes any reasonable model fit to the intensity
profile (see e.g. the dot-dashed $p=1.0$ model in Fig.\,\ref{gl2591b}a
and b).  We attempt to find a solution by simply fitting the (sub)mm
SED and intensity profile.
The compromise that does fit the 24.5\,\micron flux and (nearly) the
intensity profile is given by a $p=1.5$ model with a large outer
radius of $\sim 5\arcmin$ (1.6\,pc). The discrepancy found between
spherical models and AFGL\,2591 is not limited to the mid and near-IR
but is still severe at far-IR wavelengths. Such models require a
substantial revision of the object's bolometric luminosity. In other
words, the source is more compact than predicted by spherical
models. A demonstration of a more appropriate 2-D modelling applied to
this source is reported in Preibisch et
al. (2003\nocite{2003A&A...412..735P}). These authors using speckle
interferometry probing scales of 170\,AU. They are able to model a
resolved structure either as the inner rim of a circumstellar disk or
as the dust sublimation radius of the MYSO envelope.

\begin{figure}[t]
  \center{\includegraphics[height=8.5cm,width=8.5cm]{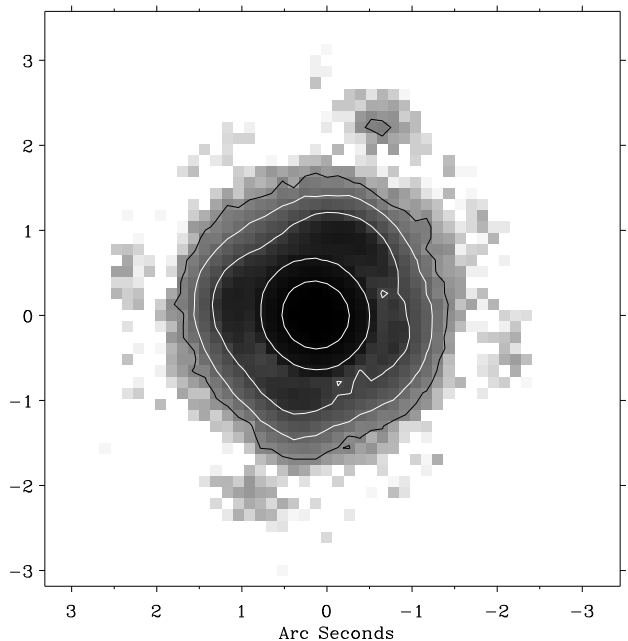}}
  \caption[]{COMICS 24.5\,\micron image of NGC\,2264 IRS1. Contour
  levels are at 1\%, 2\%, 5\%, 10\%, and 40\% of peak flux
  density ($\rm 4.1\,10^{2}\,Jy\,arcsec^{-2}$). North is up, East is
  to the left. }
  \label{ngc2264a}
\end{figure}

\begin{figure}[t]
  \center{\includegraphics[height=12cm,width=8.5cm]{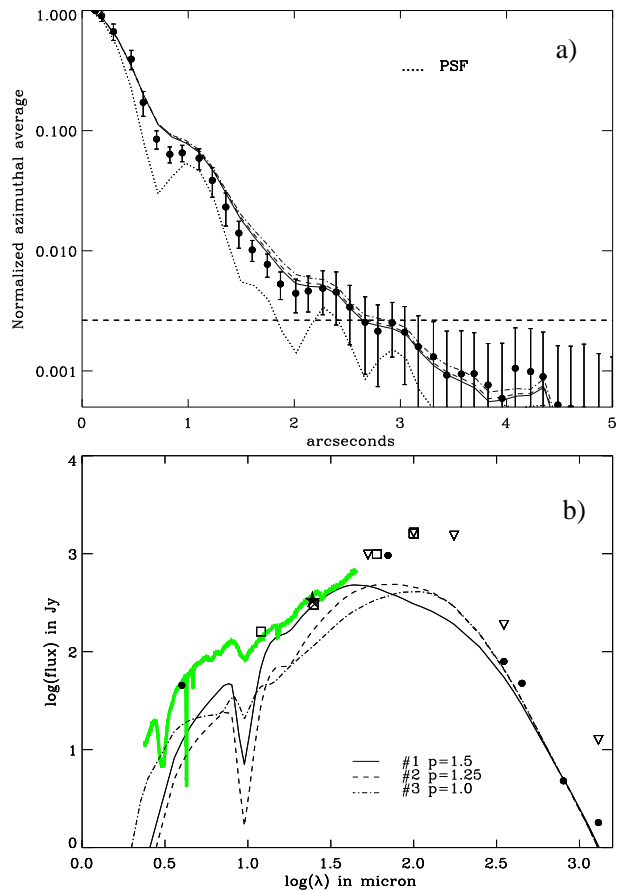}}
  \caption[]{As Fig.\,\ref{s140b}, but for NGC\,2264\,IRS1. The SED data is described in Sect.\,\ref{ngc2264}. The best fitting model has $p=1.5$ radial density profile.}
  \label{ngc2264b}
\end{figure}

\subsubsection{NGC\,2264\,IRS1 (\object{AFGL\,989}, Figs.\,\ref{ngc2264a} and \ref{ngc2264b})}
\label{ngc2264}

{\it Description:} A multi-wavelength study by Schreyer et al. (2003)
suggests that NGC\,2264\,IRS1 is a young B-type star surrounded by
low-mass companions located in a low-density cavity of a
clumpy, shell-like, and dense cloud remnant. 
NGC\,2264\,IRS1 shows
evidence for a CO outflow oriented along our line of sight (Schreyer
et al. 1997\nocite{1997A&A...326..347S}). Attempts to resolve sub-arcsecond structure
related to the CO outflow cavities in the near-IR with HST (Thompson
et al. 1998\nocite{1998ApJ...492L.177T}) and using speckle techniques (Alvarez et
al. 2004\nocite{2004A&A...427..505A}) failed. A reflection nebula on scales of arcseconds
is clearly visible in the optical (Scarrott \& Warren-Smith
1989\nocite{1989MNRAS.237..995S}) and near-IR (Schreyer et al. 1997).\newline
{\it Mid-IR morphology:} The COMICS image shows a single, symmetric object (IRS1). 
Most of the substructure seen in the image are diffraction patterns.\newline
{\it Model results:}   The model fits to SED and intensity profile are
presented in Fig.\,\ref{ngc2264b}. IRS1 coincides with submm source MMS5 (Ward-Thompson et
al. 2000; Peretto et al. 2006\nocite{2006A&A...445..979P}) if one takes the correct near-IR
coordinates (from 2MASS: $\alpha=06^{\rm h}41^{\rm m}10.16^{\rm s}$ and $\delta=+09\degr29\arcmin33.7\arcsec$) and the phase-referenced map by Nakano et al. (2003\nocite{2003PASJ...55....1N}).
We emphasise that the often reported lack of (sub)mm emission of IRS1 is due to wrong coordinates for IRS1 
combined with poor astrometry (Ward-Thompson et
al. 2000\nocite{2000A&A...355.1122W}; Schreyer et
al. 2003\nocite{2003ApJ...599..335S}; Nakano et
al. 2003\nocite{2003PASJ...55....1N}; Peretto et
al. 2006\nocite{2006A&A...445..979P}). Far-IR (25\,\micron and onwards)
observations presented in the literature are done with beams larger
than 20\arcsec~(Harvey, Campbell \& Hoffmann 1977\nocite{1977ApJ...215..151H}; Chini et
al. 1986\nocite{1986A&A...167..315C}; IRAS) and are considered to be upper
limits. The model fit uses the JCMT and IRAM (sub)mm data presented in Ward-Thompson et al. (2000) with 
beam FWHM of 6\arcsec, 8\arcsec, 13\arcsec~and 12\arcsec~for the 350, 450, 800 and 1300\,\micron data, respectively.
Archive {\it Spitzer} 70\,\micron MIPS  data are available for IRS1. We extracted the photometry applying the non-linearity correction recipe of Dale 
et al. (2007)\nocite{2007ApJ...655..863D}. A  70\,\micron flux value of 960\,Jy for IRS1 is used in the model fits.

The modelling indicates that bolometric luminosities that generate reasonable model fits
to the intensity profile, find a rough correspondence with the
mid-IR flux levels only if the powerlaw exponent equals $p=1.5$. The 70\,\micron MIPS is not attained by any model. It is
clear that any spherical model aimed at fitting the far-IR or (sub)mm
part of the SED would need much larger luminosities, which is incompatible
with the 24.5\,\micron intensity profile.

\begin{figure}[t]
  \center{\includegraphics[height=8.5cm,width=8.5cm]{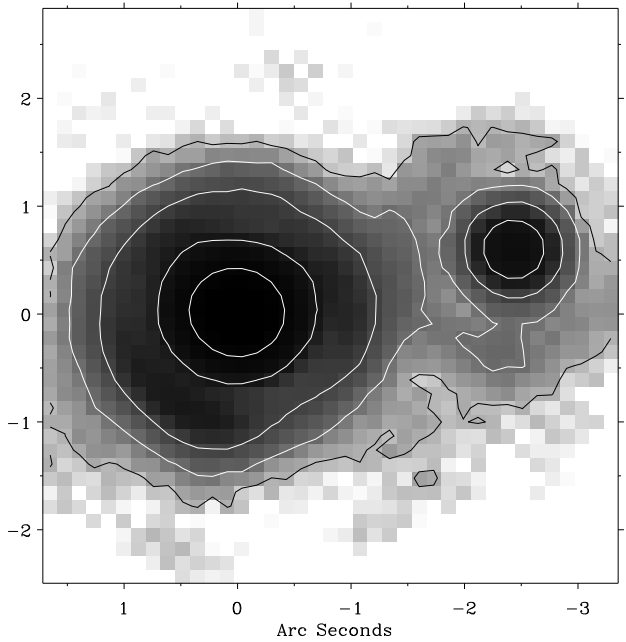}}
  \caption[]{COMICS 24.5\,\micron image of the S255 region. Contour
  levels are at 1\%, 2\%, 5\%, 10\%, and 40\% of peak flux
  density ($\rm 1.9\,10^{2}\,Jy\,arcsec^{-2}$). North is up, East is
  to the left.}
  \label{s255a}
\end{figure}

\begin{figure}[t]
  \center{\includegraphics[height=12cm,width=8.5cm]{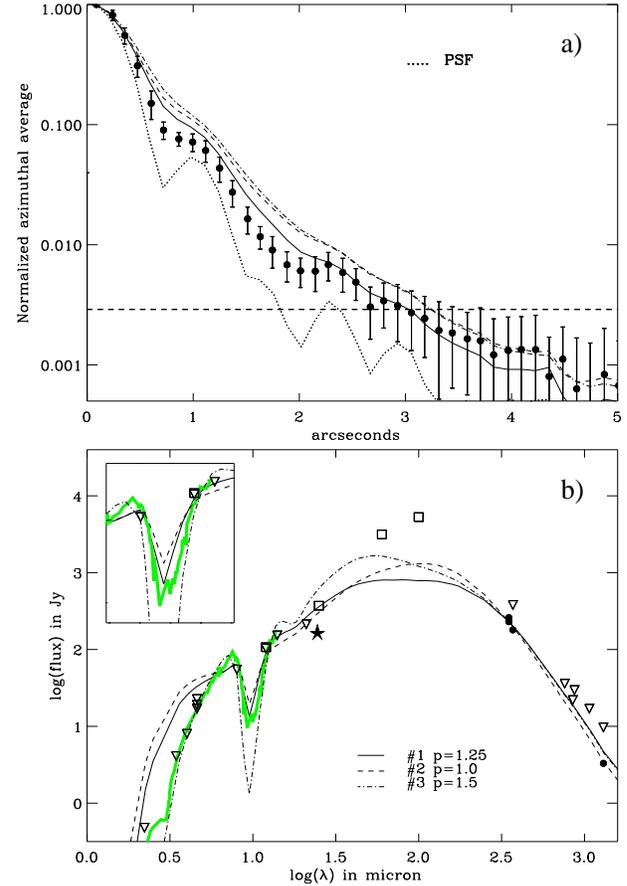}}
  \caption[]{As Fig.\,\ref{s140b}, but for S255 IRS3. The SED data is described in Sect.\,\ref{subs255}. The best fitting model has $p=1.25$ radial density profile.}
  \label{s255b}
\end{figure}

\subsubsection{S255 (Figs.\,\ref{s255a} and \ref{s255b})}
\label{subs255}

{\it Description:} The region is dominated by two near-IR 
sources (NIRS1 and 3)
and their bipolar IR reflection nebulae that 
were initially identified as the single source IRS1 (Beichman et al. 1979\nocite{1979ApJ...232L..47B}; Tamura et al. 1991\nocite{1991ApJ...378..611T}; Itoh et
al. 2001\nocite{2001PASJ...53..495I}). High resolution near-IR speckle observations by
Alvarez \& Hoare (2004) resolve the bipolar reflection nebula of
NIRS1 (the Western source), revealing it to be twisted in an ``S'' shape. 
High resolution Subaru near-IR polarimetric data indicate a depolarisation plane
perpendicular to the near-IR bipolar nebula, suggesting the presence of a disk 
(Jiang et al. 2008). \newline
{\it Mid-IR morphology:} The COMICS image shows NIRS3 to be the dominant source at 
24.5\,$\mu$m. Both mid-IR sources are found to be symmetric and resolved.\newline
{\it Model results:} We focus our analysis on the bright source
NIRS3 (the Eastern source) for the SED and intensity profile fit. The spectral range of the 
ISO spectrum does not include the 9.7\,\micron silicate feature.  Instead we use the spectrum
published in Willner et al. (1982\nocite{1982ApJ...253..174W}), that describes the depth
of the feature, but does not constrain the slope of the continuum
longwards of it. The short wavelength range of the Willner et al. spectrum
corresponds well with photometry taken by Evans et
al. (1977\nocite{1977ApJ...217..448E}).  The SED uses 
(sub)mm data from the combined ``core-envelope'' emission at
350\,\micron with a FWHP of 30\arcsec from Metzger et al. (1988\nocite{1988A&A...191...44M}), and
measurements (FWHP of 40\arcsec) by Richardson et al. (1985\nocite{1985MNRAS.216..713R}).
The mm measurement is by Chini et
al. (1986a)\nocite{1986A&A...154L...8C} obtained with the IRTF with a 90\arcsec~beamsize. Other sets of (sub)mm points
with higher fluxes are available in the literature and are represented
by open triangles (Richardson et al. 1985; Metzger et al. 1988; Klein
et al. 2005\nocite{2005ApJS..161..361K}). The total COMICS flux level (i.e. NIRS1 and 3 taken together)
is within 5\% of the MSX flux. More 
emission structure is hidden in the IRAS beam (see e.g. Longmore et al. (2006) 
for larger scale 18.7\,\micron images).

S255\,IRS3 constitutes a source that is mildly resolved at
24.5\,\micron. The intensity profile for IRS3 has been
obtained by excluding the region between  P.A. $-$45\degr and $-$135\degr,
where emission from IRS1 dominates. In Fig.\,\ref{s255b} the observations are
compared to three spherical envelope models, the parameters of which
are listed in Table\,\ref{tabtar}. The model with a 
radial density distribution with power exponent $p=1.25$ best fits both the SED and intensity profile. The $p=1.5$ model
is clearly excluded as it requires high optical depths not
substantiated by the silicate absorption profile.  The case of S255 IRS3 is similar to the one encountered for M8E-IR.
The models are required to have large envelopes which increase the long wavelength flux. 
This can be achieved without changing the bolometric flux, keeping the intensity 
profile unchanged.

\begin{figure}[t]
  \center{\includegraphics[height=7cm,width=8.5cm]{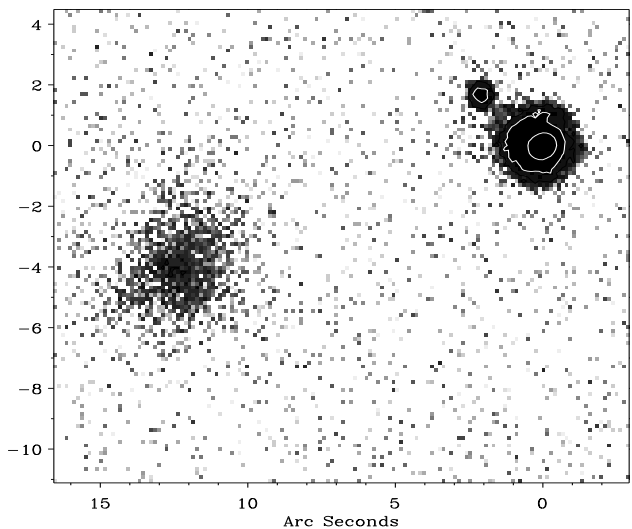}}
  \caption[]{COMICS 24.5\,\micron image of the AFGL\,5180 region. The principal Western source corresponds to
    NIRS1 (Tamura et al. 1991). Contour
  levels are at 5\%, 10\%, and 40\% of peak flux
  density ($\rm 4.5\,10^{2}\,Jy\,arcsec^{-2}$). North is up, East is
  to the left.}
  \label{gl5180a}
\end{figure}

\begin{figure}[t]
  \center{\includegraphics[height=12cm,width=8.5cm]{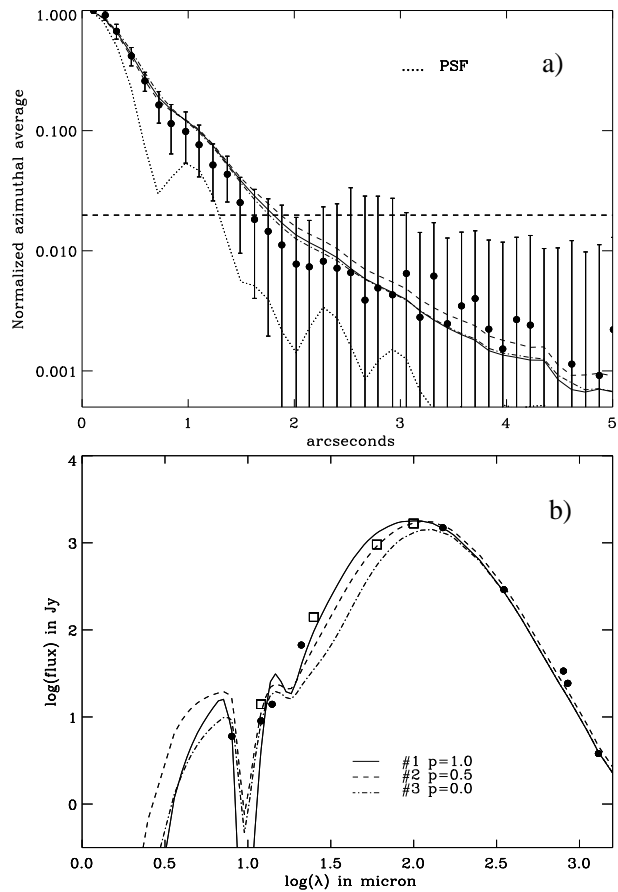}}
  \caption[]{As Fig.\,\ref{s140b}, but for AFGL\,5180. The SED data is described in Sect.\,\ref{gl5180}. The best fitting model has $p=1.0$ radial density profile.}
  \label{gl5180b}
\end{figure}

\subsubsection{AFGL\,5180 (Figs.\,\ref{gl5180a} and \ref{gl5180b})}
\label{gl5180}

{\it Description:} AFGL\,5180 has
a blueshifted CO flow with a P.A. of $\sim 130\degr$ (Snell et al. 1988\nocite{1988ApJ...325..853S}). Saito et
al. (2006\nocite{2006PASJ...58..343S}) show that the IRAS point source breaks up into two main
(sub)mm cores (see also Minier et al. 2005\nocite{2005A&A...429..945M}).  One of these
cores is centred on the near-IR source NIRS1 as identified by Tamura et
al. (1991\nocite{1991ApJ...378..611T}).  The region reveals an intricate collection of mm
cores mixed in with various near-IR sources (Tamura et
al. 1991; Saito et al. 2007\nocite{2007ApJ...659..459S}).\newline 
{\it Mid-IR morphology:} At 24.5\,\micron and at the resolution of Subaru, 
AFGL\,5180 consists of three sources: one extended source and two point-like sources. We follow
the nomenclature by Tamura et al. (1991), in which the
main discrete source can be identified as NIRS1 and the diffuse source as NIRS2.
The region was recently imaged by Longmore et al. (2006\nocite{2006MNRAS.369.1196L}) in
the mid-IR. Their deconvolved images at 7.9\,\micron shows that NIRS1 itself
consists of three components that have a mutual distance of about 1\arcsec. 
The COMICS image shows that NIRS1 is somewhat asymmetric, 
but it is not resolved in multiple sources.\newline
{\it Model results:} We concentrate our discussion on NIRS1, the main
mid-IR source. Model comparisons to the observations
are less well constrained than in the previous cases, because NIRS1 lacks
a 10\,\micron spectrum. An independent measure of the total optical depth is
therefore absent. We use the (sub)mm photometry from Gear et al. (1988\nocite{1988MNRAS.231P..55G}) and
Thompson et al. (2006\nocite{2006A&A...453.1003T}). The Gear et al. data have FWHM beamsizes of about an arcminute, 
whereas the Thompson et al. data are taken with SCUBA with FWHM beamsizes of 8\arcsec~and 14\arcsec~for the 450\,\micron and 850\,\micron observations. 
The far-IR datapoint is from Ghosh et
al. (2000\nocite{2000BASI...28..515G}). For the mid-IR we use the MSX photometry, that
have a good correspondence with the IRAS data points. We find
reasonable fits to the intensity profile for $p=0.0$, $p=0.5$ and
$p=1.0$ radial density distributions (see Fig.\,\ref{gl5180b}),
but the MSX and IRAS points in the SED are better reproduced by the $p=1.0$
model. Any steeper distributions may provide equally good fits, except
that they would require optical depths exceeding $A_{\rm V}=200^{m}$
and much larger outer radii then adopted in our grid, which are
typical for most sources in our sample.  The $p=1.0$ model
already requires a comparatively large optical depth with respect to the other
MYSO envelopes for it to fit the MSX data.

\begin{figure}[t]
  \center{\includegraphics[height=8.5cm,width=8.5cm]{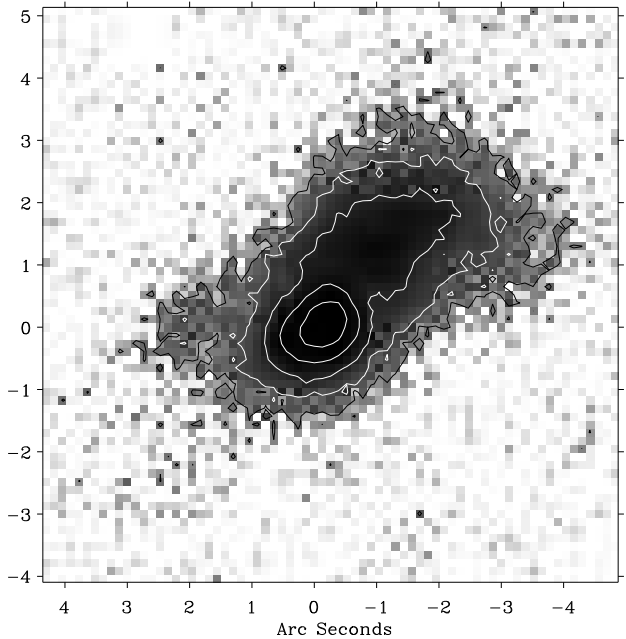}}
  \caption[]{COMICS 24.5\,\micron image of IRAS\,20126+4104. Contour
  levels are at 5\%, 10\%, 20\%, 40\%, and 70\% of peak flux
  density ($\rm 4.7\,10^{1}\,Jy\,arcsec^{-2}$). North is up, East is
  to the left.}
  \label{20126a}
\end{figure}

\begin{figure}[t]
  \center{\includegraphics[height=12cm,width=8.5cm]{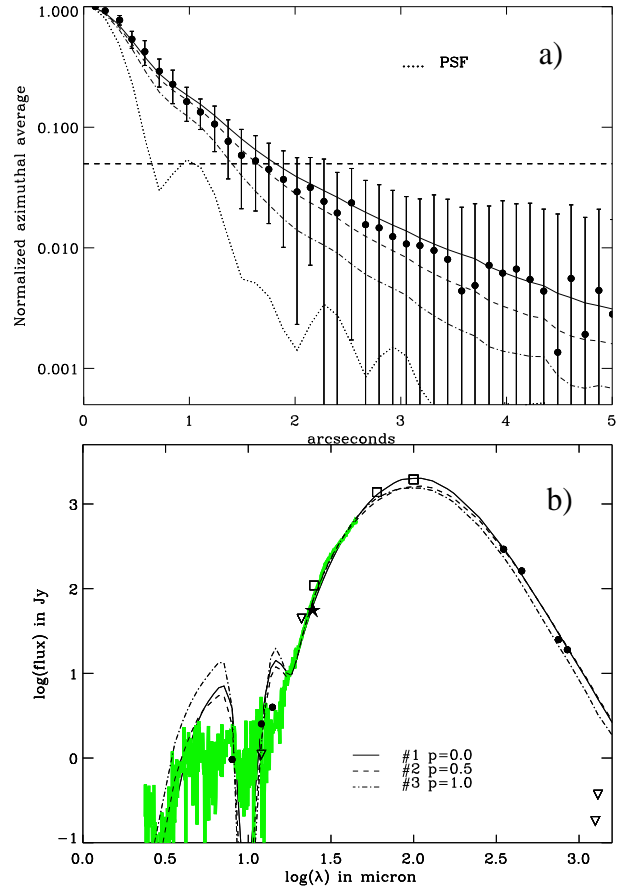}}
  \caption[]{As Fig.\,\ref{s140b}, but for IRAS\,20126+4104. The SED data is described in Sect.\,\ref{iras20126}. The best fitting models have $p=0.0$ or $p=0.5$ radial density profiles.}
  \label{20126b}
\end{figure}

\subsubsection{\object{IRAS\,20126+4104} (Figs.\,\ref{20126a} and \ref{20126b})}
\label{iras20126}


{\it Description:} IRAS\,20126 is one of the best examples of a relatively massive young star that 
is actively fed by an accretion disk. The star is thought to have a mass of approximately $7\,M_{\odot}$ 
(Cesaroni et al. 2005\nocite{2005A&A...434.1039C}). The system is known to show
jet/outflow phenomena at a P.A. of about $-60\degr$ (see Cesaroni et
al. 2005\nocite{2005A&A...434.1039C}; Su et al.
2007\nocite{2007ApJ...671..571S}), with the jet oriented 
almost perpendicular to the line of sight (Cesaroni et al. 1999). At wavelengths shorter than
20\,\micron a ``dark lane'' separates two emission regions (Sridharan et
al. 2005\nocite{2005ApJ...631L..73S}; De Buizer 2007). This morphology
could be either due to the presence of two tight clusters (in which
case the dark lane does not correspond to a physical structure) or
due to a silhouetted disk with emission emanating from the cavity walls of the outflow either side of
the disk (De Buizer 2007).\newline
{\it Mid-IR morphology:} The COMICS image consists of two
emission regions. The morphology of the North-West source (source ``5'' in De Buizer 2007) corresponds to the one 
presented in Shepherd et al. (2000\nocite{2000ApJ...535..833S}) at 17.9\,$\mu$m and coincides with the
direction of the outflow.  The South-Eastern peak emission overlaps the two
sources that are seen separated by a dark lane at shorter wavelengths. No
separation due to a dark lane is detected, although it is extended
perpendicular to the dark lane seen at shorter wavelengths. 
The two
sources have a separation of 0.7\arcsec~in 18.3\,\micron images by De
Buizer (2007). The somewhat lower resolution at 24.5 micron is probably partly
responsible for the failure to separate the two sources. In addition, 
it seems likely that lower extinction and thermal emission from the putative dark lane
itself could be blurring the distinction between the two emission
components. It would argue in favour of the dark lane being a physical
structure rather than a clearance of emitting material as suggested in
De Buizer (2007).\newline
{\it Model results:} 
The ISO-SWS spectrum has a low SNR at wavelengths shorter than
10\,$\mu$m. This includes the silicate absorption profile, inhibiting a
independent measure of the envelope's optical depth. A sharp rise of
the spectrum towards longer wavelengths indicates a very high
extinction. 
In the model fits we use the photometry collected by
Hofner et al. (2007\nocite{2007A&A...465..197H}, and references therein). We fit only
the large beam data presumably corresponding to the dusty halo. The submm observations are obtained
with the JCMT by Cesaroni et al. (1999\nocite{1999A&A...345..949C}) with HPBW between 7\arcsec~and 14\arcsec. The intensity profile
is obtained from the region that excludes emission from the North-Western source.
We find good simultaneous fits to both intensity profile and SED, with a preference
for flat radial density distributions with powerlaw indices $p=0.0$ and $p=0.5$. 
Steeper density distributions are not able to fit the 24.5\,\micron 
intensity profile. These values for the density distribution power index are at odds
with Hofner et al. (2007) and van der Tak et al. (2000). The former 
find consistency between the observed SED and what one would expect from an 
accretion disk embedded in a spherical infalling halo (on scales of 10\arcsec)
with $p=1.5$. Van der Tak et al. (2000) find $p=1.75$, on scales larger than
the ones probed by the 24.5\,\micron image.

\begin{figure}[t]
  \center{\includegraphics[height=8.cm,width=8.5cm]{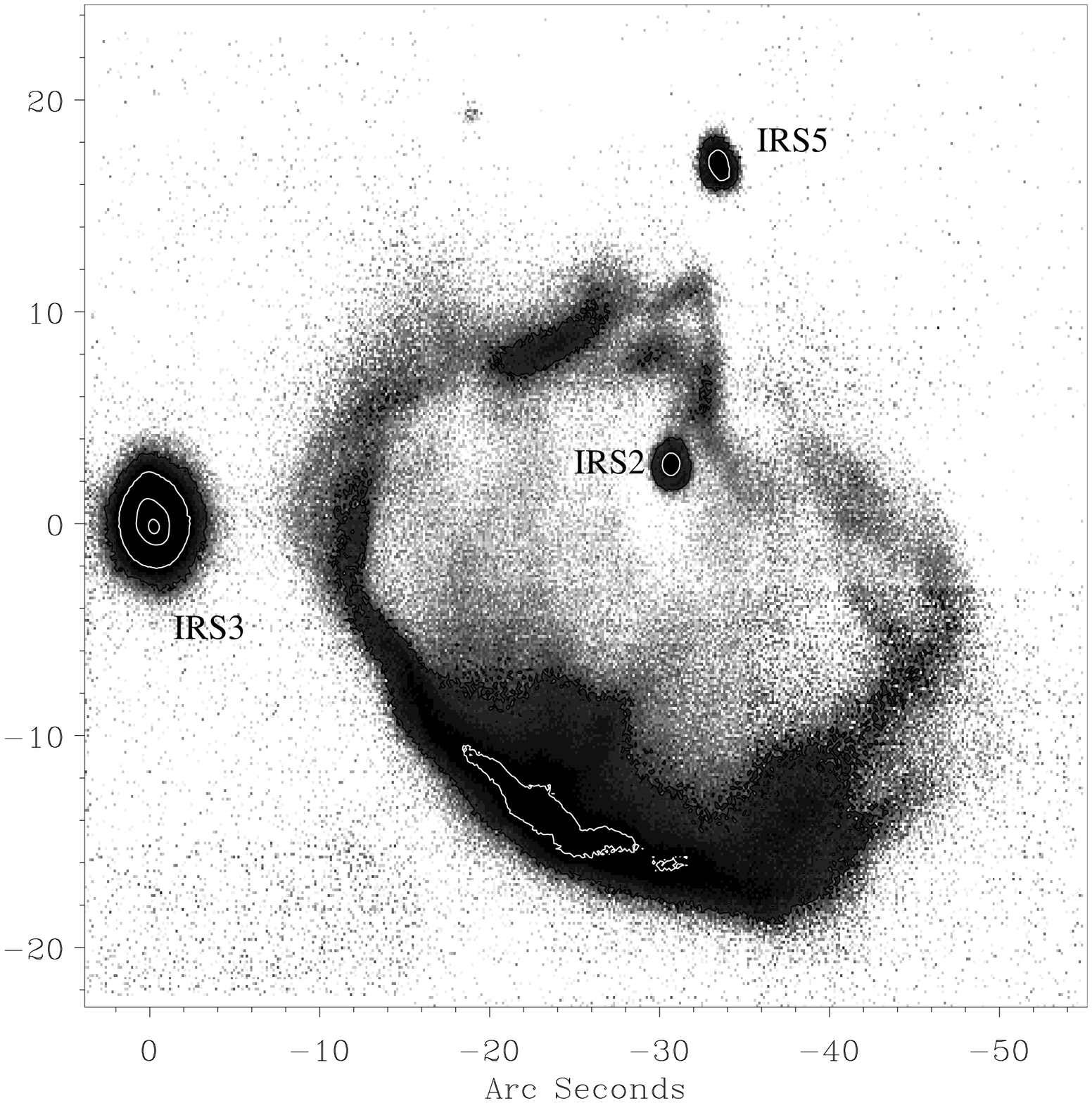}}
  \caption[]{COMICS 24.5\,\micron image of the Mon R2 region. Contour
  levels are at 1\%, 3\%, 20\%, and 70\% of peak flux
  density ($\rm 5.0\,10^{2}\,Jy\,arcsec^{-2}$). North is up, East is
  to the left.}
  \label{monr2a}
\end{figure}

\begin{figure}[t]
  \center{\includegraphics[height=12cm,width=8.5cm]{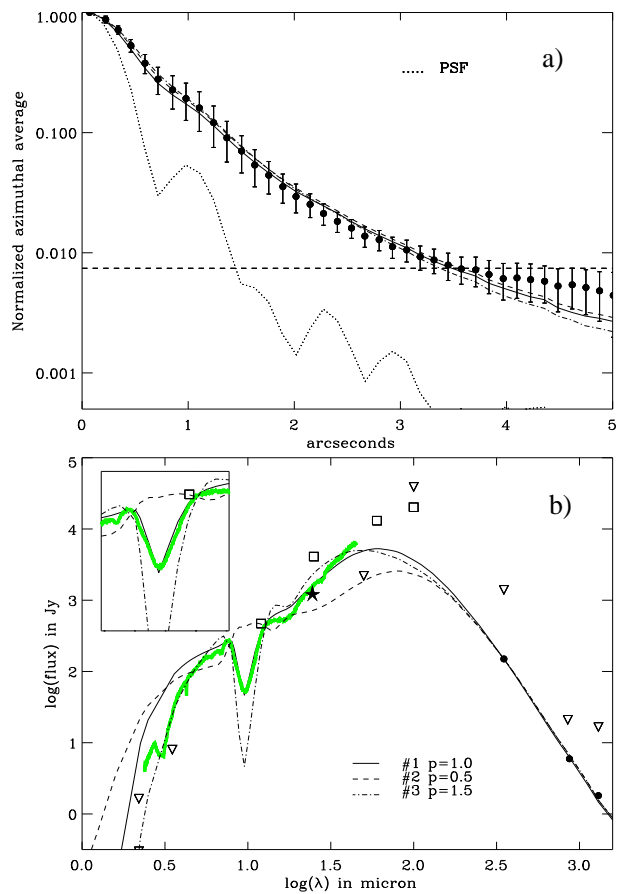}}
  \caption[]{As Fig.\,\ref{s140b}, but for Mon R2 IRS3. The SED data is described in Sect.\,\ref{monr2}. The best fitting model has $p=1.0$ radial density profile.}
  \label{monr2b}
\end{figure}

\subsubsection{Mon\,R2 (Figs.\,\ref{monr2a} and \ref{monr2b})}
\label{monr2}

{\it Description:}
The central region of the Mon\,R2 cloud consists of at least five
bright IR sources within $\sim 0.25$\,pc (Beckwith et
al. 1976\nocite{1976ApJ...208..390B}). The region shows complex submm
dust continuum emission which is well illustrated by the 850\,\micron
map by Giannakopoulou et al. (1997\nocite{1997ApJ...487..346G}).
High resolution near-IR imaging (0.075\arcsec~resolution)
resolves the brightest mid-IR source (IRS3) into a triple system surrounded by strong diffuse
nebulosity (Preibisch et al. 2002\nocite{2002A&A...392..945P}).\newline
{\it Mid-IR morphology:}
The COMICS mosaic in Fig.\,\ref{monr2a} shows IRS3 to be the dominant discrete source at 24.5\,$\mu$m.
The large, 30\arcsec~scale mid-IR shell structure is identifiable in near-IR
continuum emission in which case it is due to dust scattering
(Howard et al. 1994\nocite{1994ApJ...425..707H}).  Its morphology corresponds
roughly to the extent of the blister \ion{H}{ii} region (Massi et
al. 1985\nocite{1985A&A...152..387M}).  If we compare the mid-IR image with these maps,
we may identify the 24.5\,\micron emission with the walls of the ionized
region. Maximum intensity
along the mid-IR ridge corresponds to the location of the source IRS1. The image
does not reveal any point source at this position. IRS2, IRS3 and IRS5 are all
observed to be extended sources. IRS3 is elongated with a P.A. along the 
direction of the main  binary (Preibisch et al. 2002).\newline
{\it Model results:} Submm and mm imaging at 870\,\micron and 1.3\,mm 
(HPBW of 18\arcsec~and 23\arcsec, respectively) by Henning et al. (1992\nocite{1992A&A...263..285H})
resolve and identify the various components of the cloud with
the IR sources. We show in Fig.\,\ref{monr2b} the intensity
profile and SED fits to IRS3, the most luminous mid-IR source of the
region.  Fig.\,\ref{monr2b}b also shows the Mueller et al. (2002) compilation of
continuum measurements at the long wavelengths,  including the 350\,\micron CSO measurement (HPBW 14\arcsec). The flux level of the
ISO spectrum corresponds closely to the COMICS flux measurement. The
intensity profile levels out at distances $>5\arcsec$ which is probably
due to contribution in flux by the diffuse emission in the region.

The case of Mon R2 IRS3 is similar to S140 IRS1. Simultaneous model
fit to the intensity profile and SED shows that models with a radial
density profile of $p=1.0$ are preferred. Again steep radial profiles
are excluded as they require high optical depths in order to provide
a good fit to the intensity profile.

\begin{figure}[t]
  \center{\includegraphics[height=8.5cm,width=8.5cm]{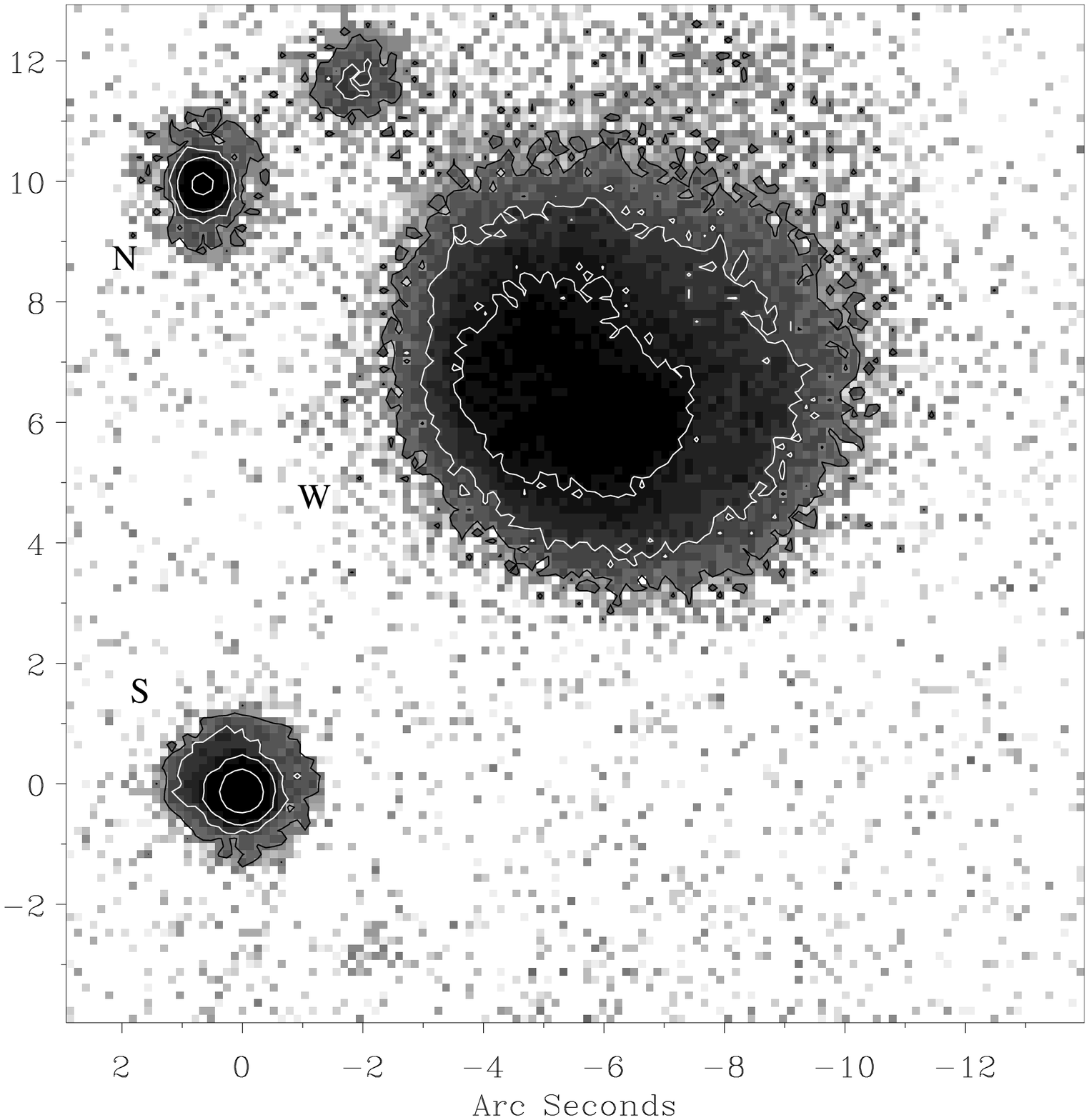}}
    \caption[]{COMICS 24.5\,\micron image of the AFGL\,437 region. Contour
  levels are at 5\%, 10\%, 20\%, and 50\% of peak flux
  density ($\rm 3.2\,10^{1}\,Jy\,arcsec^{-2}$). North is up, East is
  to the left.}
    \label{gl437sa}
\end{figure}

\begin{figure}[t]
  \center{\includegraphics[height=12cm,width=8.5cm]{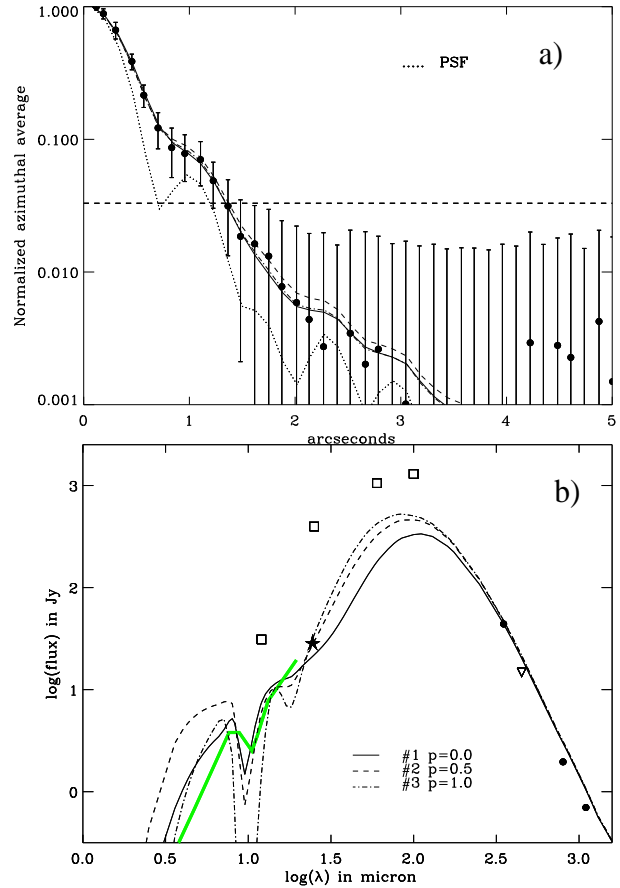}}
    \caption[]{As Fig.\,\ref{s140b}, but for AFGL\,437S. The SED data is described in Sect.\,\ref{gl437s}. The best fitting model has $p=0.0$ radial density profile.}
    \label{gl437sb}
\end{figure}

\subsubsection{AFGL 437S (Figs.\,\ref{gl437sa} and \ref{gl437sb})}
\label{gl437s}

{\it Description:} AFGL\,437 is a compact cluster of few dozen near-IR
sources located at a distance of 2.7 kpc (Wynn-Williams et al. 1981;
Weintraub \& Kastner 1996\nocite{1996ApJ...458..670W}). The cluster is
dominated by four bright sources 437N, S, E and W. The source 
437W dominates the radio emission, whereas only weak radio emission is
measured towards the 437S.  (Kurtz, Churchwell, \& Wood
1994\nocite{1994ApJS...91..659K}; Torrelles et
al. 1992\nocite{1992ApJ...392..616T}). Alvarez et al. (2004) resolve a
monopolar sub-arcsecond near-IR nebula from 437S. Water masers have
been detected towards 437W and 437N (Torrelles et al. 1992).
Weintraub \& Kastner (1996) find that 437N actually breaks up into two
components (see also Rayner \& McClean
1987\nocite{1987iawa.conf..272R}). The South-Eastern source of the two (named
WK34) is the most embedded and found to be responsible for the
region's molecular outflow and the dominant source of the near-IR
reflection nebula (see also Meakin et
al. 2005\nocite{2005ApJ...634.1146M}).\newline
{\it Mid-IR morphology:}
The COMICS image is dominated by the UC\ion{H}{ii} region associated
with 437W.  No discrete counterpart is found for 437W, only a diffuse
emission region.  The 437N source is resolved into two components,
with the driving source WK34 the most luminous of the two. Of the 4
sources detected, 437S is the most luminous and marginally resolved by
our observations at 24.5\,$\mu$m. \newline 
{\it Model results:} We
concentrate our analysis on 437S. Submm and mm observations were
performed by Dent et al. (1998\nocite{1998MNRAS.301.1049D}) with the JCMT at four wavelengths with
beamsizes between 16\arcsec~and 19\arcsec.  The ISO-SWS spectrum of the
region is dominated by emission from the \ion{H}{ii} region and is
discarded. Instead we use the 10\micron~photometry of 437S taken by
Wynn-Williams et al. (1981\nocite{1981ApJ...246..801W}). The model
fitting procedure discards the IRAS photometry, but takes into
account the COMICS flux measurement.

The spatial information derived from the marginally resolved source is
not enough to strongly constrain the various radial density
distributions. We show in Fig.\,\ref{gl437sb} three models that
reproduce the intensity profile and the continuum emission in the
SED. We find that $p=1.0$ profiles require large optical depth in
order to fit the (sub)mm and the 24.5\,\micron data. This seems to be
excluded by the silicate absorption profile. We therefore prefer
rather shallow radial density distributions with $p=0.0$ or $p=0.5$,
but with considerable uncertainty.

\subsection{Complex sources}
\label{compl}
MYSO sources presented in this section show evidence for multiple condensations within 
1\arcsec~of the profile centre (AFGL\,4029, AFGL\,961, and W3\,IRS5). Azimuthally
averaged intensity profiles yield consequently large uncertainties, which 
constrain the models only little. In the case of the fourth source presented
in this section, the Cep\,A star forming region, no 
discrete central source could be identified at all. We discuss the 24.5\,\micron
images alongside some literature background for each of these four objects.

\begin{figure}[t]
  \center{\includegraphics[height=8.5cm,width=8.5cm]{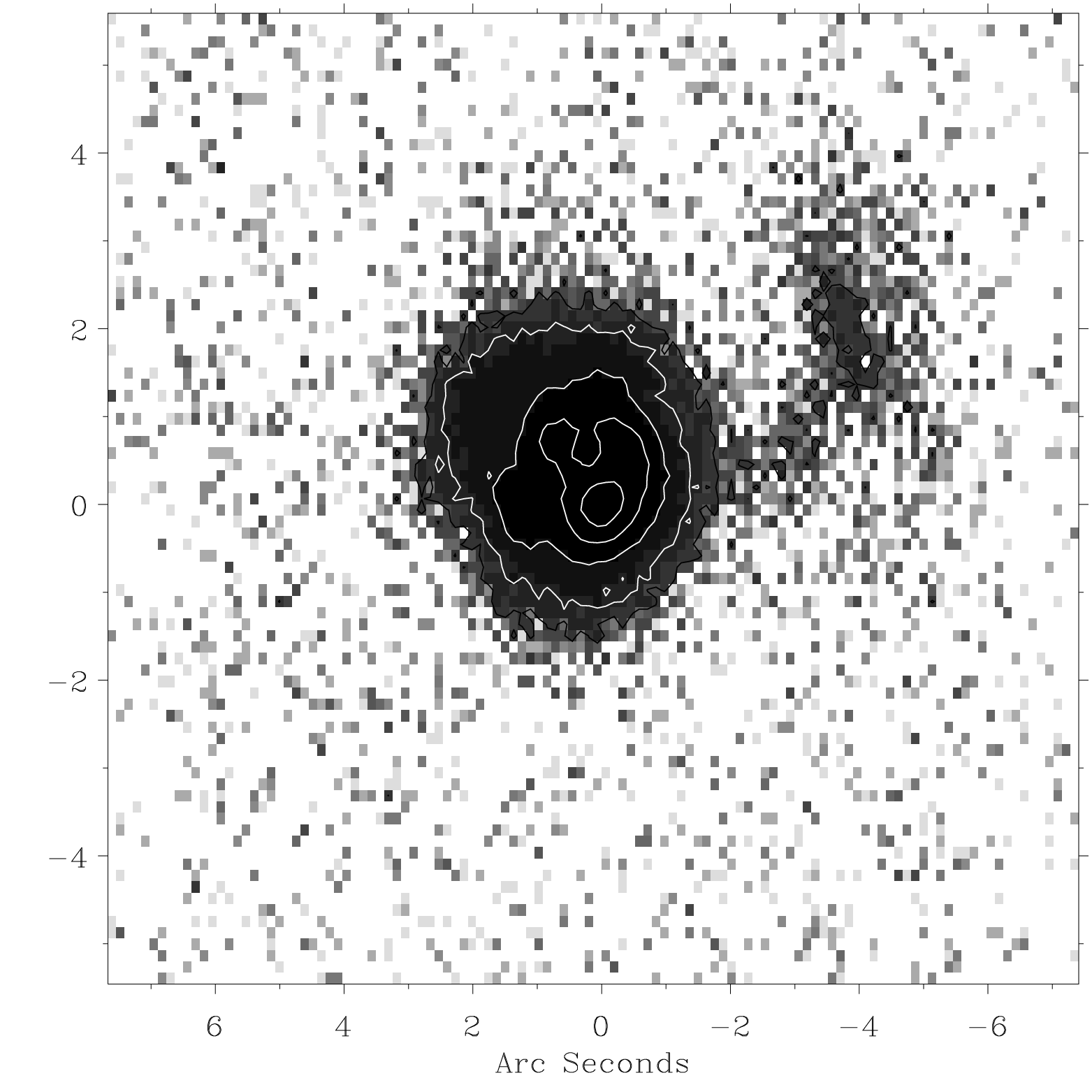}}
  \caption[]{COMICS 24.5\,\micron image of the AFGL\,4029 region. Contour
  levels are at 5\%, 10\%, 25\%, 50\%, and 80\% of peak flux
  density ($\rm 2.2\,10^{1}\,Jy\,arcsec^{-2}$). IRS2 is 20\arcsec~to the East to IRS1, and not shown in the image. 
North is up, East is to the left.}
\label{4029a}
\end{figure}

\begin{figure}[t]
  \center{\includegraphics[height=12cm,width=8.5cm]{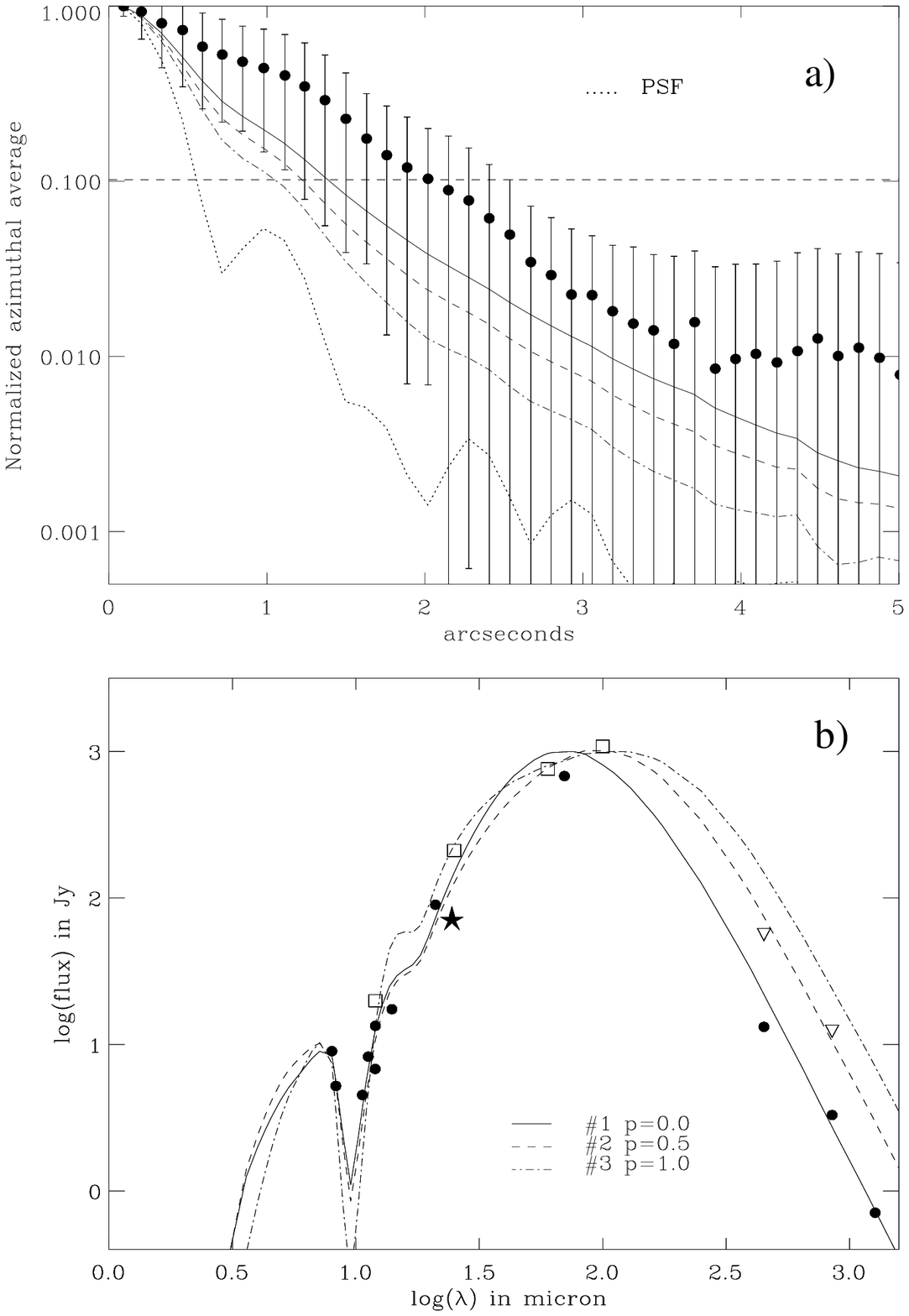}}
  \caption[]{As Fig.\,\ref{s140b}, but for AFGL\,4029. The SED data is described in Sect.\,\ref{gl4029}. The best fitting model has a $p=0.0$ radial density profile.}
  \label{4029b}
\end{figure}

\subsubsection{AFGL\,4029 (Figs.\,\ref{4029a} and \ref{4029b}) }
\label{gl4029}
{\it Description:} Beichman et al. (1979\nocite{1979ApJ...232L..47B}) have 
shown that the region consists of two sources at 10 and 20\,$\mu$m,
separated by about 20\arcsec. Both mid-IR sources have radio emission
(Kurtz et al. 1994\nocite{1994ApJS...91..659K}). IRS1 shows evidence for outflow
activity. Zapata et al. (2001\nocite{2001RMxAA..37...83Z}) find IRS1 to consist of a
double radio source, interpreted to be a binary object. They identify the Southern of the two sources
responsible for the outflow activity. \newline
{\it Mid-IR morphology:}
The COMICS image reveals two emission regions. A compact emission
region identified as IRS1 which is shown in Fig.\,\ref{4029a}, and a
diffuse emission region associated with IRS2 located at 20\arcsec~from 
IRS1 (not shown in Fig.\,\ref{4029a}). IRS1 shows a
discrete source with evidence for at least two condensations, the
Southern being the brightest one. In addition at least two bands of emission
are located within the inner 2\arcsec, rendering IRS\,1 a rather patchy
source at 24.5\,$\mu$m. Finally, IRS\,1 has an extended wing ($\sim$4\arcsec)
to the West. Scattered light emission and thermal emission due to extended nebulosity 
to the West of IRS1 is discussed in Deharveng et al. (1997)\nocite{1997A&A...317..459D} and Zavagno et al. (1999)\nocite{1999A&A...344..499Z}.
The 24.5\,\micron morphology of IRS1 follows closely the
radio morphology as presented in Zapata et al. (2001\nocite{2001RMxAA..37...83Z}). The
mid-IR wing is possibly the counterpart to the East-West extension seen
in high-resolution radio maps. \newline
{\it Model results:}
The mid-IR part of the SED is adopted from data presented in Zavagno et al. (1999). IRAM 1.27 mm flux (beamsize $\sim$15\arcsec)
is taken from Klein et al. (2005)\nocite{2005ApJS..161..361K}, and submm SCUBA
fluxes (beamsizes of 8\arcsec~and 14\arcsec~for the 450\,\micron and 850\,$\mu$m) are presented in Di Francesco et al. (2008)\nocite{2008ApJS..175..277D} . Finally
a {\it Spitzer} MIPS flux measurement gives a flux density of 680\,Jy for this source.

AFGL\,4029 presents an intensity profile that is characterized by a
large scatter in intensity, as represented by the errorbars in
Fig.\,\ref{4029b}, due to the presence of patchy emission within
$2\arcsec$. The lack of a 10\micron spectrum for the source also
complicates the analysis. The data of Zavagno et al. (1999) cover the
wings of the silicate absorption profile, providing some
constraints. Although the intensity profile shows a large scatter,
simultaneous model fits to SED and intensity profile are hard to find;
the basic problem being the extent of the source.  This
translates into a very luminous object, incompatible with the SED. In
Fig.\,\ref{4029b}, we show models with relatively flat density
profiles ($p=0.0$ or $0.5$), steeper profiles are incompatible
with the intensity profile.

\begin{figure}[t]
  \center{
    \includegraphics[height=8.5cm,width=8.5cm]{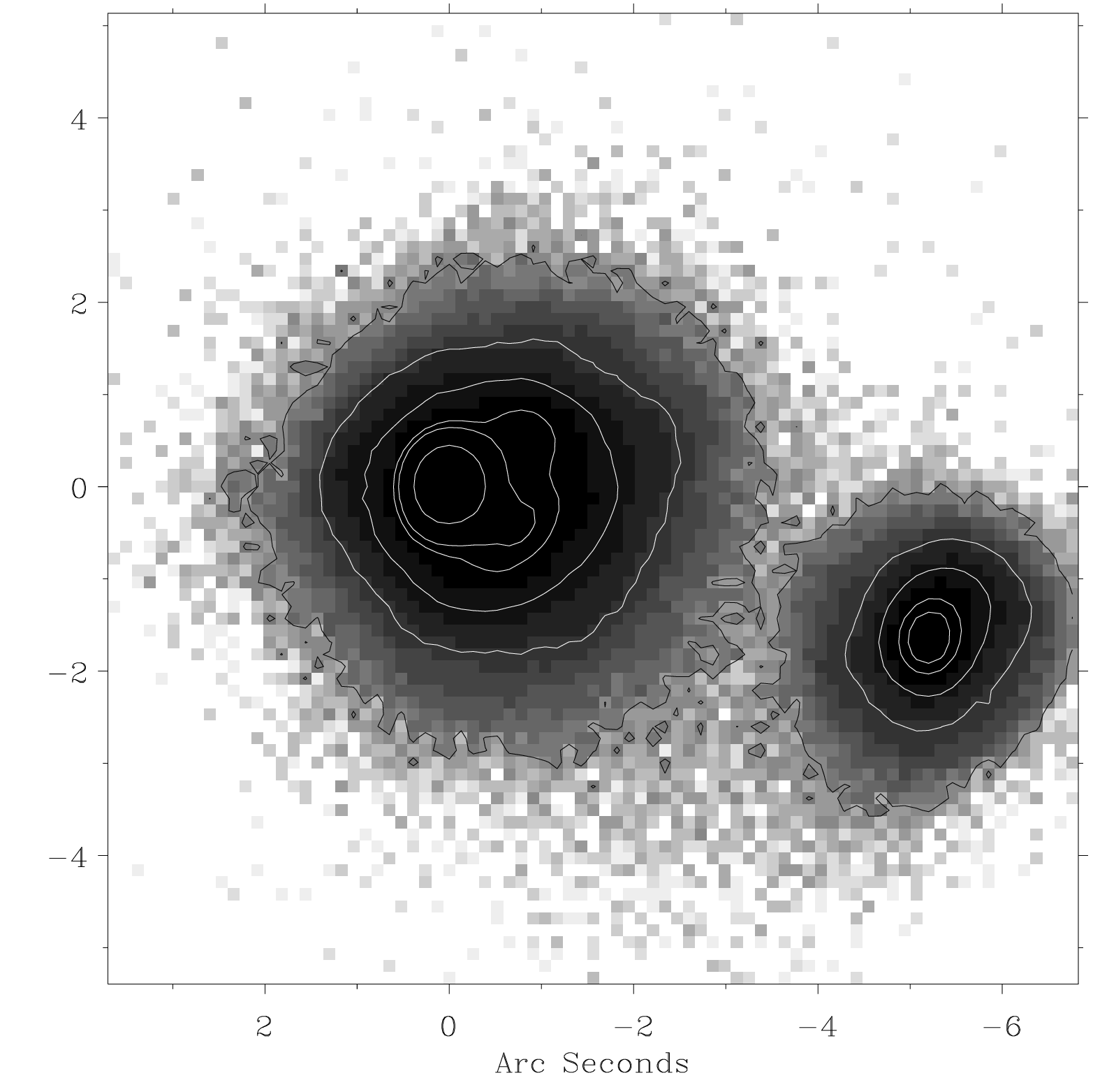}}
  \caption[]{COMICS 24.5\,\micron image of the AFGL\,961 region.  Contour
  levels are at 1\%, 4\%, 10\%, 20\%, 25\%, and 50\% of peak flux
  density ($\rm 1.3\,10^{2}\,Jy\,arcsec^{-2}$). North is up, East is
  to the left.}
\label{961a}
\end{figure}

\begin{figure}[t]
  \center{\includegraphics[height=12cm,width=8.5cm]{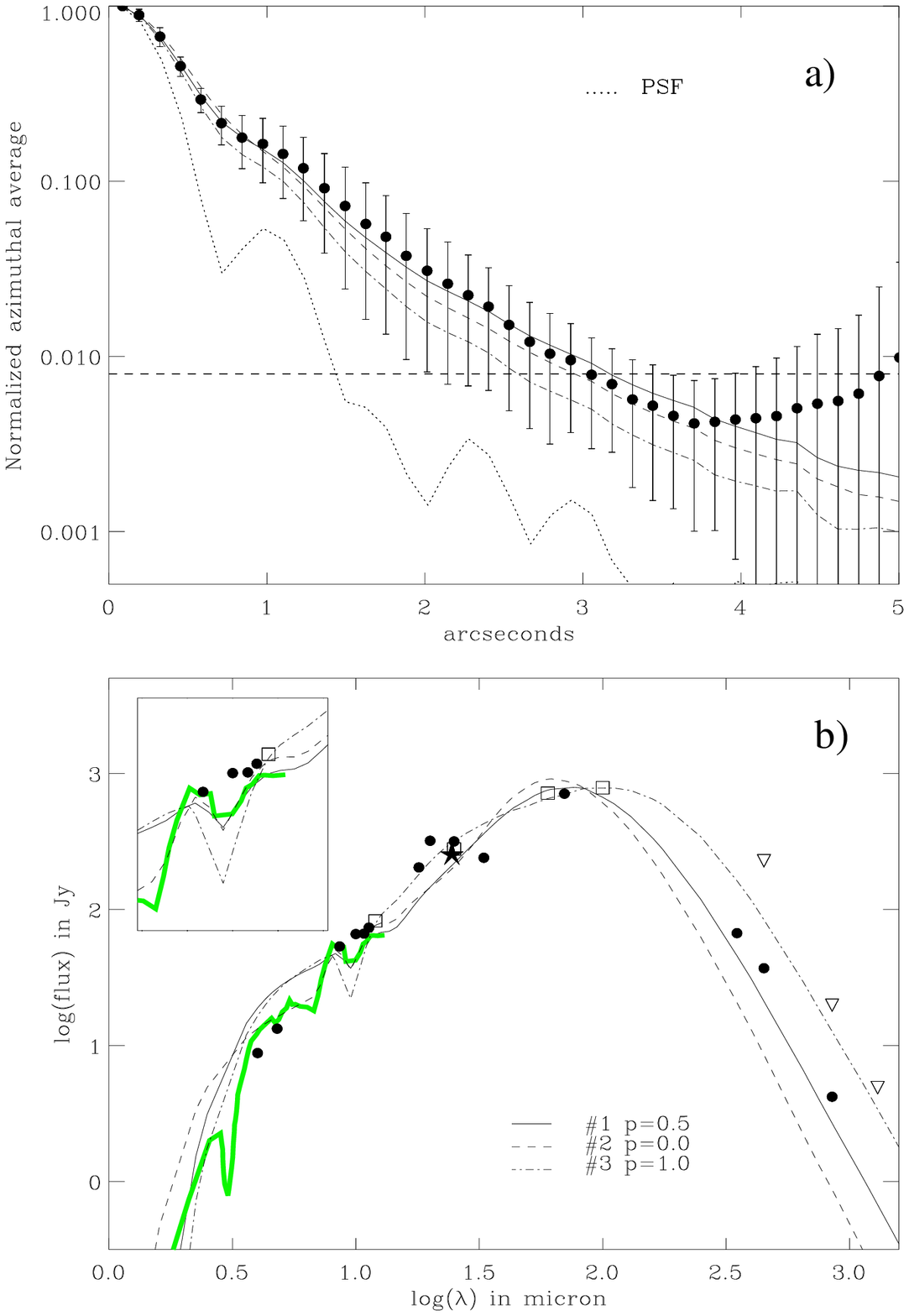}}
  \caption[]{As Fig.\,\ref{s140b}, but for AFGL\,961E. The SED data is described in Sect.\,\ref{gl961}. The best fitting model has a $p=0.5$ radial density profile.}
  \label{961b}
\end{figure}

\subsubsection{AFGL\,961 (Figs.\,\ref{961a} and \ref{961b})}
\label{gl961}

{\it Description:} The source is a well-known double object located in
the outskirts of the Rosette Nebula (Lenzen et al. 1984\nocite{1984A&A...137..365L}) and
surrounded by a stellar cluster (Aspin 1998\nocite{1998A&A...335.1040A}).  The East and
West components have a separation of about 6\arcsec and
display outflow phenomena (Lada \& Gautier 1982\nocite{1982ApJ...261..161L}). 
Snell and Bally (1986)\nocite{1986ApJ...303..683S} determined the radio spectral index, finding it to be 
consistent with an ionized wind.\newline
{\it Mid-IR morphology:} The dominant source in the COMICS image is
AFGL\,961E. The Western source 961W is discrete and somewhat extended in
a North-Westerly direction. AFGL\,961E displays two diffuse blobs to the
North-West and South-West of the peak emissions.\newline
{\it Model results:} 
Chini et al. (1986) model this source and find good correspondence
with a spherical model for wavelengths shorter than 100\,$\mu$m. An
emission excess with respect to their spherical model is found for
wavelengths longer than this.  We use SCUBA measurements at
450\,\micron and 850\,$\mu$m (beamsizes of 8\arcsec~and 14\arcsec), which are significantly lower.  Further,
we use the spectrum taken by Willner et al. (1982), mid-IR photometry from Cohen
(1973), Simon \& Dyck (1977\nocite{1977AJ.....82..725S}), and the
350\,\micron IRTF  measurement (HPBW 90\arcsec) is from G\"{u}rtler et al. (1991). We measured the
70\,\micron flux from archival {\it Spitzer} MIPS data, finding a flux of 720\,Jy for the source.
The broad SED and intensity profile are well fit by $p=0.5$ density powerlaws
with a moderate optical depth. Steeper density profiles do not fit the
intensity profile, and would additionally require either a higher
optical depth or a much larger envelope outer radius in order to fit
the submm points.

\begin{figure}[t]
  \center{
    \includegraphics[height=8.5cm,width=8.5cm]{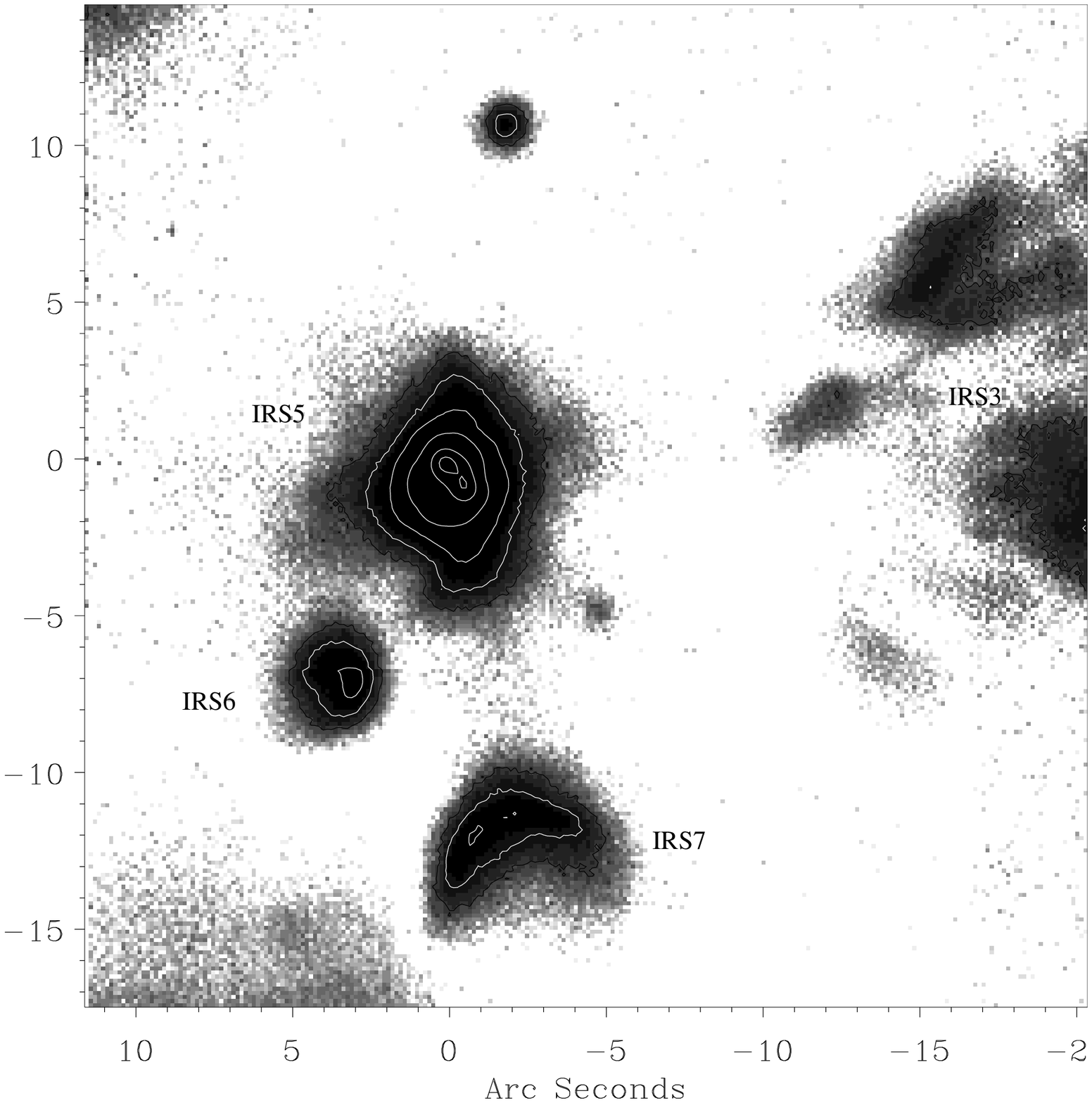}}
  \caption[]{COMICS 24.5\,\micron mosaic of the W3 region. Contour
  levels are at 1\%, 2\%, 5\%, 15\%, 50\%, and 85\% of peak flux
  density ($\rm 3.8\,10^{2}\,Jy\,arcsec^{-2}$). North is up, East is
  to the left.}
\label{w3a}
\end{figure}

\begin{figure}[t]
  \center{\includegraphics[height=12cm,width=8.5cm]{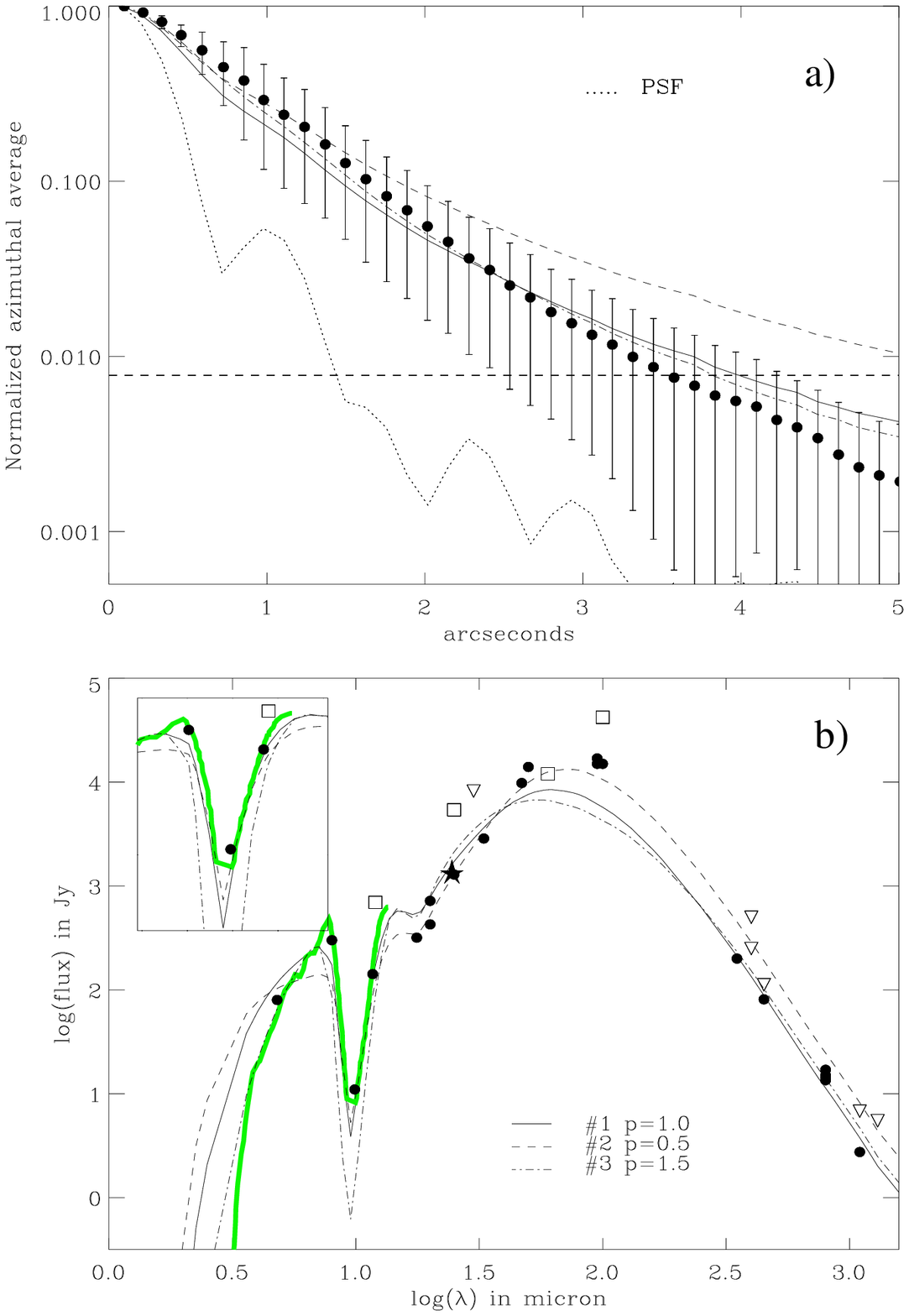}}
  \caption[]{As Fig.\,\ref{s140b}, but for W3\,IRS\,5. The SED data is described in Sect.\,\ref{w3}. The best fitting model has a $p=1.0$ radial density profile.}
  \label{w3b}
\end{figure}

\subsubsection{W3 (Fig.\,\ref{w3a} and \ref{w3b})}
\label{w3}

{\it Description:}
W3 is an important and complex region containing objects in various
stages of the formation process (Wynn-Williams et al. 1972\nocite{1972MNRAS.160....1W};
Claussen et al. 1994\nocite{1994ApJ...424L..41C}).  One prominent source in the region is
IRS5, which consists of a double source (1\arcsec~separation) at
infrared wavelengths (Howell et al. 1981\nocite{1981ApJ...251L..21H}; van der Tak et
al. 2005\nocite{2005A&A...431..993V}). IRS5 was discovered to
harbour at least seven very compact radio continuum sources within a
radius of 0.03\,pc (Claussen et al. 1994\nocite{1994ApJ...424L..41C}). A number of these 
are likely to be shock excited clumps in the surrounding molecular
material, driven by embedded OB stars (Wilson et
al. 2003\nocite{2003ApJ...597..434W};van der Tak et
al. 2005\nocite{2005A&A...431..993V}). Water masers in IRS5 have been
shown to trace two outflows, both in roughly a North-South direction
(Imai et al. 2000\nocite{2000ApJ...538..751I}), a direction similar to the overall CO
outflow of the region.\newline
{\it Mid-IR morphology:}
The main 24.5\,\micron component in W3 region is the bright source IRS5. The
double source at its centre is resolved, the Northern one is the brightest
of the two sources. IRS5 shows prominent diffuse emission that
surrounds the double source. It extends in both North-South and
East-West direction. IRS6 is a resolved source with the peak
emission somewhat offset to the West of centre. IRS7 is a diffuse,
cometary shaped emission region. IRS3 is resolved in various
24.5\,\micron patches of diffuse emission. An unidentified source is found
10\arcsec~North of IRS5.\newline
{\it Model results:} 
Data for W3\,IRS5 are from van der Tak et al. (2005), the compilation made
by Campbell et al. (1995)\nocite{1995ApJ...454..831C}, and the 10\,\micron spectrum 
presented in Willner et al. (1982). The (sub)mm data have been obtained with beamsizes between 15\arcsec~and 19\arcsec.

Shallow radial density models provide good fits to the SED, but fail to reproduce
the intensity profile. Models with $p=1.5$ powerlaw produce to much mid-IR flux requiring
a high optical depth, leading to a too deep silicate absorption feature. The best fitting
models require a $p=1.0$ powerlaw.

\begin{figure}[t]
  \center{
    \includegraphics[height=8.5cm,width=8.5cm]{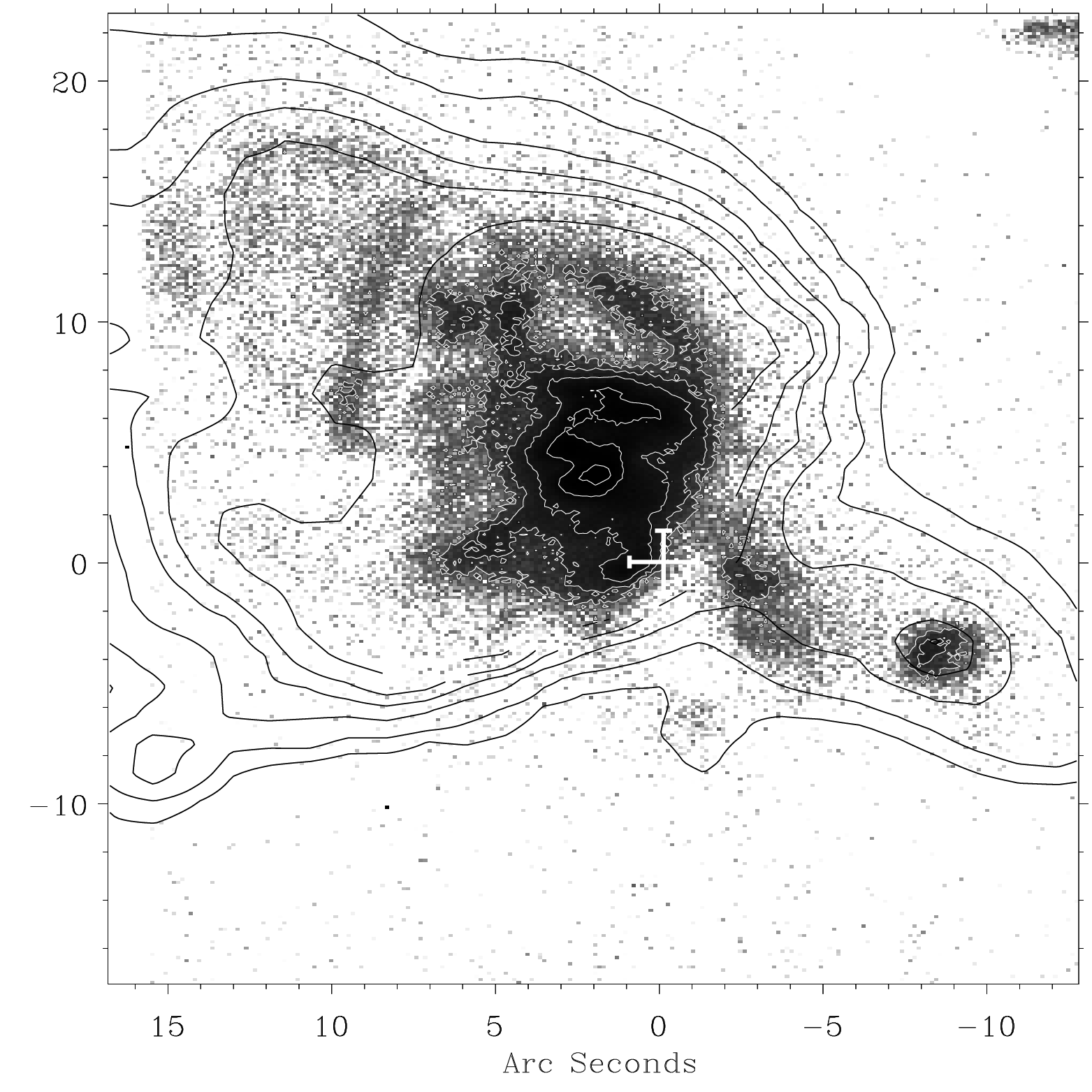}}
  \caption[]{COMICS 24.5\,\micron mosaic of the Cep A region. Dark contours represent {\it Spitzer} 8\,\micron emission. Centre of the coordinate system
corresponds to the phase centre of the VLA 7\,mm continuum map in Jim\'{e}nez-Serra et al. (2007): $\rm \alpha=22^{h}56^{m}18.0^{s} ,\delta=62\degr01\arcmin49.5\arcsec$ (indicated by a white cross).
Contour levels of the 24.5\,\micron emission are at 10\%, 15\%, 25\%, 50\%, and 80\% of peak flux
  density ($\rm 2.7\,10^{1}\,Jy\,arcsec^{-2}$). North is up, East is
  to the left.}
\label{cepa}
\end{figure}

\subsubsection{Cep A (Fig.\,\ref{cepa})}
{\it Description:}
Patel et al. (2005\nocite{2005Natur.437..109P}) reported a flattened HCN structure to be a circumstellar disk 
near the HW2 radio jet, the brightest radio source of the region (Hughes \& Wouterloot 1984\nocite{1984ApJ...276..204H}).
The radio jet is known to be the driving force of a large-scale molecular outflow 
in a North-East, South-West direction (G\'{o}mez et al. 1999\nocite{1999ApJ...514..287G}). Jim\'{e}nez-Serra
et al. (2007\nocite{2007ApJ...661L.187J}) showed that the HCN disk is resolved into a much smaller
disk and a hot core (Mart{\'{\i}}n-Pintado et al. 2005\nocite{2005ApJ...628L..61M}). The centre of 
their disk falls close to the centre of the ionized HW2 jet as determined in Curiel et al. (2006\nocite{2006ApJ...638..878C}).\newline
{\it Mid-IR morphology:}
The region shows a complex morphology comprising various arcs and patches.
The bright 24.5\micron emission corresponds to the bright near-IR reflection nebula that occupies the 
blue-shifted outflow cavity (Lenzen et al. 1984\nocite{1984A&A...137..202L}).  The COMICS image does not show any point source and we performed a cross matching
with a lower angular resolution, archival {\it Spitzer} 8\,\micron image in order to determine a rough astrometric solution. The result shown in 
Fig.\,\ref{cepa} is a reasonable match between 8 and 24.5\,$\mu$m. Note that the 8\,\micron emission
corresponding to the 24.5\,\micron emission peaks is saturated. HW2 appears to be hidden by high extinction just at the bottom 
tip of a dark lane. \newline 
{\it Model results:} Cep A is not modelled because direct radiation from the source is obscured at 24.5\,$\mu$m.

\section{Summary of the model fitting}
\label{summ}
We give a summary of the best-fitting models in Table\,\ref{modov}. Whether 
the best model reproduces certain important SED wavelength intervals and the 24.5\,\micron intensity profile has been indicated 
by tickmarks and crosses. In some cases no such assessment could be made due to
lack of data or data of insufficient quality. 
The table  makes clear that spherical models are capable of reproducing
at least part of the SED and intensity profile simultaneously. 
The models systematically fail to reproduce the short wavelength range, which is a well-known
shortcoming of spherical geometries (e.g. G\"{u}rtler et
al. 1991). 

A pattern emerges from Table\,\ref{modov} in which the objects that 
are satisfactorily reproduced by spherical models (S140
IRS1, AFGL\,2136, AFGL\,5180, Mon\,R2\,IRS3) are described by 
radial density distribution with a $p=1.0$ power index. Four objects
that require steeper ($p>1.0$) density distributions, {\it viz.} AFGL\,2591, NGC\,2264\,IRS1,
M8E-IR, S255\,IRS3, fail in reproducing
either the silicate absorption or the intensity profile.  These
objects seem to show excess mid-IR flux with respect to spherical
models that can only be accounted for if higher bolometric
luminosities are adopted, a solution denied by the
24.5\,\micron intensity profile. The MYSO that poses the biggest
problem is AFGL\,2591. Van der Tak et al. (2000) forwarded the idea
that the star's outflow activity is directed nearly along the
line-of-sight, and the problems with spherical models encountered here
could be a manifestation of this. Schreyer et al. (1997) have argued
for NGC\,2264\,IRS1 that again there is outflow activity along the
line-of-sight towards this source. M8E-IR shows evidence of a CO
outflow with the blue and red shifted components projected on top of
each other, arguing for a small inclination angle (Simon et
al. 1984\nocite{1984ApJ...278..170S}; see also Linz et al. 2008\nocite{2008ASPC..387..132L}). If these
objects were viewed down the outflow cavities, we would directly view
the warm dust at the base of the cavity and perhaps from the accretion
disk itself.
This extinction free view of the central regions would produce artificially steep laws from spherical models that cannot account
for the geometry.

The remaining objects have power indices that are shallower with power indices between $p=0.0$ and
$p=0.5$. The best-studied of these objects is IRAS\,20126. As discussed
in Sect.\,\ref{iras20126} this object is a case where there is clear
evidence that we are viewing the object close to edge-on. A dark lane
appears at wavelengths shorter than 20\,$\mu$m.  As a result, there will
still be significant optical depth at 24.5\,\micron which suppresses the
envelope emission. Instead the weaker, but more extended cavity wall
emission becomes significant. This leads to an apparent need for
shallower density profiles in our fits.

We therefore conclude that in those MYSO cases where we view the
objects at intermediate inclinations, the emission from the envelope 
dominates. Spherical models are then capable of reproducing
simultaneously the 24.5\,\micron intensity profile and the SED. For these cases, radial
density profiles with a power index of $p=1.0$ are preferred.

\section{Discussion}
\label{discus}

We have presented resolved 24.5\,\micron images of a sample of 14
MYSOs.  In most cases, the MYSO is discrete, single (within $<2\arcsec$) and has a
circularly symmetric profile on the sky.  Simultaneous modelling of
the spatial profile and the SED with simple spherical envelope models
shows that those objects that can be adequately modelled in this way have radial
density powerlaws $n \propto r^{-p}$ with exponent $p=1.0$.

A number of studies have analysed the density structure of MYSO (and UC\ion{H}{ii})
envelopes by means of 1-D radiative transfer modelling. The approaches consist
in reproducing the observed SED and the submm intensity profiles (Mueller et al. 2002; Beuther et al. 2002; 
Williams et al. 2005) and molecular line emission (van der Tak et al. 2000;
Hatchell \& van der Tak 2003).
A few important differences exist between these studies and the present one.
First, the COMICS observations provide information on scales about ten
times smaller, i.e. 1000\,AU. These scales could correspond to the transition region 
between the material reservoir (the envelope) and a putative accretion disk.
Secondly, the above mentioned studies ease or ignore spectral constraints at
wavelengths shorter than 100\,$\mu$m. In contrast, our approach is to
be more cautious with the IRAS measurements if they cannot be
reproduced by any spherical model, but keep the mid-IR and submm measurements.  This assumption derives from the
likelihood of source confusion and background contamination within the
large IRAS beam, and our experience with {\it Spitzer} 70\micron fluxes 
which are often significantly lower than the interpolated IRAS fluxes, even
after correction for detector non-linearity effects.
In practice, this resulted in somewhat lower bolometric luminosities than 
usually adopted for our MYSO sample, primarily caused by the 
luminosity sensitive intensity profile. 

As far as the radial density distribution is concerned, we find a
preference for $p=1.0$ models, as discussed in Sect.\,\ref{summ}. 
The present analysis shows that radial profiles as steep as $p=2.0$ are
incompatible with the observations, and even less steep
$p=1.5$ cannot always be justified. Our values are consistent with the ones
found by van der Tak et al. (2000) from the submm dust continuum and
molecular line emission, albeit on much larger size scales. These authors 
find a range of $1.0-1.5$ for the powerlaw index.  Williams et al. (2005) draw a similar 
conclusion for a large sample of candidate MYSOs.
They find an average value of the powerlaw index of
$1.3\pm0.4$. However, one should note that the models in the latter study require very high optical depths
(generally optically thick at 100\,$\mu$m), which is a factor of a few
to an order of magnitude larger than derived by us. It goes to show
that the DUSTY models presented by Williams et al. would probably
never fit the silicate feature nor the mid-IR part of the SED, and
indeed they do not attempt to fit the mid-IR nor the far-IR data
points.  Mueller et al. (2002) claim a value for the power index ``p'' which is consistent with
the previous two but slightly higher: $1.8\pm0.4$. Importantly, they
argue that the van der Tak (2000) results would favour higher values
for the power index if they had convolved their models with the actual
telescope beam instead of a Gaussian profile. Finally, Beuther et al. (2002)
claim $1.6\pm0.5$ from powerlaw fits to the inner 32\arcsec~of their
resolved 1.2\,mm images.  In summary, the far-IR and submm
studies generally prefer somewhat steeper density profiles than the ones derived from
24.5\,$\mu$m images, recalling the caveats regarding the
van der Tak et al. (2000) and Williams et al. (2005) results. This apparent inconsistency 
can however be brought into agreement when allowing for the different spatial
scales probed. 

On small angular scales, various studies in the (sub)mm find
evidence for shallower density distributions or flat intensity
distributions, which are suggested to be due to an unresolved core or
possibly a collection of cores.  For example, Hatchell et al. (2000)
find better model fits to their submm intensity profiles of a number
of UC\ion{H}{ii} regions, when they include an unresolved, high optical depth, central
core.  Beuther et al. (2002) describe the 1.2\,mm intensity profiles
of a large sample of massive cores with models that require an
unresolved, inner, constant intensity distribution with radii
between $2\,000$--$60\,000$\,AU.
Van der Tak et al. (1999, 2000) conclude from their
interferometric mm data (probing $\sim$1000\,AU scales) that the
detected emission is caused by a compact structure with an estimated
angular size of 0.3\arcsec, and which is different from the larger scale spherical
envelope structure.

Compact substructures thus seem to be required in order to explain the (sub)mm observations.
Flat intensity distributions are suggestive of an optically thick dust component, 
such as a dense shell or a disk (van der Tak et al. 2000), or a collection of subcores created by the 
ongoing fragmentation of the large-scale molecular core (Beuther et al. 2002).
In the latter case, the low angular resolution single dish submm observations would find 
the integrated emission of the various subcores to be equivalent to 
a shallow or a flat density distribution. 
It can therefore be argued that the flattening of the radial density
distribution seen in the submm close to the unresolved central source
is consistent with the relatively shallow density distribution as
traced by our high-resolution 24.5\,\micron images.

The images are generally dominated by one single, resolved
principal source, identified with the MYSO, which has a circular
appearance on the sky. Only in a small minority of cases do we find
more than one condensation on scales $\lta 2\arcsec$. This
observation goes against the idea of fragmentation as the
main cause of the flattening of the inner density profile. Rod\'{o}n et al. (2008)\nocite{2008A&A...490..213R}
present 0.35\arcsec~resolution mm images of the near-IR double source W3\,IRS5, in which 
the dominant component does not resolve into multiple mm sources. Given 
that this source is the most luminous source in our sample, any fragmentation 
into a cluster of protostars would be expected in particular in this source, but this is not 
observed. Instead, it seems that the 24.5\micron could indicate that it is 
the density distribution of the ambient core 
material itself that actually flattens out at
scales of an arcsecond. An alternative explanation why this could be so 
is rotation. The
density structure for a rotating and infalling envelope (TSC
envelopes, Terebey et al. 1984\nocite{1984ApJ...286..529T}) becomes on
average significantly shallower within the so-called centrifugal radius,
i.e. where rotational motion dominates over infall. The evolution of
the density distribution from relatively steep on large scales as seen by single-dish
the (sub)mm observations to shallow on smaller scales at shorter wavelengths could
be a manifestation of this effect.

\begin{table}
  {
    \begin{center}
      \caption[]{Overview of the quality of the simultaneous model fits to the SED and the intensity profile. The second column gives the power index of the 
radial density distribution corresponding to the best fitting model. A colon is added to uncertain power index values.
Model parameters are given in table\,\ref{tabtar}. A tickmark indicates a reasonable
correspondence between model and observations; an ``x'' means no correspondence, and ``..'' indicates that not enough data or only 
data of insufficient quality is available.}
      \begin{tabular}{llcccccc}
        \hline
        \hline
	MYSO         & $p$  &   4-8        & 9.7                 & 24.5       & 100         & mm  & Intensity \\
	             &    &  \micron     & \micron             & \micron    & \micron     &        &  profile      \\
	\hline 	     				
	S140\,IRS1   & 1.0  &     x        &      $\surd$        &  $\surd$   &  $\surd$    &$\surd$ &  $\surd$  \\
	AFGL2136     & 1.0  &     x        &      $\surd$        &  $\surd$   &   ..        &$\surd$ &  $\surd$ \\
	AFGL\,5180   & 1.0  &    ..        &      ..             &  $\surd$   &  $\surd$    &$\surd$ &  $\surd$   \\
	Mon R2\,IRS3 & 1.0  &     x        &      $\surd$        &  $\surd$   &   ..        &$\surd$ &  $\surd$   \\ 
	W3\,IRS5     & 1.0: &     x        &      $\surd$        &  $\surd$   &   x         &$\surd$ &  $\surd$   \\ 
	M8E-IR       & 1.25 &     x        &      x              &  $\surd$   &   ..        &$\surd$ &  $\surd$ \\
	S255\,IRS3   & 1.25 &     x        &      $\surd$        &  $\surd$   &   ..        &$\surd$ &  x       \\
	AFGL\,2591   & 1.5  &     x        &        x            &  $\surd$   &   ..        &$\surd$ &  x\\ 
	NGC\,2264\,IRS1& 1.5  &     x        &        x            &  $\surd$   &   ..        &$\surd$ &  $\surd$  \\
	IRAS\,20126  & 0.0  &     x        &      ..             &  $\surd$   &  $\surd$    &$\surd$ &  $\surd$     \\ 
	AFGL\,437S   & 0.0: &    ..        &      $\surd$        &  $\surd$   &   ..        &$\surd$ &  $\surd$     \\ 
	AFGL\,4029   & 0.0: &	 ..        &     ..              &  $\surd$   &  $\surd$    &$\surd$ &  $\surd$ \\
	AFGL\,961E   & 0.5: &	 $\surd$   &      $\surd$        &  $\surd$   &  $\surd$    &$\surd$ &  $\surd$ \\
	\hline
      \end{tabular}
      \label{modov}
    \end{center}
  }
  \label{qual}
\end{table}

The inability of models with spherical and smooth density
distributions to reproduce the near-IR and mid-IR part of the SED was
demonstrated again in the analysis presented in this paper. The
development of more sophisticated models (e.g. Yorke \& Sonnhalter 2002\nocite{2002ApJ...569..846Y}; Whitney et al. 2003\nocite{2003ApJ...591.1049W}; Indebetouw et
al. 2006\nocite{2006ApJ...636..362I}) is driven in part to reproduce this emission. Radiative transfer models that incorporate the
mentioned rotating TSC envelopes and implement a flared equatorial
accretion disk and a bipolar outflow cavity are presented in Whitney
et al. (2003).
The geometrical features are inspired mainly by detailed observations
of low-mass class I YSOs (Whitney et al. 2003; see also Tobin et
al. 2008\nocite{2008ApJ...679.1364T}) and may therefore be considered more realistic than the
simple spherical models we have adopted here.

Recent work has applied the more sophisticated envelope models to MYSO
SEDs and concluded that high-mass star formation proceeds similar to
low-mass star formation (e.g. De Buizer, Osorio \& Calvet 2005; Fazal
et al. 2007\nocite{2007arXiv0711.2261F}). Such work is becoming
popular especially after the publication of an extensive grid of SEDs
(Robitaille et al. 2006\nocite{2006ApJS..167..256R}) that are based on
the Whitney et al. models.  The rationale behind the Robitaille et
al. SED grid is that high-mass stars may form similarly to low-mass
stars, or at least that the SEDs of MYSOs are determined by similar
geometrical structures as found in low-mass objects.  Inferring 2D (or
3D) information from SED fits only is prone to be subjective and
misleading. The quoted conclusion therefore that massive SF is similar
to low-SF based on SED fits to such models is a circular argument.
What the Whitney et al. models in particular show is that one should be
careful interpreting SEDs, without any independent additional
data. Large scale spatial information of the dust emission in MYSOs can be retrieved from
the single-dish (sub)mm studies previously mentioned, intermediate
scales can be probed in the mid-IR as our study presented
here, and with (sub)mm interferometry. Finally, scales down to 100\,AU in the mid-IR can be reached
with mid-IR interferometry using e.g. the VLTI (de Wit et
al. 2007\nocite{2007ApJ...671L.169D}; Linz et al
2008\nocite{2008ASPC..387..132L}). Eventually the question of how a
massive star forms can only be settled with such 
spatially resolved information.

\section{Conclusions}
\label{concl}We have presented a study of resolved 24.5\,\micron images of 14
massive star forming regions. The images probe linear size scales of
1000\,AU. Emission at 24.5\,\micron is dominated by the MYSOs in these
regions.  They are discrete sources and most have a
circular profile to first order. In three cases we find multiple condensations embedded 
in a larger mid-IR envelope. Various regions display extended, diffuse emission.  
This emission is associated with UC\ion{H}{ii} regions (e.g. Mon R2, AFGL\,437,
AFGL\,2591) in which case the dust is heated by Ly$\alpha$ within the ionized zone. Shock excited material (e.g. S\,140) also 
seems to produce diffuse emission at 24.5\,$\mu$m.

Simple 1-D spherical model fits to the MYSO 24.5\,\micron spatial
profile and SED show that radial density powerlaws of the form
$n=n_{0}\,(r/r_{\rm subl})^{-p}$ with a power $p=1.0$ are preferred. When there is 
evidence that we are viewing the MYSO either face-on down the outflow cavity 
or edge-on through a torus we find steeper or  shallower density laws respectively .
These density laws are more likely to be due to the inadequacies of the 
spherical models in these cases than a true representation of a different density law, and 
we expect that $p=1.0$ also applies there to.

We find that the spatial profile of the dust emission on scales of 1000\,AU is
shallower than that from larger 10\,000\,AU scales probed by (sub)mm dust
emission. Inner flattenings seen in the submm are consistent with our
results here. This flattening is not likely to be due to fragmentation of the core,
but due to the actual distribution of the emitting material. This is
supported by the relatively small multiplicity of condensations seen
on sub-arcsecond scales. The continuous flattening from large to small scales 
could be the manifestation of rotation, but this requires further study.
The application of more sophisticated, multi-dimensional models in relation to the data
presented here and at even higher resolution using mid-IR
interferometric observations will be the subject of a future
paper. 

\begin{acknowledgements}
RDO is grateful for the support from the Leverhulme Trust for awarding
a Research Fellowship.  The authors would like to thank I. Jim\'{e}nez-Serra and E.R. Parkin for 
fruitful discussions, and an anonymous referee for valuable comments. The version of the ISO data presented in this
paper correspond to the Highly Processed Data Product (HPDP) by
W.F. Frieswijk et al., available for public use in the ISO Data
Archive.  This research has made use of the SIMBAD database and the
VizieR catalogue access tool, operated at CDS, Strasbourg, France. 
This research has made use of NASA's Astrophysics Data System. We acknowledge
the use of the RMS database, that can be found at URL \verb www.ast.leeds.ac.uk/RMS/ .
\end{acknowledgements}

\bibliographystyle{aa}

\begin{thebibliography}{158}
\expandafter\ifx\csname natexlab\endcsname\relax\def\natexlab#1{#1}\fi

\bibitem[{{Alvarez} {et~al.}(2004){Alvarez}, {Hoare}, {Glindemann}, \&
  {Richichi}}]{2004A&A...427..505A}
{Alvarez}, C., {Hoare}, M., {Glindemann}, A., \& {Richichi}, A. 2004, \aap,
  427, 505

\bibitem[{{Alvarez} \& {Hoare}(2005)}]{2005A&A...440..569A}
{Alvarez}, C. \& {Hoare}, M.~G. 2005, \aap, 440, 569

\bibitem[{{Aspin}(1998)}]{1998A&A...335.1040A}
{Aspin}, C. 1998, \aap, 335, 1040

\bibitem[{{Becker} \& {Fenkart}(1971)}]{1971A&AS....4..241B}
{Becker}, W. \& {Fenkart}, R. 1971, \aaps, 4, 241

\bibitem[{{Beckwith} {et~al.}(1976){Beckwith}, {Evans}, {Becklin}, \&
  {Neugebauer}}]{1976ApJ...208..390B}
{Beckwith}, S., {Evans}, II, N.~J., {Becklin}, E.~E., \& {Neugebauer}, G. 1976,
  \apj, 208, 390

\bibitem[{{Beichman} {et~al.}(1979){Beichman}, {Becklin}, \&
  {Wynn-Williams}}]{1979ApJ...232L..47B}
{Beichman}, C.~A., {Becklin}, E.~E., \& {Wynn-Williams}, C.~G. 1979, \apjl,
  232, L47

\bibitem[{{Beltr{\'a}n} {et~al.}(2006){Beltr{\'a}n}, {Brand}, {Cesaroni},
  {Fontani}, {Pezzuto}, {Testi}, \& {Molinari}}]{2006A&A...447..221B}
{Beltr{\'a}n}, M.~T., {Brand}, J., {Cesaroni}, R., {et~al.} 2006, \aap, 447,
  221

\bibitem[{{Beuther} {et~al.}(2002){Beuther}, {Schilke}, {Menten}, {Motte},
  {Sridharan}, \& {Wyrowski}}]{2002ApJ...566..945B}
{Beuther}, H., {Schilke}, P., {Menten}, K.~M., {et~al.} 2002, \apj, 566, 945

\bibitem[{{Campbell} {et~al.}(1995){Campbell}, {Butner}, {Harvey}, {Evans},
  {Campbell}, \& {Sabbey}}]{1995ApJ...454..831C}
{Campbell}, M.~F., {Butner}, H.~M., {Harvey}, P.~M., {et~al.} 1995, \apj, 454,
  831

\bibitem[{{Castelaz} {et~al.}(1985){Castelaz}, {Grasdalen}, {Hackwell},
  {Capps}, \& {Thompson}}]{1985AJ.....90.1113C}
{Castelaz}, M.~W., {Grasdalen}, G.~L., {Hackwell}, J.~A., {Capps}, R.~W., \&
  {Thompson}, D. 1985, \aj, 90, 1113

\bibitem[{{Cesaroni} {et~al.}(1999){Cesaroni}, {Felli}, {Jenness}, {Neri},
  {Olmi}, {Robberto}, {Testi}, \& {Walmsley}}]{1999A&A...345..949C}
{Cesaroni}, R., {Felli}, M., {Jenness}, T., {et~al.} 1999, \aap, 345, 949

\bibitem[{{Cesaroni} {et~al.}(1997){Cesaroni}, {Felli}, {Testi}, {Walmsley}, \&
  {Olmi}}]{1997A&A...325..725C}
{Cesaroni}, R., {Felli}, M., {Testi}, L., {Walmsley}, C.~M., \& {Olmi}, L.
  1997, \aap, 325, 725

\bibitem[{{Cesaroni} {et~al.}(2005){Cesaroni}, {Neri}, {Olmi}, {Testi},
  {Walmsley}, \& {Hofner}}]{2005A&A...434.1039C}
{Cesaroni}, R., {Neri}, R., {Olmi}, L., {et~al.} 2005, \aap, 434, 1039

\bibitem[{{Chini} {et~al.}(1986{\natexlab{a}}){Chini}, {Kreysa}, {Mezger}, \&
  {Gemuend}}]{1986A&A...154L...8C}
{Chini}, R., {Kreysa}, E., {Mezger}, P.~G., \& {Gemuend}, H.-P.
  1986{\natexlab{a}}, \aap, 154, L8

\bibitem[{{Chini} {et~al.}(1986{\natexlab{b}}){Chini}, {Kruegel}, \&
  {Kreysa}}]{1986A&A...167..315C}
{Chini}, R., {Kruegel}, E., \& {Kreysa}, E. 1986{\natexlab{b}}, \aap, 167, 315

\bibitem[{{Churchwell} {et~al.}(1990){Churchwell}, {Wolfire}, \&
  {Wood}}]{1990ApJ...354..247C}
{Churchwell}, E., {Wolfire}, M.~G., \& {Wood}, D.~O.~S. 1990, \apj, 354, 247

\bibitem[{{Claussen} {et~al.}(1994){Claussen}, {Gaume}, {Johnston}, \&
  {Wilson}}]{1994ApJ...424L..41C}
{Claussen}, M.~J., {Gaume}, R.~A., {Johnston}, K.~J., \& {Wilson}, T.~L. 1994,
  \apjl, 424, L41

\bibitem[{{Cohen}(1973)}]{1973ApJ...185L..75C}
{Cohen}, M. 1973, \apjl, 185, L75+

\bibitem[{{Cohen} {et~al.}(1999){Cohen}, {Walker}, {Carter}, {Hammersley},
  {Kidger}, \& {Noguchi}}]{1999AJ....117.1864C}
{Cohen}, M., {Walker}, R.~G., {Carter}, B., {et~al.} 1999, \aj, 117, 1864

\bibitem[{{Cohen} {et~al.}(1995){Cohen}, {Witteborn}, {Walker}, {Bregman}, \&
  {Wooden}}]{1995AJ....110..275C}
{Cohen}, M., {Witteborn}, F.~C., {Walker}, R.~G., {Bregman}, J.~D., \&
  {Wooden}, D.~H. 1995, \aj, 110, 275

\bibitem[{{Crampton} \& {Fisher}(1974)}]{1974QB4.V5v14n12...}
{Crampton}, D. \& {Fisher}, W.~A. 1974, {Spectroscopic observations of stars in
  HII regions Crampton and W. A. Fisher.} (Victoria : National Research Council
  of Canada, Radio and Electrical Engineering Division, Astrophysics Branch,
  1974.), 12--+

\bibitem[{{Curiel} {et~al.}(2006){Curiel}, {Ho}, {Patel}, {Torrelles},
  {Rodr{\'{\i}}guez}, {Trinidad}, {Cant{\'o}}, {Hern{\'a}ndez}, {G{\'o}mez},
  {Garay}, \& {Anglada}}]{2006ApJ...638..878C}
{Curiel}, S., {Ho}, P.~T.~P., {Patel}, N.~A., {et~al.} 2006, \apj, 638, 878

\bibitem[{{Dale} {et~al.}(2007){Dale}, {Gil de Paz}, {Gordon}, {Hanson},
  {Armus}, {Bendo}, {Bianchi}, {Block}, {Boissier}, {Boselli}, {Buckalew},
  {Buat}, {Burgarella}, {Calzetti}, {Cannon}, {Engelbracht}, {Helou},
  {Hollenbach}, {Jarrett}, {Kennicutt}, {Leitherer}, {Li}, {Madore}, {Martin},
  {Meyer}, {Murphy}, {Regan}, {Roussel}, {Smith}, {Sosey}, {Thilker}, \&
  {Walter}}]{2007ApJ...655..863D}
{Dale}, D.~A., {Gil de Paz}, A., {Gordon}, K.~D., {et~al.} 2007, \apj, 655, 863

\bibitem[{{De Buizer}(2007)}]{2007ApJ...654L.147D}
{De Buizer}, J.~M. 2007, \apjl, 654, L147

\bibitem[{{De Buizer} {et~al.}(2005{\natexlab{a}}){De Buizer}, {Osorio}, \&
  {Calvet}}]{2005ApJ...635..452D}
{De Buizer}, J.~M., {Osorio}, M., \& {Calvet}, N. 2005{\natexlab{a}}, \apj,
  635, 452

\bibitem[{{De Buizer} {et~al.}(2000){De Buizer}, {Pi{\~n}a}, \&
  {Telesco}}]{2000ApJS..130..437D}
{De Buizer}, J.~M., {Pi{\~n}a}, R.~K., \& {Telesco}, C.~M. 2000, \apjs, 130,
  437

\bibitem[{{De Buizer} {et~al.}(2005{\natexlab{b}}){De Buizer}, {Radomski},
  {Telesco}, \& {Pi{\~n}a}}]{2005ApJS..156..179D}
{De Buizer}, J.~M., {Radomski}, J.~T., {Telesco}, C.~M., \& {Pi{\~n}a}, R.~K.
  2005{\natexlab{b}}, \apjs, 156, 179

\bibitem[{{de Graauw} {et~al.}(1996){de Graauw}, {Haser}, {Beintema},
  {Roelfsema}, {van Agthoven}, {Barl}, {Bauer}, {Bekenkamp}, {Boonstra},
  {Boxhoorn}, {Cote}, {de Groene}, {van Dijkhuizen}, {Drapatz}, {Evers},
  {Feuchtgruber}, {Frericks}, {Genzel}, {Haerendel}, {Heras}, {van der Hucht},
  {van der Hulst}, {Huygen}, {Jacobs}, {Jakob}, {Kamperman}, {Katterloher},
  {Kester}, {Kunze}, {Kussendrager}, {Lahuis}, {Lamers}, {Leech}, {van der
  Lei}, {van der Linden}, {Luinge}, {Lutz}, {Melzner}, {Morris}, {van Nguyen},
  {Ploeger}, {Price}, {Salama}, {Schaeidt}, {Sijm}, {Smoorenburg}, {Spakman},
  {Spoon}, {Steinmayer}, {Stoecker}, {Valentijn}, {Vandenbussche}, {Visser},
  {Waelkens}, {Waters}, {Wensink}, {Wesselius}, {Wiezorrek}, {Wieprecht},
  {Wijnbergen}, {Wildeman}, \& {Young}}]{1996A&A...315L..49D}
{de Graauw}, T., {Haser}, L.~N., {Beintema}, D.~A., {et~al.} 1996, \aap, 315,
  L49

\bibitem[{{de Wit} {et~al.}(2007){de Wit}, {Hoare}, {Oudmaijer}, \&
  {Mottram}}]{2007ApJ...671L.169D}
{de Wit}, W.~J., {Hoare}, M.~G., {Oudmaijer}, R.~D., \& {Mottram}, J.~C. 2007,
  \apjl, 671, L169

\bibitem[{{de Wit} {et~al.}(2008){de Wit}, {Oudmaijer}, {Fujiyoshi}, {Hoare},
  {Honda}, {Kataza}, {Miyata}, {Okamoto}, {Onaka}, {Sako}, \&
  {Yamashita}}]{2008ApJ...685L..75D}
{de Wit}, W.~J., {Oudmaijer}, R.~D., {Fujiyoshi}, T., {et~al.} 2008, \apjl,
  685, L75

\bibitem[{{Deharveng} {et~al.}(1997){Deharveng}, {Zavagno}, {Cruz-Gonzalez},
  {Salas}, {Caplan}, \& {Carrasco}}]{1997A&A...317..459D}
{Deharveng}, L., {Zavagno}, A., {Cruz-Gonzalez}, I., {et~al.} 1997, \aap, 317,
  459

\bibitem[{{Dent} {et~al.}(1998){Dent}, {Matthews}, \&
  {Ward-Thompson}}]{1998MNRAS.301.1049D}
{Dent}, W.~R.~F., {Matthews}, H.~E., \& {Ward-Thompson}, D. 1998, \mnras, 301,
  1049

\bibitem[{{Di Francesco} {et~al.}(2008){Di Francesco}, {Johnstone}, {Kirk},
  {MacKenzie}, \& {Ledwosinska}}]{2008ApJS..175..277D}
{Di Francesco}, J., {Johnstone}, D., {Kirk}, H., {MacKenzie}, T., \&
  {Ledwosinska}, E. 2008, \apjs, 175, 277

\bibitem[{{Draine} \& {Lee}(1984)}]{1984ApJ...285...89D}
{Draine}, B.~T. \& {Lee}, H.~M. 1984, \apj, 285, 89

\bibitem[{{Evans} {et~al.}(1977){Evans}, {Beckwith}, \&
  {Blair}}]{1977ApJ...217..448E}
{Evans}, II, N.~J., {Beckwith}, S., \& {Blair}, G.~N. 1977, \apj, 217, 448

\bibitem[{{Evans} {et~al.}(1989){Evans}, {Mundy}, {Kutner}, \&
  {Depoy}}]{1989ApJ...346..212E}
{Evans}, II, N.~J., {Mundy}, L.~G., {Kutner}, M.~L., \& {Depoy}, D.~L. 1989,
  \apj, 346, 212

\bibitem[{{Evans} {et~al.}(1981){Evans}, {Slovak}, {Becklin}, {Beichman},
  {Gatley}, {Werner}, {Hildebrand}, {Keene}, \&
  {Whitcomb}}]{1981ApJ...244..115E}
{Evans}, II, N.~J., {Slovak}, M.~H., {Becklin}, E.~E., {et~al.} 1981, \apj,
  244, 115

\bibitem[{{Faison} {et~al.}(1998){Faison}, {Churchwell}, {Hofner}, {Hackwell},
  {Lynch}, \& {Russell}}]{1998ApJ...500..280F}
{Faison}, M., {Churchwell}, E., {Hofner}, P., {et~al.} 1998, \apj, 500, 280

\bibitem[{{Fazal} {et~al.}(2007){Fazal}, {Sridharan}, {Qiu}, {Robitaille},
  {Whitney}, \& {Zhang}}]{2007arXiv0711.2261F}
{Fazal}, F.~M., {Sridharan}, T.~K., {Qiu}, K., {et~al.} 2007, ArXiv e-prints

\bibitem[{{Forrest} \& {Shure}(1986)}]{1986ApJ...311L..81F}
{Forrest}, W.~J. \& {Shure}, M.~A. 1986, \apjl, 311, L81

\bibitem[{{Gear} {et~al.}(1988){Gear}, {Robson}, \&
  {Griffin}}]{1988MNRAS.231P..55G}
{Gear}, W.~K., {Robson}, E.~I., \& {Griffin}, M.~J. 1988, \mnras, 231, 55P

\bibitem[{{Gezari} {et~al.}(1999){Gezari}, {Pitts}, \&
  {Schmitz}}]{1999yCat.2225....0G}
{Gezari}, D.~Y., {Pitts}, P.~S., \& {Schmitz}, M. 1999, VizieR Online Data
  Catalog, 2225, 0

\bibitem[{{Ghosh} {et~al.}(2000){Ghosh}, {Iyengar}, {Karnik}, {Rengarajan},
  {Tandon}, \& {Verma}}]{2000BASI...28..515G}
{Ghosh}, S.~K., {Iyengar}, K.~V.~K., {Karnik}, A.~D., {et~al.} 2000, Bulletin
  of the Astronomical Society of India, 28, 515

\bibitem[{{Giannakopoulou} {et~al.}(1997){Giannakopoulou}, {Mitchell},
  {Hasegawa}, {Matthews}, \& {Maillard}}]{1997ApJ...487..346G}
{Giannakopoulou}, J., {Mitchell}, G.~F., {Hasegawa}, T.~I., {Matthews}, H.~E.,
  \& {Maillard}, J.-P. 1997, \apj, 487, 346

\bibitem[{{Gibb} \& {Hoare}(2007)}]{2007MNRAS.380..246G}
{Gibb}, A.~G. \& {Hoare}, M.~G. 2007, \mnras, 380, 246

\bibitem[{{G{\'o}mez} {et~al.}(1999){G{\'o}mez}, {Sargent}, {Torrelles}, {Ho},
  {Rodr{\'{\i}}guez}, {Cant{\'o}}, \& {Garay}}]{1999ApJ...514..287G}
{G{\'o}mez}, J.~F., {Sargent}, A.~I., {Torrelles}, J.~M., {et~al.} 1999, \apj,
  514, 287

\bibitem[{{Greenhill} {et~al.}(2004){Greenhill}, {Gezari}, {Danchi}, {Najita},
  {Monnier}, \& {Tuthill}}]{2004ApJ...605L..57G}
{Greenhill}, L.~J., {Gezari}, D.~Y., {Danchi}, W.~C., {et~al.} 2004, \apjl,
  605, L57

\bibitem[{{Guertler} {et~al.}(1991){Guertler}, {Henning}, {Kruegel}, \&
  {Chini}}]{1991A&A...252..801G}
{Guertler}, J., {Henning}, T., {Kruegel}, E., \& {Chini}, R. 1991, \aap, 252,
  801

\bibitem[{{Hackwell} {et~al.}(1982){Hackwell}, {Grasdalen}, \&
  {Gehrz}}]{1982ApJ...252..250H}
{Hackwell}, J.~A., {Grasdalen}, G.~L., \& {Gehrz}, R.~D. 1982, \apj, 252, 250

\bibitem[{{Harvey} {et~al.}(2000){Harvey}, {Butner}, {Colom{\'e}}, {Di
  Francesco}, \& {Smith}}]{2000ApJ...534..846H}
{Harvey}, P.~M., {Butner}, H.~M., {Colom{\'e}}, C., {Di Francesco}, J., \&
  {Smith}, B.~J. 2000, \apj, 534, 846

\bibitem[{{Harvey} {et~al.}(1977){Harvey}, {Campbell}, \&
  {Hoffmann}}]{1977ApJ...215..151H}
{Harvey}, P.~M., {Campbell}, M.~F., \& {Hoffmann}, W.~F. 1977, \apj, 215, 151

\bibitem[{{Hatchell} {et~al.}(2000){Hatchell}, {Fuller}, {Millar}, {Thompson},
  \& {Macdonald}}]{2000A&A...357..637H}
{Hatchell}, J., {Fuller}, G.~A., {Millar}, T.~J., {Thompson}, M.~A., \&
  {Macdonald}, G.~H. 2000, \aap, 357, 637

\bibitem[{{Hatchell} \& {van der Tak}(2003)}]{2003A&A...409..589H}
{Hatchell}, J. \& {van der Tak}, F.~F.~S. 2003, \aap, 409, 589

\bibitem[{{Hayashi} {et~al.}(1987){Hayashi}, {Hasegawa}, {Omodaka}, {Hayashi},
  \& {Miyawaki}}]{1987ApJ...312..327H}
{Hayashi}, M., {Hasegawa}, T., {Omodaka}, T., {Hayashi}, S.~S., \& {Miyawaki},
  R. 1987, \apj, 312, 327

\bibitem[{{Henning} {et~al.}(1992){Henning}, {Chini}, \&
  {Pfau}}]{1992A&A...263..285H}
{Henning}, T., {Chini}, R., \& {Pfau}, W. 1992, \aap, 263, 285

\bibitem[{{Henning} {et~al.}(1990){Henning}, {Pfau}, \&
  {Altenhoff}}]{1990A&A...227..542H}
{Henning}, T., {Pfau}, W., \& {Altenhoff}, W.~J. 1990, \aap, 227, 542

\bibitem[{{Henning} {et~al.}(2000){Henning}, {Schreyer}, {Launhardt}, \&
  {Burkert}}]{2000A&A...353..211H}
{Henning}, T., {Schreyer}, K., {Launhardt}, R., \& {Burkert}, A. 2000, \aap,
  353, 211

\bibitem[{{Herbst} \& {Racine}(1976)}]{1976AJ.....81..840H}
{Herbst}, W. \& {Racine}, R. 1976, \aj, 81, 840

\bibitem[{{Hoare}(2006)}]{2006ApJ...649..856H}
{Hoare}, M.~G. 2006, \apj, 649, 856

\bibitem[{{Hoare} \& {Franco}(2007)}]{2007dmsf.book...61H}
{Hoare}, M.~G. \& {Franco}, J. 2007, {Massive Star Formation} (Diffuse Matter
  from Star Forming Regions to Active Galaxies), 61--+

\bibitem[{{Hoare} {et~al.}(2007){Hoare}, {Kurtz}, {Lizano}, {Keto}, \&
  {Hofner}}]{2007prpl.conf..181H}
{Hoare}, M.~G., {Kurtz}, S.~E., {Lizano}, S., {Keto}, E., \& {Hofner}, P. 2007,
  in Protostars and Planets V, ed. B.~{Reipurth}, D.~{Jewitt}, \& K.~{Keil},
  181--196

\bibitem[{{Hoare} {et~al.}(1991){Hoare}, {Roche}, \&
  {Glencross}}]{1991MNRAS.251..584H}
{Hoare}, M.~G., {Roche}, P.~F., \& {Glencross}, W.~M. 1991, \mnras, 251, 584

\bibitem[{{Hofner} {et~al.}(2007){Hofner}, {Cesaroni}, {Olmi},
  {Rodr{\'{\i}}guez}, {Mart{\'{\i}}}, \& {Araya}}]{2007A&A...465..197H}
{Hofner}, P., {Cesaroni}, R., {Olmi}, L., {et~al.} 2007, \aap, 465, 197

\bibitem[{{Howard} {et~al.}(1994){Howard}, {Pipher}, \&
  {Forrest}}]{1994ApJ...425..707H}
{Howard}, E.~M., {Pipher}, J.~L., \& {Forrest}, W.~J. 1994, \apj, 425, 707

\bibitem[{{Howell} {et~al.}(1981){Howell}, {McCarthy}, \&
  {Low}}]{1981ApJ...251L..21H}
{Howell}, R.~R., {McCarthy}, D.~W., \& {Low}, F.~J. 1981, \apjl, 251, L21

\bibitem[{{Hughes} \& {Wouterloot}(1984)}]{1984ApJ...276..204H}
{Hughes}, V.~A. \& {Wouterloot}, J.~G.~A. 1984, \apj, 276, 204

\bibitem[{{Imai} {et~al.}(2000){Imai}, {Kameya}, {Sasao}, {Miyoshi}, {Deguchi},
  {Horiuchi}, \& {Asaki}}]{2000ApJ...538..751I}
{Imai}, H., {Kameya}, O., {Sasao}, T., {et~al.} 2000, \apj, 538, 751

\bibitem[{{Indebetouw} {et~al.}(2006){Indebetouw}, {Whitney}, {Johnson}, \&
  {Wood}}]{2006ApJ...636..362I}
{Indebetouw}, R., {Whitney}, B.~A., {Johnson}, K.~E., \& {Wood}, K. 2006, \apj,
  636, 362

\bibitem[{{Itoh} {et~al.}(2001){Itoh}, {Tamura}, {Suto}, {Hayashi}, {Murakawa},
  {Oasa}, {Nakajima}, {Kaifu}, {Kosugi}, {Usuda}, \&
  {Doi}}]{2001PASJ...53..495I}
{Itoh}, Y., {Tamura}, M., {Suto}, H., {et~al.} 2001, \pasj, 53, 495

\bibitem[{{Ivezic} \& {Elitzur}(1997)}]{1997MNRAS.287..799I}
{Ivezic}, Z. \& {Elitzur}, M. 1997, \mnras, 287, 799

\bibitem[{{Jaffe} {et~al.}(1984){Jaffe}, {Davidson}, {Dragovan}, \&
  {Hildebrand}}]{1984ApJ...284..637J}
{Jaffe}, D.~T., {Davidson}, J.~A., {Dragovan}, M., \& {Hildebrand}, R.~H. 1984,
  \apj, 284, 637

\bibitem[{{Jenness} {et~al.}(1995){Jenness}, {Scott}, \&
  {Padman}}]{1995MNRAS.276.1024J}
{Jenness}, T., {Scott}, P.~F., \& {Padman}, R. 1995, \mnras, 276, 1024

\bibitem[{{Jiang} {et~al.}(2008){Jiang}, {Tamura}, {Hoare}, {Yao}, {Ishii},
  {Fang}, \& {Yang}}]{2008ApJ...673L.175J}
{Jiang}, Z., {Tamura}, M., {Hoare}, M.~G., {et~al.} 2008, \apjl, 673, L175

\bibitem[{{Jim{\'e}nez-Serra} {et~al.}(2007){Jim{\'e}nez-Serra},
  {Mart{\'{\i}}n-Pintado}, {Rodr{\'{\i}}guez-Franco}, {Chandler}, {Comito}, \&
  {Schilke}}]{2007ApJ...661L.187J}
{Jim{\'e}nez-Serra}, I., {Mart{\'{\i}}n-Pintado}, J.,
  {Rodr{\'{\i}}guez-Franco}, A., {et~al.} 2007, \apjl, 661, L187

\bibitem[{{Kastner} {et~al.}(1992){Kastner}, {Weintraub}, \&
  {Aspin}}]{1992ApJ...389..357K}
{Kastner}, J.~H., {Weintraub}, D.~A., \& {Aspin}, C. 1992, \apj, 389, 357

\bibitem[{{Kastner} {et~al.}(1994){Kastner}, {Weintraub}, {Snell}, {Sandell},
  {Aspin}, {Hughes}, \& {Baas}}]{1994ApJ...425..695K}
{Kastner}, J.~H., {Weintraub}, D.~A., {Snell}, R.~L., {et~al.} 1994, \apj, 425,
  695

\bibitem[{{Kataza} {et~al.}(2000){Kataza}, {Okamoto}, {Takubo}, {Onaka},
  {Sako}, {Nakamura}, {Miyata}, \& {Yamashita}}]{2000SPIE.4008.1144K}
{Kataza}, H., {Okamoto}, Y., {Takubo}, S., {et~al.} 2000, in Society of
  Photo-Optical Instrumentation Engineers (SPIE) Conference Series, Vol. 4008,
  Society of Photo-Optical Instrumentation Engineers (SPIE) Conference Series,
  ed. M.~{Iye} \& A.~F. {Moorwood}, 1144--1152

\bibitem[{{Kessler} {et~al.}(1996){Kessler}, {Steinz}, {Anderegg}, {Clavel},
  {Drechsel}, {Estaria}, {Faelker}, {Riedinger}, {Robson}, {Taylor}, \&
  {Xim{\'e}nez de Ferr{\'a}n}}]{1996A&A...315L..27K}
{Kessler}, M.~F., {Steinz}, J.~A., {Anderegg}, M.~E., {et~al.} 1996, \aap, 315,
  L27

\bibitem[{{Klein} {et~al.}(2005){Klein}, {Posselt}, {Schreyer}, {Forbrich}, \&
  {Henning}}]{2005ApJS..161..361K}
{Klein}, R., {Posselt}, B., {Schreyer}, K., {Forbrich}, J., \& {Henning}, T.
  2005, \apjs, 161, 361

\bibitem[{{Kurtz} {et~al.}(1994){Kurtz}, {Churchwell}, \&
  {Wood}}]{1994ApJS...91..659K}
{Kurtz}, S., {Churchwell}, E., \& {Wood}, D.~O.~S. 1994, \apjs, 91, 659

\bibitem[{{Lada} \& {Gautier}(1982)}]{1982ApJ...261..161L}
{Lada}, C.~J. \& {Gautier}, III, T.~N. 1982, \apj, 261, 161

\bibitem[{{Lada} {et~al.}(1984){Lada}, {Thronson}, {Smith}, {Schwartz}, \&
  {Glaccum}}]{1984ApJ...286..302L}
{Lada}, C.~J., {Thronson}, Jr., H.~A., {Smith}, H.~A., {Schwartz}, P.~R., \&
  {Glaccum}, W. 1984, \apj, 286, 302

\bibitem[{{Ladd} {et~al.}(1993){Ladd}, {Deane}, {Sanders}, \&
  {Wynn-Williams}}]{1993ApJ...419..186L}
{Ladd}, E.~F., {Deane}, J.~R., {Sanders}, D.~B., \& {Wynn-Williams}, C.~G.
  1993, \apj, 419, 186

\bibitem[{{Lenzen} {et~al.}(1984{\natexlab{a}}){Lenzen}, {Hodapp}, \&
  {Reddmann}}]{1984A&A...137..365L}
{Lenzen}, R., {Hodapp}, K.-W., \& {Reddmann}, T. 1984{\natexlab{a}}, \aap, 137,
  365

\bibitem[{{Lenzen} {et~al.}(1984{\natexlab{b}}){Lenzen}, {Hodapp}, \&
  {Solf}}]{1984A&A...137..202L}
{Lenzen}, R., {Hodapp}, K.-W., \& {Solf}, J. 1984{\natexlab{b}}, \aap, 137, 202

\bibitem[{{Lester} {et~al.}(1986){Lester}, {Harvey}, {Joy}, \&
  {Ellis}}]{1986ApJ...309...80L}
{Lester}, D.~F., {Harvey}, P.~M., {Joy}, M., \& {Ellis}, Jr., H.~B. 1986, \apj,
  309, 80

\bibitem[{{Linz} {et~al.}(2008){Linz}, {Henning}, {Stecklum}, {Men'shchikov},
  {van Boekel}, {Follert}, \& {Feldt}}]{2008ASPC..387..132L}
{Linz}, H., {Henning}, T., {Stecklum}, B., {et~al.} 2008, in Astronomical
  Society of the Pacific Conference Series, Vol. 387, Massive Star Formation:
  Observations Confront Theory, ed. H.~{Beuther}, H.~{Linz}, \& T.~{Henning},
  132--+

\bibitem[{{Longmore} {et~al.}(2006){Longmore}, {Burton}, {Minier}, \&
  {Walsh}}]{2006MNRAS.369.1196L}
{Longmore}, S.~N., {Burton}, M.~G., {Minier}, V., \& {Walsh}, A.~J. 2006,
  \mnras, 369, 1196

\bibitem[{{Marsh} {et~al.}(2001){Marsh}, {Bloemhof}, {Koerner}, \&
  {Ressler}}]{2001ApJ...548..861M}
{Marsh}, K.~A., {Bloemhof}, E.~E., {Koerner}, D.~W., \& {Ressler}, M.~E. 2001,
  \apj, 548, 861

\bibitem[{{Mart{\'{\i}}n-Pintado} {et~al.}(2005){Mart{\'{\i}}n-Pintado},
  {Jim{\'e}nez-Serra}, {Rodr{\'{\i}}guez-Franco}, {Mart{\'{\i}}n}, \&
  {Thum}}]{2005ApJ...628L..61M}
{Mart{\'{\i}}n-Pintado}, J., {Jim{\'e}nez-Serra}, I.,
  {Rodr{\'{\i}}guez-Franco}, A., {Mart{\'{\i}}n}, S., \& {Thum}, C. 2005,
  \apjl, 628, L61

\bibitem[{{Massi} {et~al.}(1985){Massi}, {Felli}, \&
  {Simon}}]{1985A&A...152..387M}
{Massi}, M., {Felli}, M., \& {Simon}, M. 1985, \aap, 152, 387

\bibitem[{{Mathis} {et~al.}(1977){Mathis}, {Rumpl}, \&
  {Nordsieck}}]{1977ApJ...217..425M}
{Mathis}, J.~S., {Rumpl}, W., \& {Nordsieck}, K.~H. 1977, \apj, 217, 425

\bibitem[{{Meakin} {et~al.}(2005){Meakin}, {Hines}, \&
  {Thompson}}]{2005ApJ...634.1146M}
{Meakin}, C.~A., {Hines}, D.~C., \& {Thompson}, R.~I. 2005, \apj, 634, 1146

\bibitem[{{Menten} \& {van der Tak}(2004)}]{2004A&A...414..289M}
{Menten}, K.~M. \& {van der Tak}, F.~F.~S. 2004, \aap, 414, 289

\bibitem[{{Mezger} {et~al.}(1988){Mezger}, {Chini}, {Kreysa}, {Wink}, \&
  {Salter}}]{1988A&A...191...44M}
{Mezger}, P.~G., {Chini}, R., {Kreysa}, E., {Wink}, J.~E., \& {Salter}, C.~J.
  1988, \aap, 191, 44

\bibitem[{{Minchin} {et~al.}(1995){Minchin}, {Ward-Thompson}, \&
  {White}}]{1995A&A...298..894M}
{Minchin}, N.~R., {Ward-Thompson}, D., \& {White}, G.~J. 1995, \aap, 298, 894

\bibitem[{{Minier} {et~al.}(2005){Minier}, {Burton}, {Hill}, {Pestalozzi},
  {Purcell}, {Garay}, {Walsh}, \& {Longmore}}]{2005A&A...429..945M}
{Minier}, V., {Burton}, M.~G., {Hill}, T., {et~al.} 2005, \aap, 429, 945

\bibitem[{{Moffat} {et~al.}(1979){Moffat}, {Jackson}, \&
  {Fitzgerald}}]{1979A&AS...38..197M}
{Moffat}, A.~F.~J., {Jackson}, P.~D., \& {Fitzgerald}, M.~P. 1979, \aaps, 38,
  197

\bibitem[{{Mottram} {et~al.}(2007){Mottram}, {Hoare}, {Lumsden}, {Oudmaijer},
  {Urquhart}, {Sheret}, {Clarke}, \& {Allsopp}}]{2007A&A...476.1019M}
{Mottram}, J.~C., {Hoare}, M.~G., {Lumsden}, S.~L., {et~al.} 2007, \aap, 476,
  1019

\bibitem[{{Mueller} {et~al.}(2002){Mueller}, {Shirley}, {Evans}, \&
  {Jacobson}}]{2002ApJS..143..469M}
{Mueller}, K.~E., {Shirley}, Y.~L., {Evans}, II, N.~J., \& {Jacobson}, H.~R.
  2002, \apjs, 143, 469

\bibitem[{{Nakano} {et~al.}(2003){Nakano}, {Sugitani}, \&
  {Morita}}]{2003PASJ...55....1N}
{Nakano}, M., {Sugitani}, K., \& {Morita}, K. 2003, \pasj, 55, 1

\bibitem[{{Ossenkopf} \& {Henning}(1994)}]{1994A&A...291..943O}
{Ossenkopf}, V. \& {Henning}, T. 1994, \aap, 291, 943

\bibitem[{{Patel} {et~al.}(2005){Patel}, {Curiel}, {Sridharan}, {Zhang},
  {Hunter}, {Ho}, {Torrelles}, {Moran}, {G{\'o}mez}, \&
  {Anglada}}]{2005Natur.437..109P}
{Patel}, N.~A., {Curiel}, S., {Sridharan}, T.~K., {et~al.} 2005, \nat, 437, 109

\bibitem[{{Peretto} {et~al.}(2006){Peretto}, {Andr{\'e}}, \&
  {Belloche}}]{2006A&A...445..979P}
{Peretto}, N., {Andr{\'e}}, P., \& {Belloche}, A. 2006, \aap, 445, 979

\bibitem[{{Preibisch} {et~al.}(2001){Preibisch}, {Balega}, {Schertl}, {Smith},
  \& {Weigelt}}]{2001A&A...378..539P}
{Preibisch}, T., {Balega}, Y.~Y., {Schertl}, D., {Smith}, M.~D., \& {Weigelt},
  G. 2001, \aap, 378, 539

\bibitem[{{Preibisch} {et~al.}(2002){Preibisch}, {Balega}, {Schertl}, \&
  {Weigelt}}]{2002A&A...392..945P}
{Preibisch}, T., {Balega}, Y.~Y., {Schertl}, D., \& {Weigelt}, G. 2002, \aap,
  392, 945

\bibitem[{{Preibisch} {et~al.}(2003){Preibisch}, {Balega}, {Schertl}, \&
  {Weigelt}}]{2003A&A...412..735P}
{Preibisch}, T., {Balega}, Y.~Y., {Schertl}, D., \& {Weigelt}, G. 2003, \aap,
  412, 735

\bibitem[{{Rayner} \& {McLean}(1987)}]{1987iawa.conf..272R}
{Rayner}, J. \& {McLean}, I. 1987, in Infrared astronomy with arrays, ed. C.~G.
  {Wynn-Williams} \& E.~E. {Becklin}, 272--+

\bibitem[{{Richardson} {et~al.}(1985){Richardson}, {White}, {Gee}, {Griffin},
  {Cunningham}, {Ade}, \& {Avery}}]{1985MNRAS.216..713R}
{Richardson}, K.~J., {White}, G.~J., {Gee}, G., {et~al.} 1985, \mnras, 216, 713

\bibitem[{{Robitaille} {et~al.}(2006){Robitaille}, {Whitney}, {Indebetouw},
  {Wood}, \& {Denzmore}}]{2006ApJS..167..256R}
{Robitaille}, T.~P., {Whitney}, B.~A., {Indebetouw}, R., {Wood}, K., \&
  {Denzmore}, P. 2006, \apjs, 167, 256

\bibitem[{{Rod{\'o}n} {et~al.}(2008){Rod{\'o}n}, {Beuther}, {Megeath}, \& {van
  der Tak}}]{2008A&A...490..213R}
{Rod{\'o}n}, J.~A., {Beuther}, H., {Megeath}, S.~T., \& {van der Tak}, F.~F.~S.
  2008, \aap, 490, 213

\bibitem[{{Saito} {et~al.}(2006){Saito}, {Saito}, {Moriguchi}, \&
  {Fukui}}]{2006PASJ...58..343S}
{Saito}, H., {Saito}, M., {Moriguchi}, Y., \& {Fukui}, Y. 2006, \pasj, 58, 343

\bibitem[{{Saito} {et~al.}(2007){Saito}, {Saito}, {Sunada}, \&
  {Yonekura}}]{2007ApJ...659..459S}
{Saito}, H., {Saito}, M., {Sunada}, K., \& {Yonekura}, Y. 2007, \apj, 659, 459

\bibitem[{{Scarrott} \& {Warren-Smith}(1989)}]{1989MNRAS.237..995S}
{Scarrott}, S.~M. \& {Warren-Smith}, R.~F. 1989, \mnras, 237, 995

\bibitem[{{Schreyer} {et~al.}(1997){Schreyer}, {Helmich}, {van Dishoeck}, \&
  {Henning}}]{1997A&A...326..347S}
{Schreyer}, K., {Helmich}, F.~P., {van Dishoeck}, E.~F., \& {Henning}, T. 1997,
  \aap, 326, 347

\bibitem[{{Schreyer} {et~al.}(2003){Schreyer}, {Stecklum}, {Linz}, \&
  {Henning}}]{2003ApJ...599..335S}
{Schreyer}, K., {Stecklum}, B., {Linz}, H., \& {Henning}, T. 2003, \apj, 599,
  335

\bibitem[{{Shepherd} {et~al.}(2000){Shepherd}, {Yu}, {Bally}, \&
  {Testi}}]{2000ApJ...535..833S}
{Shepherd}, D.~S., {Yu}, K.~C., {Bally}, J., \& {Testi}, L. 2000, \apj, 535,
  833

\bibitem[{{Shuping} {et~al.}(2004){Shuping}, {Morris}, \&
  {Bally}}]{2004AJ....128..363S}
{Shuping}, R.~Y., {Morris}, M., \& {Bally}, J. 2004, \aj, 128, 363

\bibitem[{{Simon} {et~al.}(1984){Simon}, {Cassar}, {Felli}, {Fischer}, {Massi},
  \& {Sanders}}]{1984ApJ...278..170S}
{Simon}, M., {Cassar}, L., {Felli}, M., {et~al.} 1984, \apj, 278, 170

\bibitem[{{Simon} {et~al.}(1985){Simon}, {Peterson}, {Longmore}, {Storey}, \&
  {Tokunaga}}]{1985ApJ...298..328S}
{Simon}, M., {Peterson}, D.~M., {Longmore}, A.~J., {Storey}, J.~W.~V., \&
  {Tokunaga}, A.~T. 1985, \apj, 298, 328

\bibitem[{{Simon} \& {Dyck}(1977)}]{1977AJ.....82..725S}
{Simon}, T. \& {Dyck}, H.~M. 1977, \aj, 82, 725

\bibitem[{{Sloan} {et~al.}(2003){Sloan}, {Kraemer}, {Price}, \&
  {Shipman}}]{2003ApJS..147..379S}
{Sloan}, G.~C., {Kraemer}, K.~E., {Price}, S.~D., \& {Shipman}, R.~F. 2003,
  \apjs, 147, 379

\bibitem[{{Snell} \& {Bally}(1986)}]{1986ApJ...303..683S}
{Snell}, R.~L. \& {Bally}, J. 1986, \apj, 303, 683

\bibitem[{{Snell} {et~al.}(1988){Snell}, {Huang}, {Dickman}, \&
  {Claussen}}]{1988ApJ...325..853S}
{Snell}, R.~L., {Huang}, Y.-L., {Dickman}, R.~L., \& {Claussen}, M.~J. 1988,
  \apj, 325, 853

\bibitem[{{Sridharan} {et~al.}(2005){Sridharan}, {Williams}, \&
  {Fuller}}]{2005ApJ...631L..73S}
{Sridharan}, T.~K., {Williams}, S.~J., \& {Fuller}, G.~A. 2005, \apjl, 631, L73

\bibitem[{{Su} {et~al.}(2007){Su}, {Liu}, {Chen}, {Zhang}, \&
  {Cesaroni}}]{2007ApJ...671..571S}
{Su}, Y.-N., {Liu}, S.-Y., {Chen}, H.-R., {Zhang}, Q., \& {Cesaroni}, R. 2007,
  \apj, 671, 571

\bibitem[{{Tamura} {et~al.}(1991){Tamura}, {Gatley}, {Joyce}, {Ueno}, {Suto},
  \& {Sekiguchi}}]{1991ApJ...378..611T}
{Tamura}, M., {Gatley}, I., {Joyce}, R.~R., {et~al.} 1991, \apj, 378, 611

\bibitem[{{Tatebe} {et~al.}(2007){Tatebe}, {Hale}, {Wishnow}, \&
  {Townes}}]{2007ApJ...658L.103T}
{Tatebe}, K., {Hale}, D.~D.~S., {Wishnow}, E.~H., \& {Townes}, C.~H. 2007,
  \apjl, 658, L103

\bibitem[{{Terebey} {et~al.}(1984){Terebey}, {Shu}, \&
  {Cassen}}]{1984ApJ...286..529T}
{Terebey}, S., {Shu}, F.~H., \& {Cassen}, P. 1984, \apj, 286, 529

\bibitem[{{Thompson} {et~al.}(2006){Thompson}, {Hatchell}, {Walsh},
  {MacDonald}, \& {Millar}}]{2006A&A...453.1003T}
{Thompson}, M.~A., {Hatchell}, J., {Walsh}, A.~J., {MacDonald}, G.~H., \&
  {Millar}, T.~J. 2006, \aap, 453, 1003

\bibitem[{{Thompson} {et~al.}(1998){Thompson}, {Corbin}, {Young}, \&
  {Schneider}}]{1998ApJ...492L.177T}
{Thompson}, R.~I., {Corbin}, M.~R., {Young}, E., \& {Schneider}, G. 1998,
  \apjl, 492, L177+

\bibitem[{{Thronson} {et~al.}(1980){Thronson}, {Gatley}, {Harvey}, {Sellgren},
  \& {Werner}}]{1980ApJ...237...66T}
{Thronson}, Jr., H.~A., {Gatley}, I., {Harvey}, P.~M., {Sellgren}, K., \&
  {Werner}, M.~W. 1980, \apj, 237, 66

\bibitem[{{Thronson} {et~al.}(1983){Thronson}, {Lada}, {Smith}, {Glaccum},
  {Harper}, {Schwartz}, \& {Knowles}}]{1983ApJ...271..625T}
{Thronson}, Jr., H.~A., {Lada}, C.~J., {Smith}, H.~A., {et~al.} 1983, \apj,
  271, 625

\bibitem[{{Tobin} {et~al.}(2008){Tobin}, {Hartmann}, {Calvet}, \&
  {D'Alessio}}]{2008ApJ...679.1364T}
{Tobin}, J.~J., {Hartmann}, L., {Calvet}, N., \& {D'Alessio}, P. 2008, \apj,
  679, 1364

\bibitem[{{Tofani} {et~al.}(1995){Tofani}, {Felli}, {Taylor}, \&
  {Hunter}}]{1995A&AS..112..299T}
{Tofani}, G., {Felli}, M., {Taylor}, G.~B., \& {Hunter}, T.~R. 1995, \aaps,
  112, 299

\bibitem[{{Torrelles} {et~al.}(1992){Torrelles}, {Gomez}, {Anglada},
  {Estalella}, {Mauersberger}, \& {Eiroa}}]{1992ApJ...392..616T}
{Torrelles}, J.~M., {Gomez}, J.~F., {Anglada}, G., {et~al.} 1992, \apj, 392,
  616

\bibitem[{{Trinidad} {et~al.}(2003){Trinidad}, {Curiel}, {Cant{\'o}},
  {D'Alessio}, {Rodr{\'{\i}}guez}, {Torrelles}, {G{\'o}mez}, {Patel}, \&
  {Ho}}]{2003ApJ...589..386T}
{Trinidad}, M.~A., {Curiel}, S., {Cant{\'o}}, J., {et~al.} 2003, \apj, 589, 386

\bibitem[{{van der Tak} {et~al.}(2005){van der Tak}, {Tuthill}, \&
  {Danchi}}]{2005A&A...431..993V}
{van der Tak}, F.~F.~S., {Tuthill}, P.~G., \& {Danchi}, W.~C. 2005, \aap, 431,
  993

\bibitem[{{van der Tak} {et~al.}(1999){van der Tak}, {van Dishoeck}, {Evans},
  {Bakker}, \& {Blake}}]{1999ApJ...522..991V}
{van der Tak}, F.~F.~S., {van Dishoeck}, E.~F., {Evans}, II, N.~J., {Bakker},
  E.~J., \& {Blake}, G.~A. 1999, \apj, 522, 991

\bibitem[{{van der Tak} {et~al.}(2000){van der Tak}, {van Dishoeck}, {Evans},
  \& {Blake}}]{2000ApJ...537..283V}
{van der Tak}, F.~F.~S., {van Dishoeck}, E.~F., {Evans}, II, N.~J., \& {Blake},
  G.~A. 2000, \apj, 537, 283

\bibitem[{{Walker} {et~al.}(1990){Walker}, {Adams}, \&
  {Lada}}]{1990ApJ...349..515W}
{Walker}, C.~K., {Adams}, F.~C., \& {Lada}, C.~J. 1990, \apj, 349, 515

\bibitem[{{Walker}(1956)}]{1956ApJS....2..365W}
{Walker}, M.~F. 1956, \apjs, 2, 365

\bibitem[{{Ward-Thompson} {et~al.}(2000){Ward-Thompson}, {Zylka}, {Mezger}, \&
  {Sievers}}]{2000A&A...355.1122W}
{Ward-Thompson}, D., {Zylka}, R., {Mezger}, P.~G., \& {Sievers}, A.~W. 2000,
  \aap, 355, 1122

\bibitem[{{Weigelt} {et~al.}(2002){Weigelt}, {Balega}, {Preibisch}, {Schertl},
  \& {Smith}}]{2002A&A...381..905W}
{Weigelt}, G., {Balega}, Y.~Y., {Preibisch}, T., {Schertl}, D., \& {Smith},
  M.~D. 2002, \aap, 381, 905

\bibitem[{{Weintraub} \& {Kastner}(1996)}]{1996ApJ...458..670W}
{Weintraub}, D.~A. \& {Kastner}, J.~H. 1996, \apj, 458, 670

\bibitem[{{Whitney} {et~al.}(2003){Whitney}, {Wood}, {Bjorkman}, \&
  {Wolff}}]{2003ApJ...591.1049W}
{Whitney}, B.~A., {Wood}, K., {Bjorkman}, J.~E., \& {Wolff}, M.~J. 2003, \apj,
  591, 1049

\bibitem[{{Williams} {et~al.}(2005){Williams}, {Fuller}, \&
  {Sridharan}}]{2005A&A...434..257W}
{Williams}, S.~J., {Fuller}, G.~A., \& {Sridharan}, T.~K. 2005, \aap, 434, 257

\bibitem[{{Willner} {et~al.}(1982){Willner}, {Gillett}, {Herter}, {Jones},
  {Krassner}, {Merrill}, {Pipher}, {Puetter}, {Rudy}, {Russell}, \&
  {Soifer}}]{1982ApJ...253..174W}
{Willner}, S.~P., {Gillett}, F.~C., {Herter}, T.~L., {et~al.} 1982, \apj, 253,
  174

\bibitem[{{Wilson} {et~al.}(2003){Wilson}, {Boboltz}, {Gaume}, \&
  {Megeath}}]{2003ApJ...597..434W}
{Wilson}, T.~L., {Boboltz}, D.~A., {Gaume}, R.~A., \& {Megeath}, S.~T. 2003,
  \apj, 597, 434

\bibitem[{{Wolfire} \& {Churchwell}(1994)}]{1994ApJ...427..889W}
{Wolfire}, M.~G. \& {Churchwell}, E. 1994, \apj, 427, 889

\bibitem[{{Wynn-Williams}(1982)}]{1982ARA&A..20..587W}
{Wynn-Williams}, C.~G. 1982, \araa, 20, 587

\bibitem[{{Wynn-Williams} {et~al.}(1981){Wynn-Williams}, {Becklin}, {Beichman},
  {Capps}, \& {Shakeshaft}}]{1981ApJ...246..801W}
{Wynn-Williams}, C.~G., {Becklin}, E.~E., {Beichman}, C.~A., {Capps}, R., \&
  {Shakeshaft}, J.~R. 1981, \apj, 246, 801

\bibitem[{{Wynn-Williams} {et~al.}(1972){Wynn-Williams}, {Becklin}, \&
  {Neugebauer}}]{1972MNRAS.160....1W}
{Wynn-Williams}, C.~G., {Becklin}, E.~E., \& {Neugebauer}, G. 1972, \mnras,
  160, 1

\bibitem[{{Wynn-Williams} {et~al.}(1977){Wynn-Williams}, {Forster}, {Welch},
  {Wright}, {Matthews}, {Becklin}, \& {Neugebauer}}]{1977ApJ...211L..89W}
{Wynn-Williams}, C.~G., {Forster}, J.~R., {Welch}, W.~J., {et~al.} 1977, \apjl,
  211, L89+

\bibitem[{{Yorke} \& {Sonnhalter}(2002)}]{2002ApJ...569..846Y}
{Yorke}, H.~W. \& {Sonnhalter}, C. 2002, \apj, 569, 846

\bibitem[{{Zapata} {et~al.}(2001){Zapata}, {Rodr{\'{\i}}guez}, \&
  {Kurtz}}]{2001RMxAA..37...83Z}
{Zapata}, L.~A., {Rodr{\'{\i}}guez}, L.~F., \& {Kurtz}, S.~E. 2001, Revista
  Mexicana de Astronomia y Astrofisica, 37, 83

\bibitem[{{Zavagno} {et~al.}(1999){Zavagno}, {Lagage}, \&
  {Cabrit}}]{1999A&A...344..499Z}
{Zavagno}, A., {Lagage}, P.~O., \& {Cabrit}, S. 1999, \aap, 344, 499

\bibitem[{{Zubko} {et~al.}(1996){Zubko}, {Mennella}, {Colangeli}, \&
  {Bussoletti}}]{1996MNRAS.282.1321Z}
{Zubko}, V.~G., {Mennella}, V., {Colangeli}, L., \& {Bussoletti}, E. 1996,
  \mnras, 282, 1321

\end{thebibliography}

\end{document}